\documentclass[aps,prd,reprint,twocolumn,preprintnumbers,floatfix,nofootinbib]{revtex4-1}

\usepackage{booktabs}
\usepackage[english]{babel}
\usepackage{amsmath,amssymb,amsbsy,amstext, amsthm, simplewick}
\usepackage{hyperref}
\usepackage{graphicx}
\usepackage{amsfonts}
\usepackage{amssymb}
\usepackage[small]{caption}
\usepackage{upgreek}
\usepackage{framed}
\usepackage{enumitem}

\usepackage{booktabs}
\usepackage[english]{babel}
\usepackage{amsmath,amssymb,amsbsy,amstext, amsthm, simplewick}
\usepackage{wrapfig}
\usepackage{hyperref}
\usepackage{graphicx}
\usepackage{amsfonts}
\usepackage{amssymb}
\usepackage{subfig}
\usepackage{upgreek}
 \usepackage{exscale,relsize}
 \usepackage[makeroom]{cancel}
\usepackage{soul}
\usepackage{bbold}
\usepackage{lipsum}
\usepackage{mdframed}
\usepackage{mathtools}
\usepackage{tikz}
\usepackage{tikz-cd}
\usepackage[export]{adjustbox}

 \newcommand{\circled}[2][]{%
  \tikz[baseline=(char.base)]{%
    \node[shape = circle, draw, inner sep = 1pt]
    (char) {\phantom{\ifblank{#1}{#2}{#1}}};%
    \node at (char.center) {\makebox[0pt][c]{#2}};}}
\robustify{\circled}

\newcommand*\diff{\mathop{}\!\mathrm{d}}

\usepackage{colortbl}
\definecolor{lightgreen}{cmyk}{0.2, 0, 0.2, 0.2}
\definecolor{lightgray}{cmyk}{0.1,0.2,0,0.1}
\definecolor{lightgray2}{cmyk}{0.1,0.1,0,0.1}

\makeatletter
\newlength{\apb@width}
\newcommand{\autoparbox}[2][c]{\settowidth{\apb@width}{#2}\parbox[#1]{\apb@width}{#2}}

\newcommand{\Cen}[2]{%
  \ifmeasuring@
    #2%
  \else
    \makebox[\ifcase\expandafter #1\maxcolumn@widths\fi]{$\displaystyle#2$}%
  \fi
}
\makeatother

\newcommand{\beq}{\begin{equation}\begin{aligned}}
\newcommand{\eeq}{\end{aligned}\end{equation}}

\hypersetup{
    colorlinks,
    linkcolor={red!50!black},
    citecolor={blue!50!black},
    urlcolor={blue!80!black}
}

\def\beq{\begin{equation}}
\def\eeq{\end{equation}}

\def\Beq{\begin{equation}\begin{aligned}}
\def\Eeq{\end{aligned}\end{equation}}

\def\bea{\begin{eqnarray}}
\def\eea{\end{eqnarray}}

\def\beq{\begin{equation}}
\def\eeq{\end{equation}}
\def\bea{\begin{eqnarray}}
\def\eea{\end{eqnarray}}

\def\bp{\boldsymbol{p}}
\def\bP{\boldsymbol{P}}

\def\bK{\boldsymbol{K}}
\def\bx{\boldsymbol{x}}

\DeclareRobustCommand{\SkipTocEntry}[4]{}
\DeclareSymbolFont{extraup}{U}{zavm}{m}{n}
\DeclareMathSymbol{\varheart}{\mathalpha}{extraup}{86}
\DeclareMathSymbol{\vardiamond}{\mathalpha}{extraup}{87}

\begin{document}

\preprint{DESY-22-104}


\title{Scalar Dark Matter Production from Preheating and Structure Formation Constraints}
\author{Marcos A.~G.~Garcia$^{a}$}
\email{marcos.garcia@fisica.unam.mx}
\author{Mathias Pierre$^{b}$}
\email{mathias.pierre@desy.de}
\author{Sarunas Verner$^{c}$}
\email{nedzi002@umn.edu}
\vspace{0.5cm}

 \affiliation{
$^a$
Departamento de F\'isica Te\'orica, Instituto de F\'isica, Universidad Nacional Aut\'onoma de M\'exico, Ciudad de M\'exico C.P. 04510, Mexico} 
\affiliation{$^b$ Deutsches Elektronen-Synchrotron DESY, Notkestr. 85, 22607 Hamburg, Germany} 
\affiliation{${}^c$ William I. Fine Theoretical Physics Institute, School of Physics and Astronomy, University of Minnesota, Minneapolis, MN 55455, USA}

\date{\today}

\begin{abstract} 
We investigate the out-of-equilibrium production of scalar dark matter (DM) from the inflaton condensate during inflation and reheating. We assume that this scalar couples only to the inflaton via a direct quartic coupling and is minimally coupled to gravity. We consider all possible production regimes: purely gravitational, weak direct coupling (perturbative), and strong direct coupling (non-perturbative). For each regime, we use different approaches to determine the dark matter phase space distribution and the corresponding relic abundance. For the purely gravitational regime, scalar dark matter quanta are copiously excited during inflation resulting in an infrared (IR) dominated distribution function and a relic abundance which overcloses the universe for a reheating temperature $T_\text{reh}>34 ~\text{GeV}$. A non-vanishing direct coupling induces an effective DM mass and suppresses the large IR modes in favor of ultraviolet (UV) modes and a minimal scalar abundance is generated when the interference between the direct and gravitational couplings is maximal. For large direct couplings, backreaction on the inflaton condensate is accounted for by using the Hartree approximation and lattice simulation techniques. Since scalar DM candidates can behave as non-cold dark matter, we estimate the impact of such species on the matter power spectrum and derive the corresponding constraints from the Lyman-$\alpha$ measurements. We find that they correspond to a lower bound on the DM mass of $\gtrsim 3\times 10^{-4} \, \rm{eV}$ for purely gravitational production, and $\gtrsim 20 \, \rm {eV}$ for direct coupling production. We discuss the implications of these results.

\end{abstract}

\newpage

\maketitle

\renewcommand{\baselinestretch}{0.9}\normalsize
\tableofcontents
\renewcommand{\baselinestretch}{1.0}\normalsize

\section{Introduction}

\subsection{Motivation}
The existence of dark matter (DM) has been known since 1933, when F.~Zwicky studied the motion of galaxies in the Coma Cluster and discovered that it contained substantially more invisible dark matter than luminous matter~\cite{Zwicky:1933gu}. Zwicky's dark matter hypothesis was further confirmed by Babcock in 1939~\cite{1939LicOB..19...41B}, and by Rubin and Ford in 1970~\cite{Rubin:1970zza} from the measured rotation curve of the Andromeda galaxy. Although the existence of dark matter has been known for almost nine decades, there is still no consensus on its nature.

One of the most popular and well-studied candidates for cold dark matter (CDM) is the weakly interacting massive particle (WIMP). According to the standard WIMP paradigm, dark matter would have been produced early in the universe through thermal scattering, and subsequently decoupled from the thermal bath through the \textit{freeze-out} mechanism. Typical WIMP masses are expected to be around the electroweak scale $(1\,{\rm{GeV}} - 1\,{\rm{TeV}})$ and these particles would interact with the Standard Model (SM) particles only through weak and gravitational interactions. However, the increasingly stringent limits from direct DM detection searches, such as XENON1T~\cite{XENON:2018voc}, LUX~\cite{LUX:2016ggv}, and PandaX~\cite{PandaX-II:2020oim}, and the absence of detection from indirect DM and collider experiments serve as a motivation to consider alternative models of dark matter~\cite{Arcadi:2017kky, Mambrini}.

Among the viable dark matter production processes is the so-called \textit{freeze-in} mechanism~\cite{Hall:2009bx,Bernal:2017kxu}. Typically, in such models, the coupling between the visible and dark sectors is assumed to be sufficiently small to ensure the out-of-equilibrium production of feebly interacting massive particles (FIMPs). This feeble coupling ensures that the current DM relic abundance $\Omega_{\rm DM}h^2 =0.1198 \pm 0.0012$, determined from the measurements of the cosmic microwave background (CMB) by the \textit{Planck} collaboration~\cite{Planck:2018vyg}, is saturated, while evading all current direct detection bounds.

In various scenarios, a suppression from a large ultraviolet (UV) scale have been invoked to justify such a feeble coupling with the visible sector, from constructions inspired by Grand Unification Theory (GUT) \cite{Mambrini:2013iaa,Bhattacharyya:2018evo} to high-scale supersymmetry \cite{Benakli:2017whb,Dudas:2017rpa,Dudas:2018npp,Dudas:2017kfz}. More recently, several works have explored the possibility for such a feeble coupling to arise from the gravitational interactions~\cite{Lee:2013bua,Kang:2020huh,Garny:2015sjg,Garny:2017kha,Bernal:2018qlk,Cembranos:2019qlm, Anastasopoulos:2020gbu,Brax:2020gqg,Brax:2021gpe,Mambrini:2021zpp, Haque:2021mab, Clery:2021bwz,Clery:2022wib}, identifying this UV scale with the Planck scale $M_P$. In these cases, dark matter production is very sensitive to the highest temperature achieved in the universe and to the (post-)inflationary dynamics of the universe. Since a typical coupling to the visible sector is significantly reduced compared to standard WIMP scenarios, these models are difficult to probe using conventional direct detection or collider experiments. However, one promising possibility to test such scenarios is to exploit the common properties for these dark matter candidates of possessing a potentially sizable free-streaming length. The impact on the matter power spectrum of such non-cold dark matter (NCDM) scenarios featuring a non-vanishing pressure component have been investigated in recent phenomenological studies, constraining such feebly coupled dark matter candidates~\cite{Bae:2017dpt,Heeck:2017xbu,Boulebnane:2017fxw,DEramo:2020gpr,Decant:2021mhj}. 

In this work we explore the out-of-equilibrium production of scalar DM in a minimal setup during reheating.\footnote{We refer to the epoch of reheating as the transition between inflation and the radiation domination era. We refer to the time of reheating as the instant when radiation starts to dominate the energy budget of the universe, i.e., the end of reheating epoch. Preheating denotes the process of non-perturbative particle production during the early epoch of reheating.} Namely, we invoke a dark sector, minimally coupled to gravity, that only directly couples to the inflaton field. This sector is responsible for the accelerated expansion of the universe, which explains the present-day homogeneous, isotropic, and spatially flat state on large physical scales. We consider {\em all} regimes for this coupling: purely gravitational, weak direct coupling (perturbative) and strong direct coupling (non-perturbative). To determine the relic abundance and structure formation constraints, we compute the phase space distribution (PSD) of DM using a variety of approaches, including the integration of the Boltzmann equation, the numerical solution of the quantum equation of motion in the Hartree approximation, and the integration over a configuration-space lattice of the full classical partial differential equations of motion for the inflaton+DM+gravity system (for gravity at the background level). We show that, depending on the strength of the inflaton-DM interaction, the bulk of the DM relic abundance can be produced before, at, or after the end of inflation.

We emphasize the assumption that the dark matter scalar $\chi$ does not directly participate in the population of a thermal bath, which ultimately occurs in a reheating process. We do not consider models where $\chi$ is part of the thermal bath nor do we assume that it acts as a catalyst in the population of the visible and dark sectors~\cite{Repond:2016sol,Fan:2021otj,Garcia:2021iag}. In all cases, we assume that the population of the visible sector occurs perturbatively without the bosonic parametric resonance. The decay of the inflaton into fermions is well suited for these purposes.

\subsection{Model}
\label{sec:model}
The usual cosmic inflation scenarios assume that at the end of the inflationary epoch, when the transition to a decelerated expansion occurs, the scalar inflaton field $\phi$ begins oscillating coherently about a minimum of the potential. Ultimately, the inflaton decays into elementary particles and the universe enters the reheating epoch. In this work, we explore the possibility of direct scalar dark matter production from the inflaton decay when the inflaton oscillates about a quadratic minimum of the potential. As stated in the Introduction, we explore this process in the gravitational, perturbative, and non-perturbative regimes. The dynamics of the $\phi$ sector during inflation are determined by the following action,
\beq
\label{eq:actionphi}
\mathcal{S}_{\phi} \;=\; \int \diff ^4x\,\sqrt{-g} \left[\frac{1}{2}(\partial_{\mu}\phi)^2 - V(\phi) \right] \, ,
\eeq
where $g \equiv \det(g_{\mu \nu})$ is the metric determinant. For definiteness, we consider an inflaton potential motivated by T-model $\alpha$-attractors~\cite{Kallosh:2013hoa,Garcia:2020wiy},\footnote{The results presented in this paper are not limited to T-model $\alpha$-attractors, and generically apply to the inflationary potentials with a quadratic minimum.}
\beq
\label{eq:phipotential}
V(\phi) \;=\; \lambda M_P^{4} \left[ \sqrt{6}\tanh\left(\frac{\phi}{\sqrt{6}M_P}\right)\right]^2 \, ,
\eeq
where $M_P = 1/\sqrt{8\pi G_N}\simeq 2.435\times 10^{18}\,{\rm GeV}$ is the reduced Planck mass ($G_N$ is Newton's gravitational constant) and $\lambda$ is the normalization constant that is determined from the amplitude of the cosmic microwave background (CMB) anisotropies~\cite{Planck:2018vyg, Planck:2018jri}. From the action~(\ref{eq:actionphi}), we find that the equation of motion for the homogeneous inflaton is given by
\begin{equation}
\label{eq:eomphia}
    \ddot{\phi} + 3H \dot{\phi} + \frac{\partial V}{\partial \phi} \; = \; 0 \, ,
\end{equation}
where $H \equiv \dot{a}/a$ is the Hubble parameter and $a$ is the cosmological scale factor. Here the dot represents the derivative with respect to cosmic time. If we assume that the background dynamics are solely determined by the motion of the homogeneous inflaton field, $\phi$, from the Friedmann equation we find
\beq
\label{eq:friedinf}
\rho_{\phi} \; \equiv \; \frac{1}{2}\dot{\phi}^2 + V(\phi) \; = \; 3H^2M_P^2  \, ,
\eeq
where $\rho_{\phi}$ is the energy density of the inflaton. Inflation ends when the expansion ceases to be accelerated, $\ddot{a}=0$, which is equivalent to the condition $\dot{\phi}_{\rm end}^2=V(\phi_{\rm end})$.\footnote{The end of inflation can also be defined in terms of the slow-roll parameter $\epsilon_H \equiv - \dot{H} /H^2 = 1$.} Here the subscript ``end" denotes the quantities at the end of inflation. Subsequently, during the inflaton oscillations about the minimum of $V(\phi)$, the effective scalar potential~(\ref{eq:phipotential}) can be approximated by
\begin{equation}
    \label{eq:oscillations}
    V(\phi) \; \simeq \; \lambda \phi^2 M_P^2 \;\equiv\; \frac{1}{2} m_{\phi}^2 \phi^2\,, \quad {\rm{with}} \quad \phi \ll M_P \, ,
\end{equation}
where the inflaton mass is defined as
\begin{equation}
    \label{eq:infmass}
    m_{\phi} \; \equiv \; \sqrt{V''(\phi)} \; = \; \sqrt{2 \lambda} M_P \, .
\end{equation}
A straightforward computation in the slow-roll approximation reveals that for T-models of inflation, the normalization constant $\lambda$ can be approximated as~\cite{Garcia:2020wiy},
\beq    
\label{eq:lambda}
\lambda \;\simeq\; \frac{18\pi^2 A_{S*}}{6N_{*}^2}\,.
\eeq
Here $A_{S_*}$ denotes the amplitude of the curvature power spectrum at the pivot scale $k_{*}$ and $N_{*}$ is the number of $e$-folds between the horizon exit at this scale and the end of inflation. For T-attractors, $N_{*}$ can be determined from the solution of the following equation~\cite{Ellis:2021kad},
\beq
\label{eq:nstarfull}
N_{*} \;\simeq\; 58.36 - \frac{1}{2}\ln N_{*} + \frac{1}{6}\ln\left(\frac{ \Gamma_{\phi}}{m_{\phi}}\right) - \frac{1}{12}\ln g_{\rm reh} \, .
\eeq
In this expression $\Gamma_{\phi}$ denotes the perturbative decay rate of the inflaton into the Standard Model radiation bath during reheating, and $g_{\rm{reh}}$ is the number of relativistic degrees of freedom of the Standard Model particles at the time of reheating. The magnitude of $\lambda$ is therefore determined by the duration of the reheating epoch, parametrized by the perturbative decay rate $\Gamma_{\phi}$. Nevertheless, as we show in this work, we can scale out the value of $\lambda$ from our results. Therefore, for definiteness, we use Eq.~(\ref{eq:lambda}) with the nominal choice of $N_* = 55$ $e$-folds and the amplitude of the curvature power spectrum at the {\em Planck} pivot scale $k_*=0.05\,{\rm Mpc}^{-1}$, $A_{S*}=2.1 \times 10^{-9}$~\cite{Planck:2018vyg,Planck:2018jri}. The normalization constant and the inflaton mass are then given by $\lambda \simeq 2.05 \times 10^{-11}$ and $m_{\phi} \simeq 1.56 \times 10^{13} \; \rm{GeV}$, respectively. Using these specific values, we evaluate the energy density at the end of inflation, 
\begin{equation}
    \label{eq:rhoend}
    \rho_{\rm{end}} \;=\; \frac{3}{2} V(\phi_{\rm end}) \;\simeq\; 7.1 \times 10^{62} \, {\rm{GeV}}^4\,, 
\end{equation}
with $\phi_{\rm{end}} \simeq 0.84 \, M_P$. During its oscillation, the inflaton field is characterized by the solution of Eq.~(\ref{eq:eomphia}), and can be parametrized as
\begin{equation}
   \phi(t)\simeq \phi_0(t) \cdot \mathcal{P}(t) \, ,
   \label{eq:phioft}
\end{equation}
where the envelope $\phi_0(t)$ incorporates the redshift and decay effects, and the periodic function $\mathcal{P}(t)$ encodes the short time-scale oscillation of $\phi$ about the minimum of the potential. As is well known in models with quadratic minima, combining Eqs.~(\ref{eq:eomphia}) and (\ref{eq:oscillations}), one finds that 
$\mathcal{P}(t)\simeq \cos  \left(m_\phi t \right)$ and the envelope scales as $\phi_0(a) \propto a^{-3/2}$ during reheating. The mean energy density and pressure averaged over one oscillation can be expressed as~\cite{Garcia:2020wiy, Turner:1983he}
\begin{align}   
    \label{eq:aveosc}
    \rho_\phi \,\simeq &\, \dfrac{1}{2} \langle \dot \phi^2 \rangle + \langle V(\phi) \rangle  \simeq V(\phi_0) \,, \\
    P_\phi \,\simeq &\, \dfrac{1}{2} \langle \dot \phi^2 \rangle - \langle V(\phi) \rangle \simeq  0 \,,
\end{align}
and the equation of state parameter is $w_{\phi} \equiv P_{\phi}/\rho_{\phi} \simeq 0$. From Eqs.~(\ref{eq:friedinf}), (\ref{eq:oscillations}), and (\ref{eq:aveosc}), we see that the inflaton energy density during reheating falls off as $\rho_{\phi}(a) \sim \langle \phi(t)^2 \rangle \propto a^{-3}$ and behaves as a matter-like component at the background level. 

In what follows, we assume that the decay of the inflaton into Standard Model particles occurs perturbatively through the two-body decay processes. When this is the case, the decay rate $\Gamma_{\phi}$ can be taken as constant in time.\footnote{Even for a quadratic inflaton potential and two-body decay processes, the decay rate may obtain a time-dependence if the effective mass of the decay products, which contains a term of the form $|y\phi|$, is large with respect to $m_{\phi}$. The effect is an initial suppression of the decay rate, with the kinematic blocking becoming irrelevant in the late stages of reheating. For further details see~\cite{Garcia:2021iag}. In Section~\ref{sec_scalarpart}, we show how to account for the kinematic blocking effect for the scalar decay product $\chi$.} As it is customary, we parametrize it in terms of an effective Yukawa-like coupling $y$,
\begin{equation}
    \Gamma_{\phi} \; \equiv \; \frac{y^2}{8 \pi} m_{\phi} \,, 
    \label{eq:defeffectiveYukawa}
\end{equation}
i.e.,~reheating proceeds through a two-body decay into a pair of fermions. By construction, this rate is dark sector-independent. Under these assumptions, the inflaton energy density and the radiation bath satisfy the following Friedmann-Boltzmann system of equations,
\begin{align}
\label{eq:phidecay1}
\dot{\rho}_{\phi} + 3H\rho_{\phi} \;&=\; -\Gamma_{\phi}\rho_{\phi}  \,,\\
\label{eq:phidecay2}
\dot{\rho}_R + 4H\rho_R  \;&=\; \Gamma_{\phi}\rho_{\phi} \,,\\ \label{eq:phidecay3}
\rho_{\phi} + \rho_R \;&=\; 3H^2 M_P^2\, .
\end{align}
Here $\rho_R$ denotes the radiation energy density. We use this system of differential equations to solve numerically for $\rho_{\phi}(t)$, $\rho_R(t)$, and $a(t)$. During reheating, the evolution of the energy density of the inflaton condensate can be found by integrating the continuity equation~(\ref{eq:phidecay1})~\cite{Turner:1983he},
\begin{equation}
    \label{eq:inflevo}
    \rho_{\phi}(t) \; = \; \rho_{\rm{end}} \left(\frac{a(t)}{a_{\rm{end}}} \right)^{-3} e^{-\Gamma_{\phi} (t - t_{\rm{end}})} \, .
\end{equation}
 
Regarding the dark sector, we consider a scalar dark matter field, $\chi$, that is minimally coupled to gravity and inflaton via a four-point dimensionless coupling, $\sigma$, characterized by the action
\beq
\label{eq:actionchi}
\mathcal{S}_{\chi} \; = \; \int \diff ^4x\, \sqrt{-g}  \left[\frac{1}{2}(\partial_{\mu}\chi)^2 - \frac{1}{2}\left(m_{\chi}^2 + \sigma\phi^2\right)\chi^2 \right] \, ,
\eeq
where $m_{\chi}$ is the bare mass of the dark particle. From this action, we immediately note that the inflaton condensate generates a time-dependent contribution to the effective dark matter mass, given by
\begin{equation}
    m_{\rm{eff}}^2(t) \; = \; m_{\chi}^2 + \sigma \phi(t)^2 \,.
    \label{eq:effectiveDMmass}
\end{equation}
In this work, we cover a wide range of values of the coupling $\sigma$, always assuming that the bare mass of $\chi$ is smaller than the mass of the inflaton, $m_{\chi} < m_{\phi}$. Therefore, the decay of $\phi$ into $\chi$ is kinematically allowed at least during part of its oscillation, when $m_{\rm{eff}}(t) < m_{\phi}$.

In the following sections, we derive the equation of motion for the field $\chi$ and discuss its production in the gravitational, perturbative, and non-perturbative regimes. We also derive the lower bounds on $m_{\chi}$ from the Lyman-$\alpha$ measurement of the matter power spectrum.

\subsection{Notation and conventions}
In this work, we extensively discuss the production of scalar dark matter particles, $\chi$, arising from the effective coupling $\frac{1}{2}\sigma\phi^2\chi^2$ in the action~(\ref{eq:actionchi}),
that occurs dominantly at early times right after inflation during the initial stage of reheating. To simplify our discussion, we follow the discussion of~\cite{Lesgourgues:2011rh,Ballesteros:2020adh} and introduce the re-scaled dimensionless comoving momentum
\beq
\label{eq:qdef}
q\;\equiv\; \frac{P}{T_{\star}}\left(\frac{a}{a_0}\right)\,,
\eeq
where $P$ denotes the magnitude of the physical momentum of $\chi$, $a$ is the scale factor, with $a_0$ being its present value, and
\beq    
    \label{eq:tstar}
    T_{\star} \;\equiv\; m_{\phi}\left(\frac{a_{\rm end}}{a_0}\right)\,,
\eeq
is a time-independent energy scale, convenient for characterizing the present-day dark matter abundance. Here $a_{\rm{end}} = a(t_{\rm{end}})$ is the scale factor at the end of inflation and the product $q \, T_{\star}$ is the comoving momentum. The number density of the produced $\chi$ particles is given by
\beq
\label{eq:psd1}
n_{\chi}(t) \;=\; \int \frac{\diff ^3\bP}{(2\pi)^3}\, f_{\chi}(P,t)\, ,
\eeq
where $f_{\chi}(P,t)$ is the phase space distribution (PSD) of the field $\chi$. Using Eqs.~(\ref{eq:qdef}) and (\ref{eq:tstar}), we can rewrite the above equation in terms of the re-scaled comoving momentum, $q$, 
\beq
\label{eq:comovingnchi}
n_{\chi}\left(\frac{a}{a_{\rm end}}\right)^3 \;=\; \frac{m_{\phi}^3}{2\pi^2}\int \diff q\,q^2 f_{\chi}(q,t)\,,
\eeq
where we have integrated over the angular variables of physical momentum $\bP$ that are irrelevant in the absence of spatial inhomogeneities. The quantity that appears in Eq.~(\ref{eq:comovingnchi}) corresponds to the comoving number density, which conveniently tracks the production and decoupling regimes for the dark matter (DM) particle, since it becomes constant once DM production ends. Similarly, we define the phase space distribution of the inflaton~\cite{Garcia:2020wiy,Nurmi:2015ema},
\beq
\label{eq:psdinflaton}
f_{\phi} (P, t) \; = \; (2\pi)^3 n_{\phi}(t) \delta^{(3)} (\bf{P})\,,
\eeq
where $n_{\phi}(t)$ is the instantaneous inflaton number density, and we assumed that $\phi$ is spatially homogeneous. To ensure that it is normalized correctly, the following expression
\beq
 \int \frac{\diff^3 \bf{P}}{(2 \pi)^3 n_{\phi}} f_{\phi} (P, t) \; = \; 1\,,
\eeq
must be satisfied. 

The structure of this paper is as follows: In Section \ref{sec_scalarpart} we review the perturbative and non-perturbative production of scalar dark matter. In Section \ref{sec:three} we discuss the dark matter production with small couplings $\sigma < \lambda$, whereas in Section~\ref{sec:largecoupling} we study particle production for large coupling $\sigma > \lambda$. In Section~\ref{sec:lymalpha} we impose the Lyman-$\alpha$ constraints on the mass of warm dark matter (WDM) candidates. In Section~\ref{sec:Neff} we explore the possibility for the scalar $\chi$ to constitute a dark radiation component and compute the corresponding effective number of relativistic species. In Section~\ref{sec:discussion} we discuss the generality of our results and we compare them with the literature. Our results are summarized in Section~\ref{sec:conclusions}. Appendix~\ref{app:A} contains two different perturbative dark matter production calculations: the first one treats the inflaton as a condensate, whereas the second treats the inflaton as a collection of particles. Throughout this paper, we use natural units $k_B = \hbar = c = 1$ and the metric signature $(+,-,-,-)$.

\section{Scalar particle production during inflation and reheating}
\label{sec_scalarpart}

In this section, we describe the excitation of the momentum modes of the dark scalar field in the expanding background determined by the inflationary dynamics. First, we show how to derive the phase space distribution of produced dark matter using a perturbative approach and calculate the resulting dark matter relic abundance. Second, we discuss the non-perturbative dark matter production, including the effects arising from the oscillating inflaton condensate and Ricci curvature.

\subsection{Perturbative production of scalar dark matter}\label{sec:boltzmann}
We begin by exploring the perturbative production of scalar dark matter from the oscillating inflaton condensate that is treated as a classical field. We assume that the condensate is homogeneous, decays perturbatively, and its phase space distribution is given by Eq. (\ref{eq:psdinflaton}). In the perturbative regime, the phase space distribution of the produced dark matter scalar field, $f_{\chi}$, can be found by integrating the Boltzmann transport equation,
\begin{widetext}
\begin{align}\notag 
    \frac{\partial f_{\chi}}{\partial t} - H|\bP|\frac{\partial f_{\chi}}{\partial |\bP|} \; = \; &\frac{1}{P^0} \int \frac{\diff ^3 \bK}{(2\pi)^3 n_{\phi}} \frac{\diff^3 \bP'}{(2\pi)^32P^{\prime 0}} (2\pi)^4 \delta^{(4)}(K-P-P')|\mathcal{\overline{M}}|_{\phi \rightarrow \chi \chi}^2\\ \displaybreak[0]
    &\qquad \qquad \times \Big[ f_{\phi}(K)(1 + f_{\chi}(P))(1 + f_{\chi}(P')) - f_{\chi}(P) f_{\chi}(P') (1 + f_{\phi}(K)) \Big]\; \\
    \; = \; &\frac{1}{P^0} \sum_{n=1}^{\infty} \int \frac{\diff ^3 \bK_n}{(2\pi)^3 n_{\phi}} \frac{\diff^3 \bP'}{(2\pi)^32P^{\prime 0}} (2\pi)^4 \delta^{(4)}(K_n-P-P')|\overline{\mathcal{M}_n}|^2\notag  \\ \label{eq:boltzfull}
    & \qquad \qquad \times \Big[ f_{\phi}(K)(1 + f_{\chi}(P))(1 + f_{\chi}(P')) - f_{\chi}(P) f_{\chi}(P') (1 + f_{\phi}(K)) \Big] \, .
\end{align}
\end{widetext}
Here $K_n = \left(E_n, \bf{0} \right)$ is the four-momentum of the inflaton condensate, where $E_n = n \, m_{\phi}$ denotes the energy of the $n$-th mode of oscillation, and $\overline{\mathcal{M}_n}$ is the transition amplitude in one oscillation for each oscillating field mode of $\phi$ from the coherent condensate state $|\phi\rangle$ to the two-particle final state $|\chi\chi\rangle$~\cite{Nurmi:2015ema,Garcia:2020wiy}.\footnote{Here the overbar indicates that the symmetry factors associated with identical initial and final states are included in the definition of the transition amplitude squared, $|\overline{\mathcal{M}_n}|^2$.} We note that a factor of $2$ was included on the right-hand side to account for two identical dark matter particles $\chi$ that are produced from a single inflaton decay.  Since we are considering a quadratic potential near its minimum~(\ref{eq:oscillations}), a factor of $\phi^2$ appears in the amplitude. This factor is in turn proportional to the square of the periodic function $\mathcal{P}(t)$ defined in~(\ref{eq:phioft}), and it can be expanded in terms of Fourier modes as $\mathcal{P}^2(t) \;=\; \sum_{n=1}^{\infty} \hat{\mathcal{P}}_n e^{-in\omega_{\phi} t} = \cos^2(m_{\phi} t)$. In this case, the transition amplitude squared is proportional to $\sum_{n = 1}^{\infty} |\overline{\mathcal{M}_n}|^2 \propto \sum_{n = 1}^{\infty} |\hat{\mathcal{P}}_n|^2$, and we find that only the second mode of oscillations $n = 2$ contributes to the sum~\cite{Garcia:2020wiy, Clery:2021bwz}. The detailed computation of the dark matter production amplitude is provided in Appendix~\ref{app:A}. 

Assuming that the inflaton number density $n_{\phi}$ is large, we can approximate the Bose enhancement factor for the inflaton as $1 + f_{\phi}(K) \simeq f_{\phi}(K)$, and the Boltzmann equation~(\ref{eq:boltzfull}) reduces to 
\begin{widetext}
\begin{align}\notag 
\frac{\partial f_{\chi}}{\partial t} - H|\bP|\frac{\partial f_{\chi}}{\partial |\bP|} \;&=\; \frac{1}{P^0} \int \frac{\diff^3\bP'}{(2\pi)^3 2P^{\prime 0}}(2\pi)^4 \delta(2m_{\phi} - P^0 - P^{\prime0})\delta^{(3)}(\bP+\bP') |\overline{\mathcal{M}_2}|^2 (1+f_{\chi}(P) + f_{\chi}(P'))\\ \label{eq:boltzalmostfull}
&=\; \frac{\pi |\overline{\mathcal{M}_2}|^2}{2m_{\phi}^2\beta(t)} \delta\left(|\bP|-m_{\phi}\beta(t)\right)\left(1+2f_{\chi}(|\bP|)\right)\, ,
\end{align}
\end{widetext}
where
\beq
\beta(t) \;\equiv\; \sqrt{1- \frac{m_{\rm{eff}}^2(t)}{m_{\phi}^2}}\,
\eeq
denotes the kinematic factor and $m_{\rm{eff}}(t)$ is the time-dependent effective dark matter mass given by Eq.~(\ref{eq:effectiveDMmass}). 
The quantum mechanical Bose-enhancement factor, which appears in the right side of (\ref{eq:boltzalmostfull}), can be scaled out by defining the ``classical'' distribution $f_{\chi}^c$, as follows~\cite{Moroi:2020has, Moroi:2020bkq, Ghosh:2022hen}:
\beq\label{eq:boltzfexp}
f_{\chi}(|\bP|, t) \;\equiv\; \frac{1}{2}\Big[ \exp\big(2 f_{\chi}^c(|\bP|,t)\big) - 1\Big]\,.
\eeq
Substitution into Eq.~(\ref{eq:boltzalmostfull}) leads indeed to the simplified Boltzmann equation
\beq
\frac{\partial f_{\chi}^c}{\partial t} - H|\bP|\frac{\partial f_{\chi}^c}{\partial |\bP|} \;=\; \frac{\pi |\overline{\mathcal{M}_2}|^2}{2m_{\phi}^2\beta(t)} \delta\left(|\bP|-m_{\phi}\beta(t)\right)\,,
\eeq
which has the solution~\cite{Garcia:2018wtq,Ballesteros:2020adh}
\begin{align}\notag
f_{\chi}^c(|\bP|, t) \;&=\; \frac{\pi}{2m_{\phi}^2}\int_{t_{\rm end}}^t \diff t' \,\frac{|\overline{\mathcal{M}_2(t')}|^2}{\beta(t')}\\ \notag 
&\qquad\qquad \times \delta \left( \frac{a(t)}{a(t')}|\bP| - m_{\phi} \beta(t') \right)\\ \label{eq:boltzfc}
\notag 
&=\; \frac{\pi}{2m_{\phi}^3}\int_{t_{\rm end}}^t \diff t' \,\frac{|\overline{\mathcal{M}_2(t')}|^2}{\beta(t')^2 H(t')}\, \delta \left( t' - \hat{t}\, \right)  \\ 
&=\; \frac{\pi}{2m_{\phi}^3} \frac{|\overline{\mathcal{M}_2(\hat{t})}|^2}{\beta(\hat{t})^2 H(\hat{t})} \theta(t - \hat{t}) \, \theta(\hat{t} - t_{\rm{end}})
\,,
\end{align}
where  $\hat{t}$, the time arising from the argument of the Dirac delta function, satisfies the following algebraic constraint,
\begin{equation}
    \frac{a(t)}{a(\hat{t}\,)} = \beta(\hat{t}\,)\frac{m_{\phi}}{|\bP|} \,.
\end{equation}
In the previous derivation, we disregarded the term $\beta'(t')$ since the classical phase space distribution is computed adiabatically in perturbation theory. In general, Eq.~(\ref{eq:boltzfc}) must be solved numerically due to the time-dependence of $\beta(t)$, which encodes the decreasing value of $\phi(t)$ during reheating. Nevertheless, we can find the approximate solution in the limit when the kinematic suppression is negligible, which corresponds to $m_{\chi}\ll m_{\phi}$ and $\sigma \ll (m_{\phi}/\phi_{\rm end})^2$. 
Assuming $\beta(t) \simeq 1$ and using the transition amplitude squared~(\ref{eq:M2squarred}), we find that the classical phase space distribution~(\ref{eq:boltzfexp}) becomes~\cite{Garcia:2021iag}
\begin{equation}
    \label{eq:psdfull2}
    f_{\chi}^c(|\bP|, \, t) \; = \; \frac{\pi \hat{\sigma}^2 \rho_{\phi}^2(\hat{t}\,)}{16m_{\phi}^7 H(\hat{t} \,)} \theta(t - \hat{t}) \, \theta(\hat{t} - t_{\rm{end}}) \, ,
\end{equation}
where the effective coupling $\hat{\sigma}$ is given by 
\beq\label{eq:sigmahat0}
\hat \sigma \; \equiv \;   \sigma - \lambda\left(1+ \dfrac{m_{\rm{eff}}^2}{2 m_\phi^2} \right) \,,
\eeq
also shown in Eq.~(\ref{eq:sigmahat}).{\footnote{Here the second term in $\hat{\sigma}$ is the purely gravitational perturbative contribution for the dark matter production process $\phi \, \phi \rightarrow \chi \, \chi$ invoking the exchange of a graviton. However, this perturbative approximation does not account for the contribution arising from superhorizon modes, which requires the non-perturbative formalism discussed in the next section.}} Note that in the case when $m_{\rm eff} \ll m_{\phi}$, the effective coupling can be approximated as $\hat{\sigma} \simeq \sigma - \lambda$, and destructive interference at tree level occurs when $\sigma = \lambda$.

During reheating, when $t_{\rm{end}} \ll t \ll \Gamma_{\phi}^{-1} \simeq t_{\rm{reh}}$, the inflaton oscillations redshift approximately as pressureless matter, $\rho_{\phi}(t) \propto \phi^2_0(t) \propto a(t)^{-3}$, and from Eq.~(\ref{eq:psdfull2}), we obtain the classical phase space distribution at early times:
\begin{align}\notag
f_{\chi}^c(|\bP|, t) \;&\simeq\; \frac{\pi \hat{\sigma}^2 \rho^2_{\phi}(t)}{16m_{\phi}^7 H(t)}\left(\frac{m_{\phi}}{|\bP|}\right)^{9/2} \theta(m_{\phi}-|\bP|)\\  \notag
&\qquad \times \theta\left(|\bP|-m_{\phi}\left(\frac{a_{\rm end}}{a(t)}\right) \right)\,, \\ \label{eq:fchiapp}
& \quad (t_{\rm{end}} \ll t\ll t_{\rm reh},~\beta\simeq 1)\,,
\end{align}
where we used $a(t)/a(\hat{t}\,) \simeq (t/\hat{t}\,)^{2/3}$ with $\rho_{\phi}^2(\hat{t}\,) \propto a(\hat{t} \,)^{-6}$ and $H(\hat{t}\,) \simeq H(t) (a(\hat{t}\,)/a(t))^{-3/2}$. Rewriting this solution in terms of the re-scaled comoving momentum, $q$, given by Eq.~(\ref{eq:qdef}), we obtain
\begin{align} \notag
f_{\chi}^{c}(q, t) \;&=\; \frac{\pi \hat{\sigma}^2 \rho_{\phi}^2(t)}{16m_{\phi}^7 H(t)}\left(\frac{a(t)}{a_{\rm end}}\right)^{9/2} q^{-9/2}\\ \notag
&\qquad \times \theta(q-1)\theta\left(\tfrac{a(t)}{a_{\rm end}}-q\right)\,, \\ \label{eq:fchiappq}
& \quad (t_{\rm{end}} \ll t\ll t_{\rm reh},\, \beta\simeq 1)\,.
\end{align}
During the inflaton oscillations $\rho_{\phi}^2(t)/H(t) \propto (a(t)/a_{\rm end})^{-9/2}$, and the prefactor of the phase space distribution rapidly tends to a constant value. Hence, we can write $f_{\chi}^{c}(q,t) \propto  q^{-9/2} \theta(q-1)\theta\left(\tfrac{a(t)}{a_{\rm end}}-q\right)$, which vanishes for $q<1$ and cascades toward large $q$ at the same rate as the scale factor grows, with a distribution tail that scales as $q^{-9/2}$.

\begin{figure*}[!t]
\centering
    \includegraphics[width=0.85\textwidth]{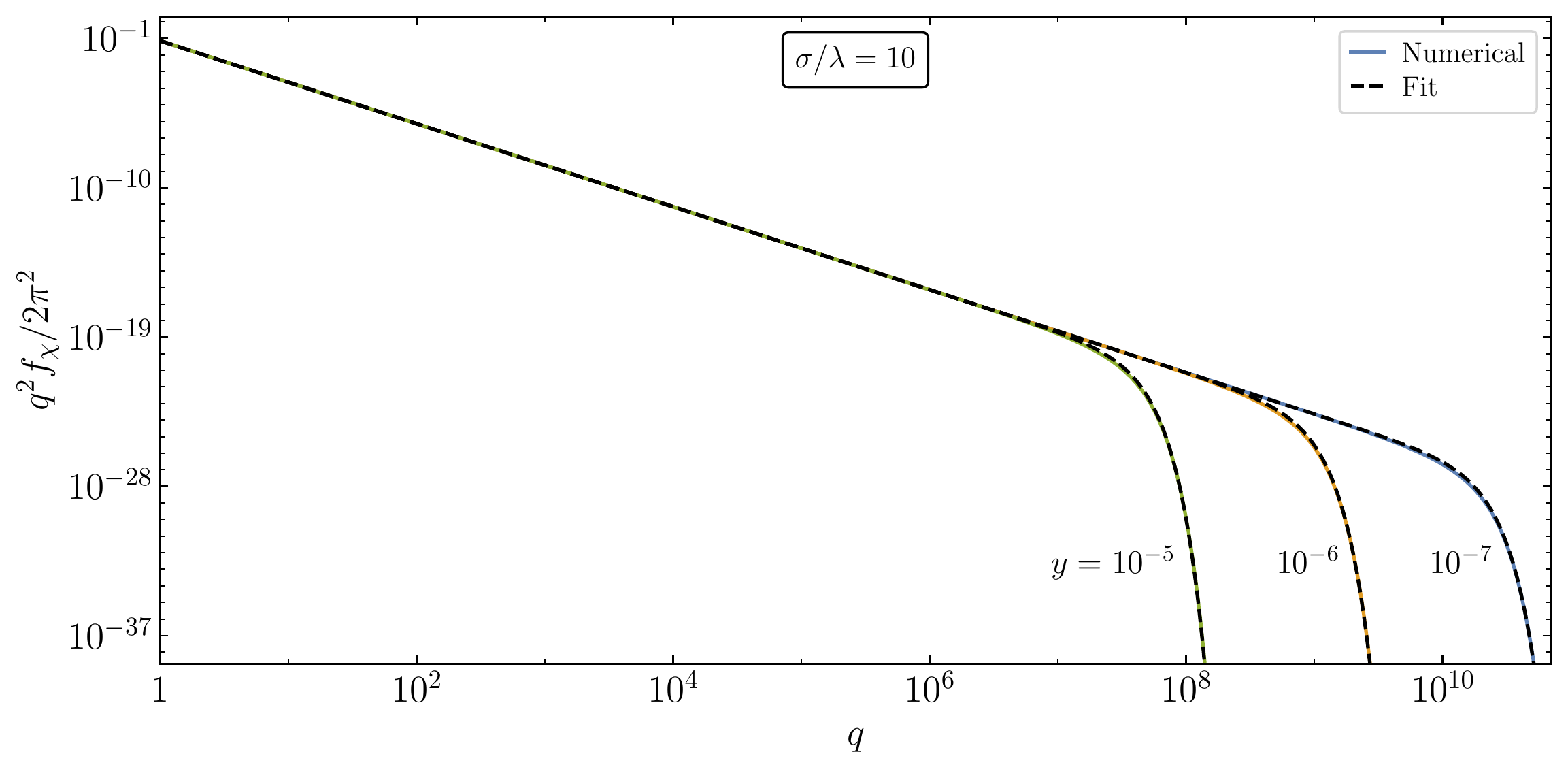}
    \caption{Perturbative phase space distribution of light scalar dark matter produced during reheating for the coupling $\sigma/\lambda=10$ and Yukawa-like couplings (defined by Eq.~(\ref{eq:defeffectiveYukawa})) $y = 10^{-5}, \, 10^{-6}$, and $10^{-7}$. The continuous curves correspond to the numerical solution of Eqs.~(\ref{eq:phidecay1})-(\ref{eq:phidecay3}) and (\ref{eq:boltzfc}) with $\beta \simeq 1$. The dashed curves are the fit approximation (\ref{eq:fchiappreh}), evaluated at the corresponding end of reheating scale factor. }
    \label{fig:PSDfluid}
\end{figure*}

To calculate the present dark matter relic abundance, one needs to extend the computation of the phase space distribution, and hence the number density, of the produced dark matter beyond the end of reheating. Reheating ends when the energy density of the inflaton is equal to the energy density of the produced radiation, $\rho_{\phi}(t_{\rm{reh}}) = \rho_R(t_{\rm{reh}})$, and we define the reheating temperature through~\cite{Garcia:2020wiy}
\begin{equation}
    \label{eq:reheating}
    \rho_R(t_{\rm{reh}}) \; = \; \frac{\pi^2 g_{\rm{reh}} T_{\rm{reh}}^4}{30} \; \equiv \; \frac{12}{25} \left(\Gamma_{\phi} M_P \right)^2 \, ,
\end{equation}
where $g_{\rm{reh}}$ denotes the effective number of relativistic degrees of freedom at reheating. The analytical expression (\ref{eq:fchiappq}) is valid only for $t \ll t_\text{reh}$ and cannot account for the transition between the matter and radiation domination era. For $t \sim t_{\rm{reh}} \simeq \Gamma_{\phi}^{-1}$, the exponential factor in Eq.~(\ref{eq:inflevo}) becomes important and the energy density of the inflaton decreases as $\rho_{\phi}(t)\propto \phi_0^2(t)\propto a^{-3}(t)e^{-\Gamma_{\phi}t}$. This results in an extra momentum dependence in the prefactor of Eq. (\ref{eq:fchiappq}), which has the form of a Gaussian tail, $\rho_{\phi}^2(\hat{t} \,) \propto e^{-\kappa(|\bP|/m_{\phi})^2}$~\cite{Ballesteros:2020adh}. Numerically solving the Friedmann-Boltzmann system of equations (\ref{eq:phidecay1})-(\ref{eq:phidecay3}), we find that the ``classical'' phase space distribution~(\ref{eq:psdfull2}) can be approximated as\footnote{Note that this PSD is a particular example of the generalized distribution $f_{\chi}(q, t) \propto q^{\alpha} \, {\rm{exp}} (- \beta q^{\gamma})$, which typically appears in connection with out-of-equilibrium particle production scenarios (see \cite{Ballesteros:2020adh} for more examples).} 

\begin{align} \notag
f_{\chi}^{c}(q, t) \;&\simeq\; \frac{\sqrt{3}\pi \hat{\sigma}^2 \rho_{\rm{end}}^{3/2} M_P}{16m_{\phi}^7 } q^{-9/2} e^{-1.56 \left(\frac{a_{\rm end}}{a_{\rm reh}}\right)^2 q^2}\\ \label{eq:fchiappreh}
&\qquad \times \theta(q-1)\, , \quad (t \gg t_{\rm reh},\, \beta\simeq 1)\,, 
\end{align}
which is time-independent and where the exponential factor is the Gaussian tail of the distribution. This fit
is in good agreement with the fully numerical solution as illustrated by Fig.~\ref{fig:PSDfluid}. The dashed black curves correspond to the fit (\ref{eq:fchiappreh}), whereas the solid lines show the numerical solutions for different Yukawa-like couplings $y = 10^{-5}, \, 10^{-6}$, and $10^{-7}$ with $\sigma/\lambda = 10$.

The comoving dark matter number density can be evaluated in a straightforward manner following Eq.~(\ref{eq:comovingnchi}). For $a_{\rm reh}\gg a_{\rm end}$, it does not depend on the Gaussian tail of the distribution, and results in
\beq
n_{\chi}\left(\frac{a}{a_{\rm end}}\right)^3 \;\simeq\; \frac{\sqrt{3} \rho_{\rm end}^{3/2} M_P}{48\pi m_{\phi}^4}  \hat{\sigma}^2\,.
\eeq
The present dark matter relic abundance is then given by
\beq\label{eq:OmegaBoltzmann}
\Omega_{\chi} \;=\; \frac{m_{\chi} n_{\chi}(a_0)}{\rho_c}\,, 
\eeq
where $\rho_c$ denotes the critical energy density at the present time. From the asymptotic value of the comoving number density (\ref{eq:OmegaBoltzmann}), we can evaluate $\Omega_{\chi}$ if the amount of expansion between the end of inflation and today, $a_0/a_{\rm end}$, is known. This quantity can be found from the standard computation~\cite{Martin:2010kz,Liddle:2003as,Ellis:2015pla}. Assuming entropy conservation after the end of reheating, we can write
\begin{align} \notag
\frac{a_0}{a_{\rm end}} \;&=\; \left(\frac{90}{\pi^2}\right)^{1/4}\left(\frac{11}{43}\right)^{1/3}g_{\rm reh}^{1/12}\left(\frac{H_{\rm end}M_P}{H_0^2}\right)^{1/2}\\ \label{eq:fulla0}
&\qquad \times \frac{H_0}{T_0} \frac{1}{R_{\rm rad}}\,,
\end{align}
where $T_0=2.7255\,{\rm K}$ and $H_0 = 100\,h \, {\rm km} \, {\rm s}^{-1}\,{\rm Mpc}^{-1}$ with $h\sim0.7$~\cite{Planck:2018vyg,Fixsen:2009ug}. Here $R_{\rm rad}$ denotes the so-called `reheating parameter'~\cite{Martin:2010kz}, for which the subscript `rad' denotes evaluation at the beginning of the radiation domination era (after the end of reheating).\footnote{In other words, quantities with a subscript `rad' are evaluated at $w\simeq 1/3$, with $w$ the equation of state parameter. On the other hand, quantities with a subscript `reh' are evaluated at inflaton-radiation equality, with $w\simeq 1/6$.} For perturbative reheating with a quadratic minimum of $V(\phi)$, $R_{\rm rad}$ can be approximated as follows~\cite{Ellis:2021kad,Garcia:2020eof}
\begin{align}\notag
R_{\rm rad} \;&\equiv\; \frac{a_{\rm end}}{a_{\rm rad}}\left(\frac{\rho_{\rm end}}{\rho_{\rm rad}}\right)^{1/4} \;\simeq\; \left(\frac{\Gamma_{\phi}}{H_{\rm end}}\right)^{1/6}\\ \label{eq:reheatingparameter}
& \simeq\; \left(\frac{5\pi^2 g_{\rm reh}}{72}\right)^{1/12} \left(\frac{T_{\rm reh}^2}{H_{\rm end}M_P}\right)^{1/6}\,,
\end{align}
where the first approximation is determined analytically, and for the final expression we used Eq.~(\ref{eq:reheating}). Substituting into (\ref{eq:OmegaBoltzmann}), including the scaling factor for the Hubble expansion rate $h$, we find
\begin{align} \notag \displaybreak[0]
\Omega_{\chi}h^2 \;&\simeq\; \frac{43\pi \hat{\sigma}^2 h^2}{3168\sqrt{30}} \frac{m_{\chi} T_{\rm reh} T_0^3 \rho_{\rm end}^{1/2}}{m_{\phi}^4 H_0^2 M_P }\\ \displaybreak[0]  \notag
& \simeq\; 0.12 \left(\frac{\lambda}{2.05\times 10^{-11}}\right)^{1/2} \left(\frac{m_{\chi}}{7\times 10^6\,{\rm GeV}}\right)\\ \label{eq:numdenchi1}
& \qquad \times \left(\frac{T_{\rm reh}}{10^{10}\,{\rm GeV}}\right) \left(\frac{\hat{\sigma}}{\lambda}\right)^2 \,.
\end{align}
In the second line, we have used the T-model of inflation. This result would be applicable for any value of the effective coupling $\hat{\sigma}$, as long as the effect of bosonic enhancement or kinematic blocking effects can be neglected. When they cannot be, the numerical solution of the Boltzmann equation would be sufficient to determine the corresponding value of $\Omega_{\chi}$. Nevertheless, as we discuss below, a significant modification to the perturbative prediction arises if one does not ignore the non-adiabaticity induced by the fast oscillation of the inflaton about the minimum of its potential.

\subsection{Non-perturbative production of scalar dark matter}\label{sec:Hartree}
In this section, we discuss dark matter produced non-perturbatively by taking into account the non-adiabatically changing effective mass of $\chi$. We start with the dark matter scalar field action~(\ref{eq:actionchi}), and find that the equation of motion for the field $\chi$ is given by
\beq
\label{boson:eom1}
\left(\frac{\diff^2}{\diff t^2} - \frac{\nabla^2}{a^2} + 3H\frac{\diff}{\diff t} + m^2_{\chi} + \sigma\phi^2 \right)\chi \;=\; 0\,.
\eeq
In order to simplify the equation of motion and quantize the scalar field $\chi$, it is convenient to switch to conformal time, $\tau$, which is related to cosmic time $t$ via $\diff t/\diff  \tau = a$. By introducing the re-scaled field
\begin{equation}
    X \equiv a \chi \, ,
\end{equation}
we can rewrite the equation of motion~(\ref{boson:eom1}) in the following form:
\beq
\label{boson:eombis}
\left[ \frac{\diff^2}{\diff \tau^2} - \nabla^2 - \frac{a''}{a} + a^2m^2_{\chi} + \sigma a^2\phi^2 \right] X \;=\; 0\, ,
\eeq
where prime represents the derivative with respect to the conformal time, $\tau$. In terms of its Fourier components, the quantum field $X$ can be written as
\beq
X(\tau,\bx) \;=\; \int \frac{\diff^3 \bp}{(2\pi)^{3/2}} e^{-i\bp\cdot \bx} \left[ X_p(\tau) \hat{a}_{\bp} + X_p^*(\tau) \hat{a}_{-\bp}^{\dagger} \right]\,,
\label{Eq:X}
\eeq
where $\bp$ is the comoving momentum and $\hat{a}_{\bp}^{\dagger}$ and $\hat{a}_{\bp}$ are the creation and annihilation operators, respectively, which obey the commutation relations $[\hat{a}_{\bp},\hat{a}_{\bp'}^{\dagger}]=\delta(\bp-\bp')$ and $[\hat{a}_{\bp},\hat{a}_{\bp'}]=[\hat{a}_{\bp}^{\dagger},\hat{a}_{\bp'}^{\dagger}]=0$. From Eq.~(\ref{boson:eombis}), we find that the mode functions $X_p$ satisfy the equation of motion
\beq\label{eq:eomX}
X_p'' + \omega_p^2 X_p \; = \; 0 \, ,
\eeq
where the angular frequency is given by
\beq
\label{eq:omS}
\omega_p^2 \; \equiv \; p^2 - \frac{a''}{a} + a^2 m_{\chi}^2 + \sigma a^2\phi^2\,.
\eeq
The canonical commutation relations between the field, $X_p$, and its conjugate momentum are fulfilled if the Wronskian constraint $X_p X_p^{*\prime}  -  X_p^*  X_p'  \;=\; i$ is imposed. 
To solve the mode equations~(\ref{eq:eomX}), we choose the positive frequency of the Bunch-Davies vacuum, 
\beq\label{eq:icondX}
X_p(\tau_0) \;=\; \frac{1}{\sqrt{2\omega_p}} \,,\qquad X_p'(\tau_0) \;=\; -\frac{i\omega_p}{\sqrt{2\omega_p}} \, ,
\eeq
where we assumed that $|\tau_0 \omega_p| \gg 1$. The particle occupation number for the scalar field is given by
\beq
n_p \;=\; \frac{1}{2\omega_p} \left|\omega_p X_p - iX'_p \right|^2 \,,
\eeq
which is consistent with the fact that the initial conditions~(\ref{eq:icondX}) correspond to the zero-particle initial state. 

The energy density of the field $\chi$ can be determined from the energy-momentum tensor $T^{\mu \nu}$, with
\begin{equation}
    T^{00} \; = \; \frac{1}{2} \dot{\chi}^2 + \frac{1}{2a^2} (\nabla \chi)^2 + \frac{1}{2} m_{\chi}^2 \chi^2 \, ,
\end{equation}
where from this expression we find that the energy of each mode, $X_p$, is given by the expression
\begin{equation}
    E_p \; = \; \omega_p \left(n_p + \frac{1}{2} \right) \; = \; \frac{1}{2} |X_p'|^2 + \frac{1}{2} \omega_p^2 |X_p|^2 \, .
\end{equation}
Using the equations above, the UV convergent expectation values of the number and energy densities of $\chi$ are given by the following expressions~\cite{Kofman:1997yn},
\begin{align}
\label{eq:nchi2}
n_{\chi} \;&=\; \frac{1}{(2\pi)^3 a^3}\int \diff^3\bp \, n_p \, , \\
\label{eq:rhochi}
\rho_{\chi} \;&=\; \frac{1}{(2\pi)^3 a^4}\int \diff^3\bp \,\omega_p n_p \, .
\end{align}
Comparing the number density expressions~(\ref{eq:psd1}) and (\ref{eq:nchi2}), we observe that 
the phase space distribution computed non-perturbatively is
\begin{equation}
f_{\chi}(P,t) \;=\; n_{aP}(t) \; = \; \, n_p(t). 
\end{equation}

By expressing the Ricci curvature scalar $R$ at the background level as $R=-6a^{\prime \prime}/a^3$, one can rewrite the angular frequency~(\ref{eq:omS}) in the form
\begin{equation}\label{eq:effmass}
    \omega_p^2 \; = \; p^2 + a^2 \hat m_{\rm{eff}}^2, \quad {\rm{with}} \quad \hat{m}_{\rm{eff}}^2 = m_{\chi}^2 + \sigma \phi^2 + \frac{1}{6}R\, .
\end{equation}
The quantity $\hat{m}_{\rm{eff}}$ is the DM effective mass generated in a curved space-time and related to the flat-space effective mass $m_{\rm{eff}}$, defined in Eq.~(\ref{eq:effectiveDMmass}), by $\hat{m}_{\rm{eff}}^2=m_{\rm{eff}}^2 + \frac{1}{6}R$, where we note that the term $\frac{1}{6}R$ is responsible for the non-perturbative gravitational particle production~\cite{Parker:1968mv, Parker:1969au, Parker:1971pt, Ford:1986sy, Ema1, Ema2, Ema3, Herranen:2015ima, Markkanen:2015xuw}. 
For a sufficiently small bare mass $m_{\chi}^2$, coupling $\sigma$, and comoving momentum $p$, the effective squared frequency becomes negative, and in general, this leads to a tachyonic excitation of the corresponding mode. 
\begin{figure*}[!t]
\centering
    \includegraphics[width=0.9\textwidth]{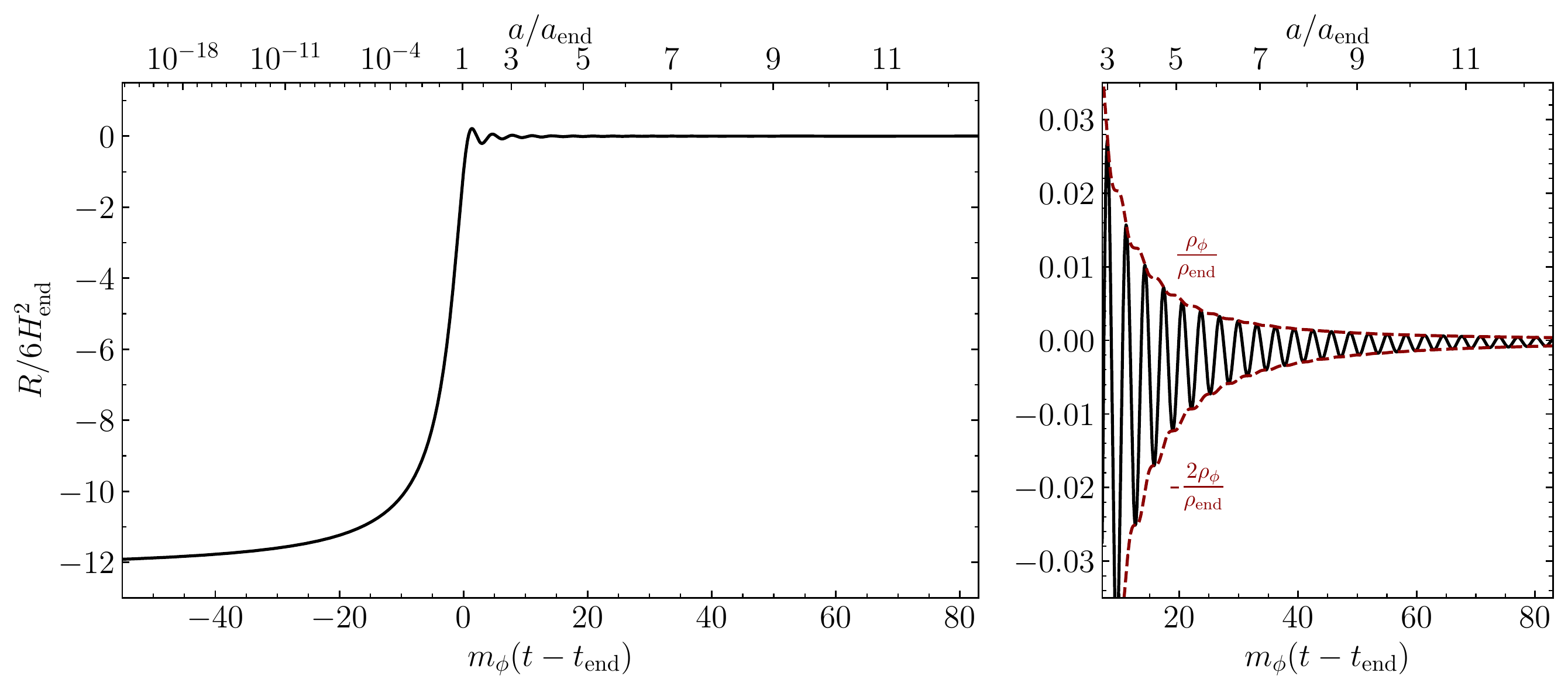}
    \caption{Gravitational contribution to the effective mass $\hat{m}_{\rm{eff}}^2$ normalized by the Hubble parameter at the end of reheating. Left panel: Evolution of $R/6H_{\rm end}^2$ during the inflation-reheating transition. Right panel: Oscillatory behavior of $R/6H_{\rm end}^2$ during reheating, together with its two envelopes.    }
    \label{fig:Rmass}
\end{figure*}
Specifically, the Ricci scalar can be expressed via the second Friedmann equation as
\begin{equation}
    \label{eq:curvature}
   \frac{1}{6} R \; = \; -\frac{a''}{a^3} \; = \; -\frac{1}{6M_P^2} \left(4V - \dot{\phi}^2 \right) \, .
\end{equation}
We show in Fig.~\ref{fig:Rmass} the evolution of the Ricci scalar during inflation and reheating. At the end of inflation, when $V_{\rm{end}} = \dot{\phi}^2_{\rm{end}}$ and $H_{\rm{end}}^2 M_P^2 = V_{\rm{end}}/2$, the angular frequency squared becomes
\beq\label{eq:omegaend}
\omega_p^2 (t_{\rm end}) \;=\; p^2 + a_{\rm{end}}^2\left(m_{\chi}^2 + \sigma \phi^2 - H_{\rm end}^2  \right)\,.
\eeq
Therefore, for $m_{\chi}^2+ \sigma \phi^2<H_{\rm end}^2$, the modes that satisfy condition $p^2 < a_{\rm{end}}^2\left(H_{\rm end}^2 - m_{\chi}^2 - \sigma \phi^2\right)$ would not be in the corresponding Bunch-Davies vacuum at the beginning of reheating. Instead, these modes have grown during inflation due to tachyonic instability. Thus, to accurately track their evolution, one must follow them from very early times, when Eq.~(\ref{eq:icondX}) applies deep inside the horizon during inflation. Before horizon crossing, $p\gg  aH$ and $\omega_p^2>0$. The amount of growth is correlated with the amount of time that each mode has spent outside the horizon. Hence, for the imaginary $\omega_p$ and very weak coupling to the inflaton, the corresponding phase space distribution is expected to be heavily red-tilted.

\section{Particle production for small couplings $(\sigma/\lambda \leq 1)$}
\label{sec:three}

In this section, we describe in detail the production of scalar DM for weak inflaton-DM couplings, namely $\sigma/\lambda\leq 1$. We begin by describing the production of $\chi$ particles in the absence of direct coupling with $\phi$, and quantify the enhancement of infrared modes due to their de Sitter growth. We then show how this growth is suppressed as the coupling $\sigma$ is increased until we reach maximal interference between direct and gravitationally mediated production at $\sigma \simeq \lambda$. 

\subsection{Purely gravitational production}\label{sec:puregravity}

We first consider the dark matter production in the limit $\sigma \rightarrow 0$, i.e., pure gravitational production.{\footnote{For a recent review on gravitational particle production, see~\cite{Ford:2021syk}.}} As discussed in the previous section, in this regime the tachyonic excitation of the $\chi$ modes will be present during inflation. Hence, to numerically solve the equation of motion (\ref{eq:eomX}) with the Bunch-Davies initial conditions (\ref{eq:icondX}), one must track the evolution of the corresponding mode starting with times $\tau_0$ when it is deep inside the horizon. Therefore, we solve the equation of motion for $X_p$ together with the inflaton equation of motion (\ref{eq:eomphia}) with the initial condition $\phi_{\rm ini} > \phi_{\rm end}$. For a given numerical infrared (IR) cutoff in the computation of DM phase space distribution, we choose $(\phi_{\rm ini}, \, \dot{\phi}_{\rm ini})$ such that the IR scale is well inside the horizon, and that these initial conditions lie in the inflationary attractor in the field-velocity phase portrait (see e.g.~\cite{Mukhanov:2005sc}).\footnote{Throughout this work we refer to the domains with $q\ll1$ and $q\gg 1$ as the infrared and ultraviolet regimes of the PSD, respectively.} In addition, unlike the previous work of~\cite{Ema3,Ling:2021zlj}, we focus only on the excitation of light scalar fields, $m_{\chi}\ll m_{\phi}$, and use the inflaton potential (\ref{eq:phipotential}). 

\begin{figure*}[!t]
\centering
    \includegraphics[width=0.85\textwidth]{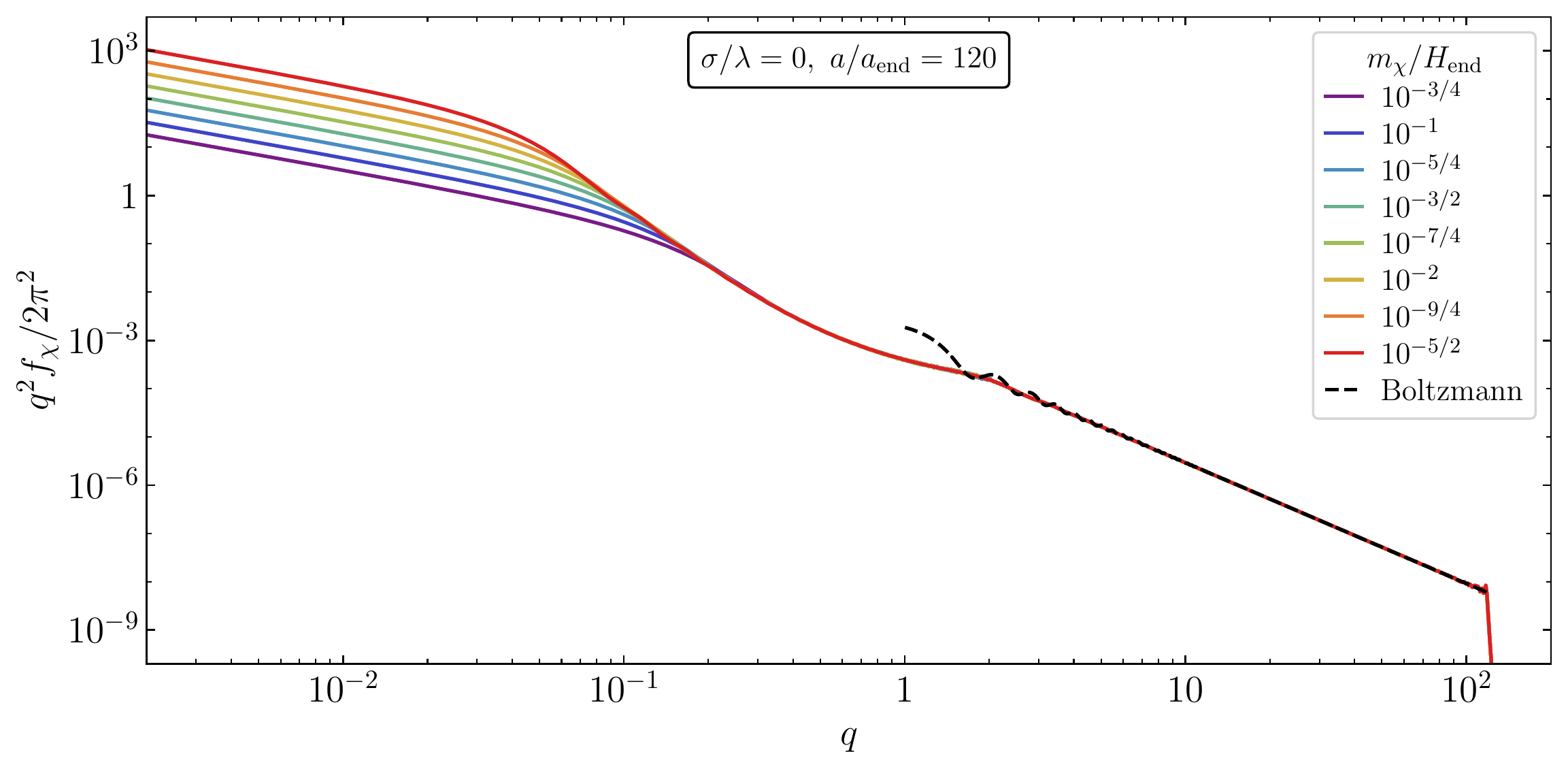}
    \caption{Phase space distribution of a gravitationally excited scalar field ($\sigma = 0$) evaluated at $a/a_{\rm end}=120$ for a range of DM masses, coded by color. The dashed black curve corresponds to the numerical integration of the Boltzmann equation (\ref{eq:boltzfc}), which is valid for $q > 1$.}
    \label{fig:PSDgrav_masses}
\end{figure*}

The resulting shape of the gravitationally sourced PSD of $\chi$ is shown in Fig.~\ref{fig:PSDgrav_masses} for a selection of scalar bare masses evaluated at $a/a_{\rm end}=120$. To determine the importance of superhorizon modes at the end of inflation, we use Eq.~(\ref{eq:omegaend}) and choose the masses relative to the value of the Hubble parameter at the end of inflation, $H_{\rm end}\simeq 0.4\, m_{\phi} \simeq  6.3 \times 10^{12} \, {\rm{GeV}}$. From the definition of $q$~(\ref{eq:qdef}), we see that the modes that were outside the horizon at the end of inflation correspond to $q\lesssim 0.4$ and the modes with $q>1$ are produced after the end of inflation. We also observe that for the modes that were outside the horizon at the beginning of reheating, the tachyonic nature of the effective frequency during inflation leads to a red-tilted distribution. For $m_{\chi}\ll H_{\rm end}$ and $q\ll 1$, this tilt corresponds to $f_{\chi}\propto q^{-3}$, in accordance with the growth of a light scalar field minimally coupled to gravity during a de Sitter phase~\cite{Herring:2019hbe}. Therefore, the corresponding number density diverges logarithmically in the IR in the absence of a cutoff scale. It is only for $m_{\chi}\gtrsim m_{\phi}$ that the distribution becomes blue-tilted for $q<1$~\cite{Ling:2021zlj}.

In the ultraviolet (UV) regime, the distribution has the behavior expected from the perturbative computation in Section~\ref{sec:boltzmann} for $m_{\chi}\ll H_{\rm end}$. Namely, the distribution falls as $f_{\chi}\propto q^{-9/2}$ and becomes sharply suppressed at $q=a/a_{\rm end}$. For comparison, the numerical integration of the Boltzmann equation~(\ref{eq:boltzfc}) is shown as the dashed black curve. Since in this regime, $\beta \simeq 1$ and $f_{\chi}\ll 1$, this dashed curve is indistinguishable from the analytical approximation (\ref{eq:fchiappq}) with the transition amplitude (\ref{eq:M2squarred}), except for some oscillations for $q\sim \mathcal{O}(1)$, which appear during the transition from inflation to reheating. We note that the ``field-picture'' (solid curves) and the ``fluid-picture'' (dashed black curve) only agree for $q\gtrsim 2$. The IR modes are not accounted for in the perturbative computation of Section~\ref{sec:boltzmann}, as there it is assumed that the scalar field modes are excited only after the end of inflation. Hence, we impose the cutoff at $q=1$ in Eq.~(\ref{eq:fchiappq}) and show it in Fig.~\ref{fig:PSDgrav_masses}.

\begin{figure*}[!t]
\centering
    \includegraphics[width=0.95\textwidth]{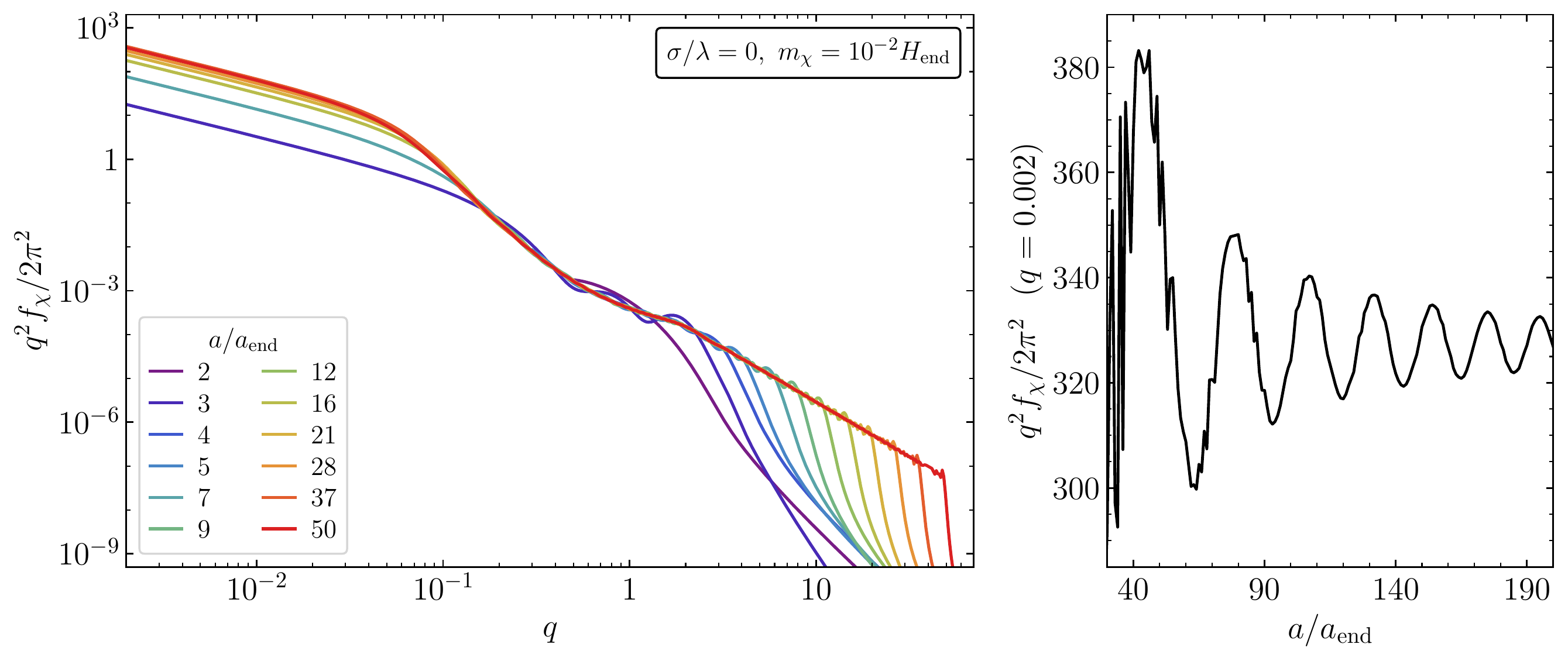}
    \caption{Phase space distribution of a gravitationally excited scalar field of mass $m_{\chi}=10^{-2}H_{\rm end}$. Left: distribution as a function of the comoving momentum for the selected values of the scale factor $a/a_{\rm{end}}$ during reheating. For small values of $a/a_{\rm{end}}$, the distribution is cut below the instantaneous horizon scale if $\omega_p^2<0$. Right: distribution as a function of the scale factor for a selected value of the comoving momentum, $q=0.002$.}
    \label{fig:PSDgrav_a}
\end{figure*}

The left panel of Fig.~\ref{fig:PSDgrav_a} shows the PSD for pure gravitational production, with $m_{\chi}=10^{-2}H_{\rm end}$, as a function of the growing scale factor during reheating. We observe that the hard cutoff of the perturbative approximation (\ref{eq:fchiappq}) at $q=a/a_{\rm end}$ becomes a better fit to the numerical distribution as the value of $a$ increases. The right panel shows the evolution of the PSD for the smallest comoving momentum in the left panel, $q=0.002$, as a function of the scale factor $a$. It shows how the distribution relaxes to its final value as reheating proceeds. This relaxation is oscillatory, mass-dependent, and asymptotes to a final value well after the end of inflation. 

The de Sitter growth of the IR modes of the PSD leads to a logarithmically-divergent comoving number density as $q\rightarrow 0$. This divergence of $n_{\chi}$ requires a natural IR cutoff scale. As argued in~\cite{Starobinsky:1994bd,Felder:1999wt,Herring:2019hbe,Ling:2021zlj}, this scale can be identified with the present comoving scale, $p_0=a_0H_0$, under the assumption that inflation began when this scale was still inside the horizon. Modes with smaller wavenumbers are outside our cosmological horizon and hence must be counted as part of the homogeneous background.\footnote{Note that, regardless of the renormalization of the background due to long-wavelength modes, a natural cutoff for the PSD always exists. Namely, long wavelength modes that were super-horizon at the beginning of inflation can only grow due to the tachyonic instability for the duration of inflation. Hence, a plateau in the PSD necessarily develops below the horizon scale at the start of inflation~\cite{Starobinsky:1994bd}.} In terms of the re-scaled comoving wavenumber, this cutoff can be expressed as
\beq
q_0 \;=\; \frac{H_0}{m_{\phi}}\left(\frac{a_0}{a_{\rm end}}\right)\,.
\eeq
Following the standard computation~\cite{Martin:2010kz,Liddle:2003as,Ellis:2015pla} and assuming entropy conservation after the end of reheating, we can equivalently write
\beq\label{eq:fullq0}
q_0 \;=\; \left(\frac{90}{\pi^2}\right)^{1/4}\left(\frac{11}{43}\right)^{1/3}\left(\frac{H_{\rm end}M_P}{m_{\phi}^2}\right)^{1/2}\frac{H_0}{T_0} \frac{g_{\rm reh}^{1/12}}{R_{\rm rad}}\,,
\eeq
where $R_{\rm rad}$ denotes the so-called reheating parameter defined in Eq.~(\ref{eq:reheatingparameter}). 
In terms of the IR cutoff $q_0$, the dark matter relic abundance can be parametrized as 
\begin{align} \notag
\Omega_{\chi} \;& \simeq \; \frac{m_{\chi} n_{\chi}}{\rho_c}\\ \label{eq:omegagrav}
&=\;\frac{1}{6 \pi^2q_0^3}\left(\frac{H_{\rm end}H_0}{M_P^2}\right)\left(\frac{m_{\chi}}{H_{\rm end}}\right)\,\int_{q_0}^{\infty} \diff q\, q^2f_{\chi}(q,t)\, ,
\end{align}
where we assumed that the produced light scalars are non-relativistic at the present time.{\footnote{For the considered mass range the difference between the fully numerical evaluation of energy density and non-relativistic matter approximation $\rho_{\chi} \simeq m_{\chi} n_{\chi}$ is smaller than $0.5 \%$. We further justify this approximation below by computing the corresponding matter power spectrum free-streaming scale.}}
Computationally, it is not necessary to numerically solve the system of Friedmann-Boltzmann equations~(\ref{eq:phidecay1})-(\ref{eq:phidecay3}) together with the mode equation~(\ref{eq:eomX}) until the end of reheating. On the one hand, the UV ($q\geq 1$) contribution to the integral (\ref{eq:omegagrav}) is rapidly saturated near the beginning of reheating, as $\int_{1}^{a/a_{\rm end}} \diff q\,q^2 f_{\chi} \propto 1-(a/a_{\rm end})^{-3/2}$. In practice, accounting for the UV cascade up to $a/a_{\rm{end}}\simeq 40$ is sufficient to obtain a sub-percent accurate result. On the other hand, as we have discussed above, in the IR the PSD relaxes to its final value upon horizon re-entry. For a given $q\ll 1$, the convergence to the asymptotic distribution becomes slower as the mass of $\chi$ decreases, and requires running the numerical integrator to later times during reheating, which in turn leads to a loss of precision. For this reason, we compute the relic abundances down to masses $m_{\chi} = 10^{-3} H_{\rm end}$, and we extrapolate our results to lower values.

\begin{figure*}[!t]
\centering
    \includegraphics[width=0.84\textwidth]{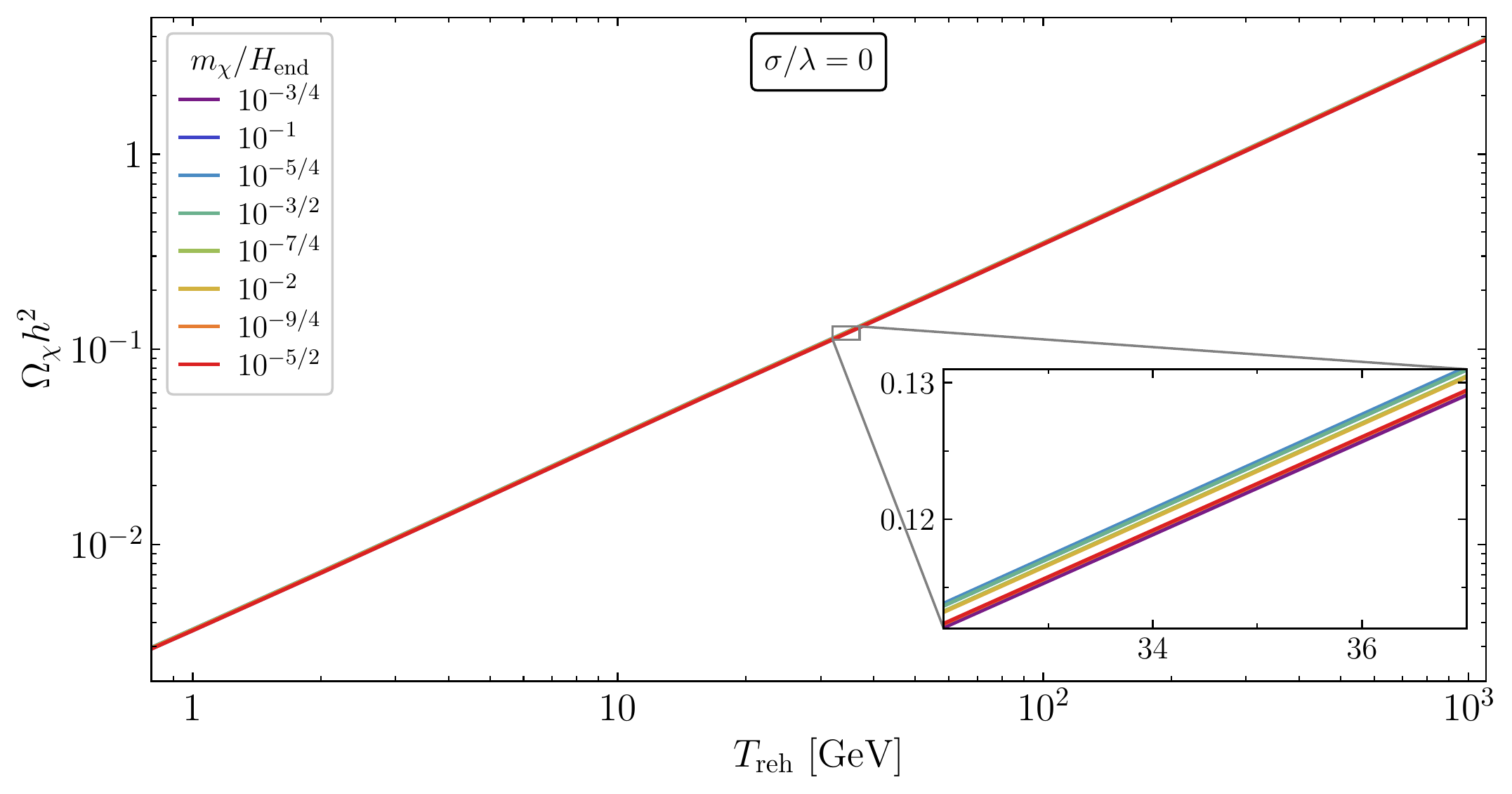}
    \caption{Relic abundance of gravitationally produced scalar dark matter. The scalar mass range is the same as that in Fig.~\ref{fig:PSDgrav_masses}. The vertical and horizontal lines intersect at $\Omega_{\chi}h^2=0.12$ and $T_{\rm reh}=34\,{\rm GeV}$, and gravitationally produced scalar dark matter is ruled out for $T_{\rm{reh}} \gtrsim 34 \, \rm{GeV}$.}
    \label{fig:OmegaGrav}
\end{figure*}

With the previously mentioned caveats taken into account, the numerical integration for $\Omega_{\chi}$ produces Fig.~\ref{fig:OmegaGrav}. We note that for all cases the relic abundance is a growing, approximately linear function of the reheating temperature. Additionally, when $m_{\chi}\ll H_{\rm end}$, a universal massless limit is reached~\cite{Ling:2021zlj}, given by
\begin{equation}
    \Omega_\chi h^2 \, \simeq \, 0.12 \left( \dfrac{T_\text{reh}}{34~\text{GeV}}\right)\,.
\end{equation}
This relation is obtained from a fit to the curve of smallest mass in Fig.~\ref{fig:OmegaGrav}. For small masses, the increased tachyonic growth of fluctuations at low momentum is compensated exactly by the decrease in the energy density of these very cold modes.  We conclude that for light scalar fields, the absence of a direct coupling to other fields (in particular to $\phi$) or the curvature $R$ is therefore ruled out if $T_{\rm reh} \gtrsim 34\,{\rm GeV}$.\footnote{This statement is true as long as additional couplings are sufficiently small and do not affect the dark matter phase space distribution and conserve the dark matter number density.}

\subsection{Weak inflaton coupling}
\label{sec:weakcoupling}

We now discuss the inflaton-DM coupling, focusing on the weak coupling regime, with $0<\sigma\leq \lambda$. In this case, the IR modes dominantly contribute to the DM relic abundance, similar 
to the purely gravitational dark matter production with $\sigma=0$. We can justify this schematically by recalling the relation between the averaged kinetic and potential energies of the inflaton over oscillations during reheating~\cite{Garcia:2021iag, Garcia:2020wiy},
\beq
 \langle \dot{\phi}^2\rangle \;\simeq\; \langle \phi V'(\phi) \rangle\, .
\eeq
If we evaluate the averaged Ricci scalar contribution~(\ref{eq:curvature}) and assume the opposite relative phases between $\dot{\phi}^2$ and $V(\phi)$ in Eq. (\ref{eq:curvature}), we can express the effective angular frequency as
\beq\label{eq:omegasup}
\omega_p^2 \;\simeq \; p^2 + a^2 m_{\chi}^2  +( \sigma - \lambda) a^2 \phi^2\,.
\eeq
Therefore, the effective frequency will become negative for the light scalar dark matter if $\sigma<\lambda$. For this reason, as mentioned previously, we track the evolution of the mode functions $X_p$ during inflation. We point out that this argument qualitatively recovers the pure gravity-$\sigma$ interference present in the perturbative computation, discussed above and in more detail in Section~\ref{sec:boltzmann} and Appendix~\ref{app:A}. 

\begin{figure*}[!t]
\centering
    \includegraphics[width=0.9\textwidth]{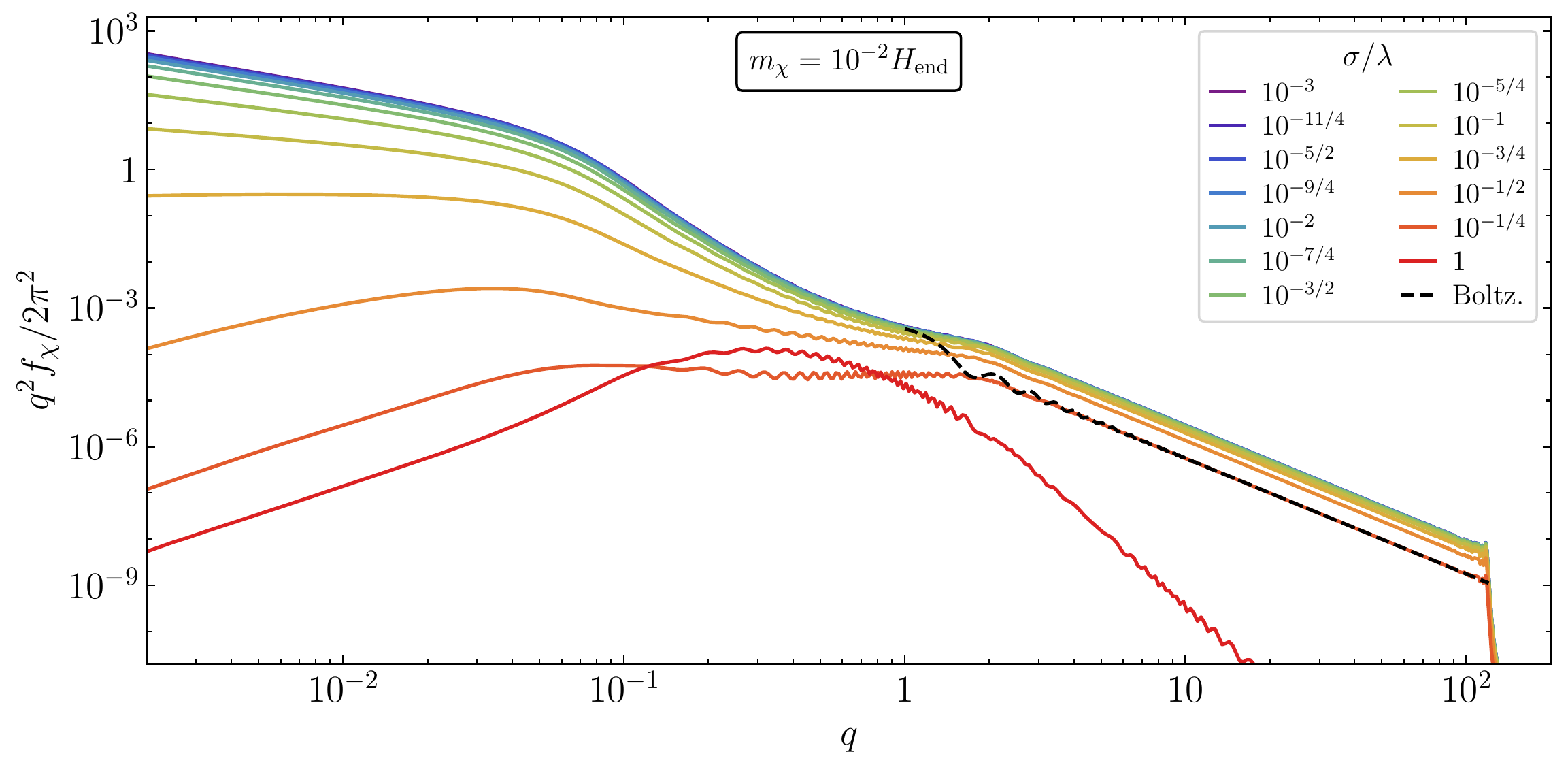}
    \caption{Phase space distribution of an excited scalar field with bare mass $m_{\chi}=10^{-2}H_{\rm end}$ evaluated at $a/a_{\rm end}=120$, for a selection of inflaton-DM couplings $\sigma/\lambda<1$, coded by color. The dashed black curve corresponds to the numerical integration of the Boltzmann equation (\ref{eq:boltzfc}) for the ratio $\sigma/\lambda=10^{-1/4}$.}
    \label{fig:PSDsmallsigma}
\end{figure*}

Fig.~\ref{fig:PSDsmallsigma} shows the phase space distribution $f_{\chi}(q,t)$ at $a/a_{\rm end}=120$ for selected values of $\sigma/\lambda\leq 1$ with $m_{\chi}=10^{-2}H_{\rm end}$. The purple line corresponds to the phase space distribution of purely gravitational production (in Fig.~\ref{fig:PSDgrav_masses} it is shown by the yellow line). In all cases the convergence of the comoving number density for $q>1$ modes is rapid, and the value of $n_{\chi}(a/a_{\rm end})^3$ can be assumed to be saturated for $a/a_{\rm end} \gtrsim 40$. This is a consequence of the fact that for most curves, the $f_{\chi}\propto q^{-9/2}$ perturbative trend at $q>1$ is present. In particular, the dashed black curve shows the result of the numerically integrated Boltzmann equation (\ref{eq:boltzfc}) for $\sigma/\lambda=10^{-1/4}$ with the DM production amplitude (\ref{eq:M2squarred}). Similar to the $\sigma=0$ case, the analytical approximation (\ref{eq:fchiappq}) is a good approximation for the solid and dashed curves with $q>2$; when $\sigma/\lambda<1$, the bosonic enhancement and the kinematic blocking effects can be ignored, and $f_{\chi}\simeq f_{\chi}^c$ . Although not shown explicitly in the plot, the same is true for all the curves with a smaller coupling. We now proceed to analyze the results from Fig.~\ref{fig:PSDsmallsigma} by addressing the differences between the various regimes of couplings $\sigma/\lambda$ and discuss the associated dark matter production.

\paragraph{Soft interference ($\sigma/\lambda\lesssim 10^{-2}$)} For the parameters in this range shown in Fig.~\ref{fig:PSDsmallsigma}, the phase space distribution for non-vanishing inflaton-DM coupling is essentially indistinguishable from the gravity-mediated scenario. For such couplings, the PSD has a red tilt that is almost as steep as for the purely gravitational case. This renders the distribution convergent, although the IR sensitivity is still present due to the smallness of the cutoff scale $q_0$.

\paragraph{Mild interference ($10^{-2}\lesssim \sigma/\lambda < 1$)}
For such couplings, the PSD becomes clearly distinguishable from the gravity-mediated scenario. The red tilt of the distribution in the IR is visibly suppressed, due to the increased effective mass of $\chi$ during inflation. Namely, the term $\sigma\phi^2$ in the effective mass can become similar or larger than the curvature term, suppressing the tachyonic growth, c.f.~(\ref{eq:effmass}). The sensitivity of the comoving number density~(\ref{eq:comovingnchi}) to the IR cutoff becomes milder as the coupling increases. As we show in more detail below, when $\sigma/\lambda \gtrsim 10^{-1/2}$, the integral (\ref{eq:omegagrav}) becomes completely insensitive to the present-horizon cutoff $q_0$, and instead the bulk of the contribution to $n_{\chi}$ comes from modes excited around the end of inflation, with $q\sim \mathcal{O}(10^{-1}-1)$.

\paragraph{Strong interference ($\sigma/\lambda \simeq 1$)} The sole exception to the $f_{\chi}\propto q^{-9/2}$ trend at $q>1$ in Fig.~\ref{fig:PSDsmallsigma} occurs when $\sigma \simeq \lambda$. The non-perturbative numerical results show an $f_{\chi}\propto q^{-15/2}$ behavior in the UV region with a clear peak at $q\sim 0.3$, corresponding to dominant particle production near the end of inflation. For this case, the perturbative analysis predicts an amplitude squared in Eq.~(\ref{eq:M2squarred}) that reaches a minimum for a coupling ratio very close to unity $\sigma/\lambda \simeq 1$. This regime corresponds to the maximal destructive interference between the gravitational and direct DM-inflaton coupling and would result in a minimal, yet not vanishing, dark matter production rate. Accounting for the effective mass term $m_{\rm eff}$ in (\ref{eq:sigmahat0}) results in $f_{\chi}\sim q^{-21/2}$, which does not match the PSD shown in the figure. This difference arises because in the Boltzmann approach the effective dark matter mass is overestimated by treating the inflaton field as an envelope and neglecting fast oscillations around the minimum of its potential. On the other hand, the non-perturbative shape of the PSD shown in Fig.~\ref{fig:PSDsmallsigma} is obtained by correctly taking into account the inflaton oscillation and the phase differences between the $\sigma\phi^2$ and the $R/6$ terms in the effective frequency of $\chi$~(\ref{eq:effmass}).

\subsection{Dark matter abundance}
We summarize the results of this section and calculate the dark matter relic abundance for purely gravitational production and production with small couplings $\sigma \lesssim \lambda$. For $\sigma/\lambda\leq 1$, we use Eq.~(\ref{eq:omegagrav}) to determine numerically the DM closure fraction. If $\sigma/\lambda\ll 1$, the relic abundance depends on the reheating temperature through the horizon-scale cutoff, as well as the factor of $q_0^{-3}$. On the other hand, if $\sigma/\lambda\gtrsim 10^{-1/2}$, $\Omega_{\chi}$ depends on $T_{\rm reh}$ through $q_0^{-3}$ only, and therefore, $\Omega_{\chi} \propto m_{\chi} T_{\rm reh}$. If we set the density parameter to its observed value, $\Omega_{\chi}h^2\simeq 0.12$, this constraint leads to a functional dependence of the reheating temperature and the scalar field mass. In general, we have found that one can parametrize this dependence as a power law
\beq\label{eq:phenofit}
\left(\frac{T_{\rm reh}}{1\,{\rm GeV}}\right) \;\simeq\; \tau\left(\frac{m_{\chi}}{1\,{\rm GeV}}\right)^{\gamma}\,.
\eeq
\begin{figure*}[!t]
\centering
    \includegraphics[width=0.9\textwidth]{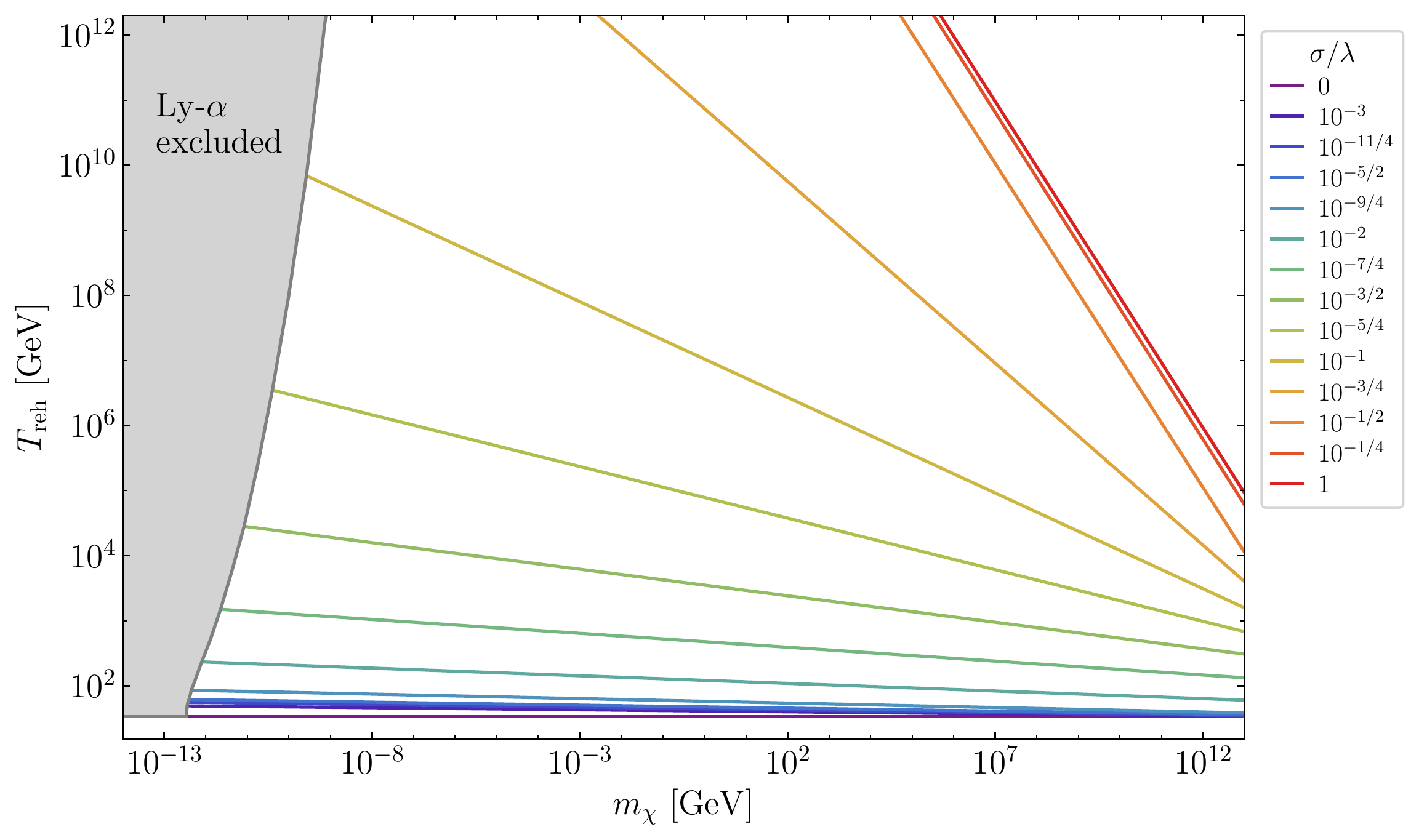}
    \caption{Parameter space for light scalar dark matter ($m_{\chi}<m_{\phi}$) with a weak coupling to the inflaton, $\sigma/\lambda\leq 1$. Each line corresponds to the observed dark matter abundance $\Omega_{\chi} h^2 \;=\; 0.12$. The gray shaded region shows the excluded parameter space by the Lyman-$\alpha$ measurement of the matter power spectrum.}
    \label{fig:abundance_small}
\end{figure*}
\begin{figure*}[!t]
\centering
    \includegraphics[width=0.95\textwidth]{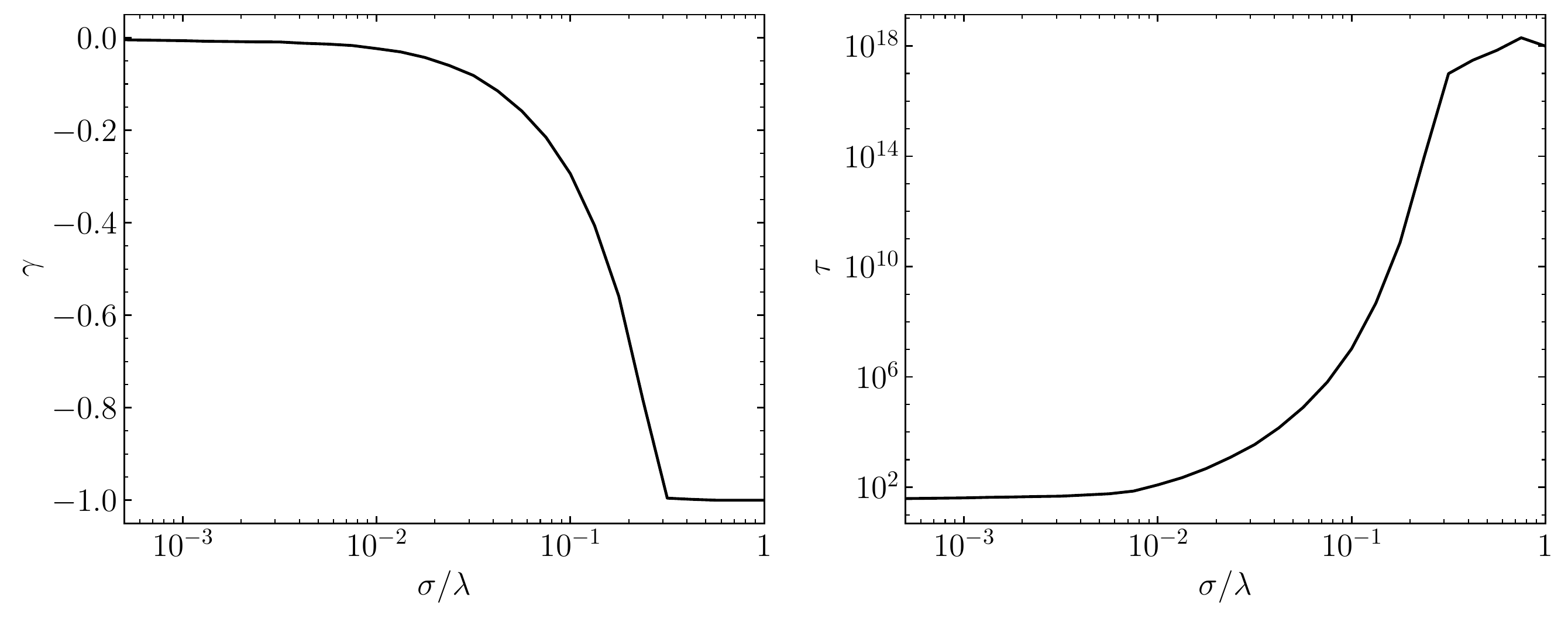}
    \caption{Fit parameters $\gamma$ and $\tau$ for Eq.~(\ref{eq:phenofit}) as a function of the ratio $\sigma/\lambda$.}
    \label{fig:fit_small}
\end{figure*}
This result is illustrated in Fig.~\ref{fig:abundance_small} for a selection of inflaton-DM couplings. To ensure that the numerical results are accurate, we have constructed this plot based on our numerical results down to $m_{\chi}=10^{-3}H_{\rm end}$ and extrapolated to smaller values of $m_{\chi}$. We work in the perturbative reheating regime, for which it is necessary to satisfy the bound $T_{\rm reh}<m_{\phi}$. We note that for $0 \leq \sigma/\lambda \leq 1$, we cover the $ -1\leq \gamma\leq0$ spectrum of possible powers in the parameter equation (\ref{eq:phenofit}). The numerical values obtained from our phenomenological fit for the parameters $\gamma$ and $\tau$ are shown in Fig.~\ref{fig:fit_small}. Therein we note that for $\sigma/\lambda\gtrsim 10^{-1/2}$, we have $\Omega_{\chi} \propto T_{\rm reh} m_{\chi}$, supporting the statement made above.
\begin{figure*}[!t]
\centering
    \includegraphics[width=0.96\textwidth]{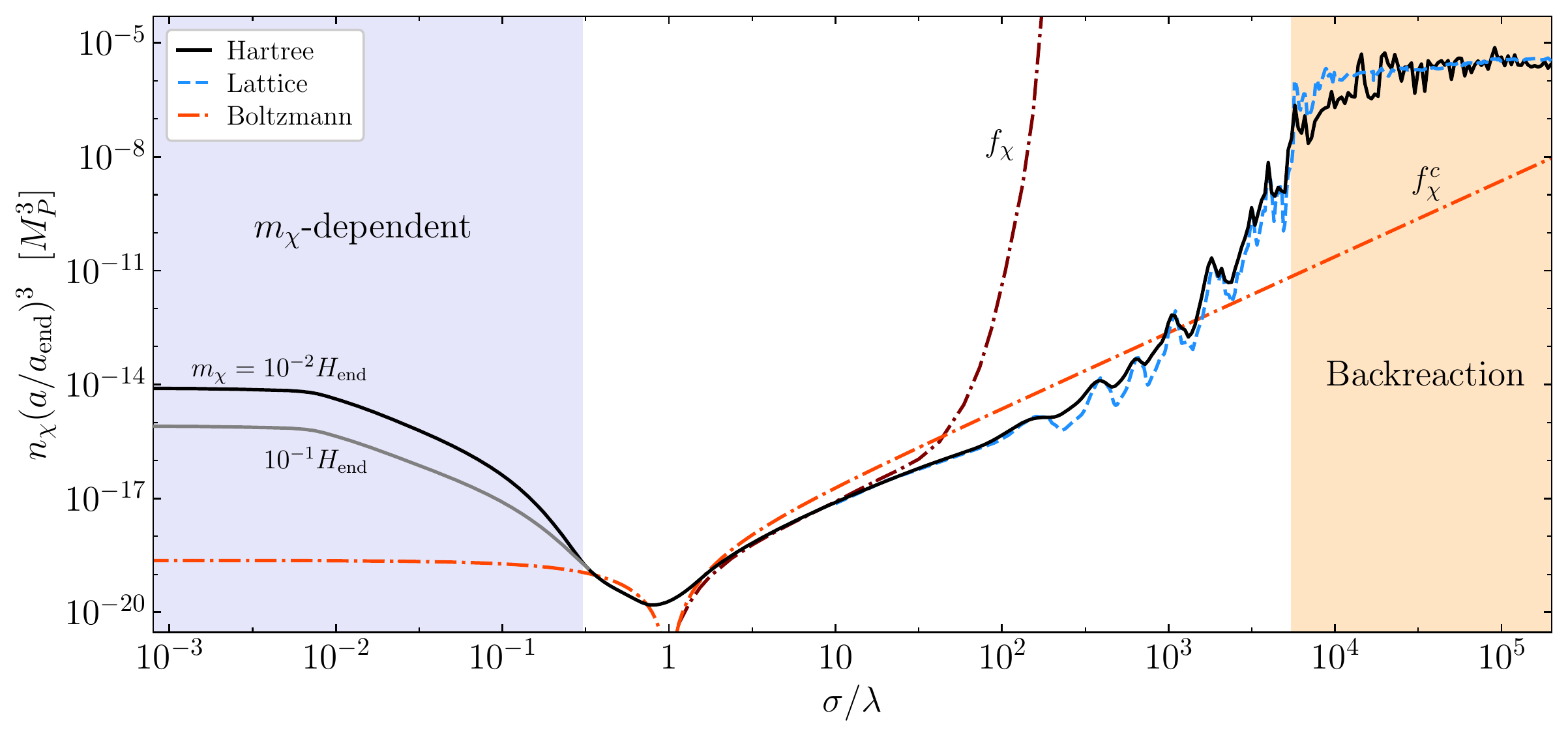}
    \caption{Late-time comoving number density of scalar dark matter $\chi$ as function of the inflaton-DM coupling, evaluated in the Hartree (solid black/gray), lattice (dashed blue), and Boltzmann (dash-dotted red/dark red) approximations. For $\sigma/\lambda\lesssim 10^{-1/2}$, the comoving number density depends on the scalar mass. In this regime the solid black (gray) line is evaluated for $m_{\chi}/H_{\rm end}=10^{-2}$ ($m_{\chi}/H_{\rm end}=10^{-1}$) at the corresponding reheating temperature which saturates the dark matter relic abundance. For the Boltzmann approximation, we show the ``classical" (dash-dotted red) and full solutions (dash-dotted dark red), given by Eqs.~(\ref{eq:boltzfc}) and (\ref{eq:boltzfexp}), respectively.}
    \label{fig:nchiAllsmall}
\end{figure*}
We mention that the limit $\gamma\rightarrow -1$ implies that the final comoving number density does not depend on the present-scale horizon cutoff $q_0$, and also that $n_{\chi}$ is independent of the mass $m_\chi$ as long as $m_{\chi}\ll m_{\phi}$. Therefore, for $\sigma/\lambda\gtrsim 10^{-1/2}$, the limit $m_{\chi}\rightarrow 0$ may be used  in the analytical/numerical computation of the PSD during reheating.

Saturating the DM relic abundance constraint, with $T_{\rm reh}$ (and therefore $q_0$) fixed through Eqs. (\ref{eq:fullq0})-(\ref{eq:omegagrav}) as a function of the mass, we determine the value of the comoving number density for each coupling. Our results are summarized in Fig.~\ref{fig:nchiAllsmall}. The non-perturbative values for $n_{\chi}$ computed using the spectral method presented in Section~\ref{sec:Hartree}, labeled as ``Hartree", are shown as the solid curves. We show the number density for $m_{\chi}/H_{\rm end}=10^{-2}$ (black curve) and $m_{\chi}/H_{\rm end}=10^{-1}$ (gray curve). As expected, for smaller masses the present value of the comoving number density is larger, an enhancement that is compensated by the factor of $m_{\chi}$ in the expression for the density parameter $\Omega_{\chi}$ (\ref{eq:omegagrav}). For $\sigma/\lambda \leq 1$, the comoving $n_{\chi}$ is maximal for pure gravitational production due to the enhancement of the superhorizon modes. For couplings $\sigma/\lambda \gtrsim 10^{-1/2}$, we observe that the mass dependence vanishes, and the black and gray curves converge. We also note the expected particle production minimum, located near (but not exactly at) $\sigma \sim \lambda$. Numerically, we find this minimum at $\sigma/\lambda\simeq 0.87$ with $n_{\chi}(a/a_{\rm end})^3\simeq 10^{-20}$.

In the same Fig.~\ref{fig:nchiAllsmall}, we show the comoving number density of $\chi$ that was calculated using the ``classical" Boltzmann approximation~(\ref{eq:boltzfc}) (dash-dotted red curve) and the full Boltzmann solution~(\ref{eq:boltzfexp}) (dash-dotted dark red curve). For the range of couplings considered in this section, we note that the classical approximation drastically underestimates the amount of produced particles due to the absence of a contribution from the IR modes. We also observe a strong suppression of the dark matter production rate at $\sigma \simeq\lambda$, well below the non-perturbative value.


\section{Particle production for large couplings $(\sigma/\lambda > 1)$}\label{sec:largecoupling}

We now turn to study the production of light scalar dark matter during reheating in the large coupling regime, $\sigma \geq \lambda$. Here it is convenient to split the discussion into two cases. First, we begin by addressing the DM production in the absence of backreaction effects. The fraction of energy transferred from the inflaton condensate into the $\chi$ quanta is below the percent level compared to $\rho_{\chi}$, and therefore one can assume that neither the equation of motion for $\phi$ nor the Hubble parameter $H$ is significantly affected by this particle production channel. Second, we address the production of $\chi$ particles in the presence of backreaction. In this regime, the formalism presented in Section~\ref{sec:Hartree} is proven to be insufficient, and we resort to the use of lattice codes to address the effect of strong DM production on the evolution and destruction of the inflaton condensate.

\subsection{No backreaction  $(1<\sigma/\lambda \lesssim 5\times 10^3)$}
\label{sec:nobackreaction}

The production of light scalar DM in the ``intermediate" regime of couplings, which are significantly large so that the gravitational effects can be ignored, but sufficiently small to avoid the backreaction effects, is arguably the simplest range to study. In this case, the scalar field $\chi$ can be treated as a spectator field, and the equations of motion (\ref{eq:eomphia}), (\ref{eq:friedinf}), and (\ref{eq:eomX}) are sufficient to determine the evolution of the PSD and hence the comoving number density of $\chi$. Moreover, as it was argued previously in Section~\ref{sec:weakcoupling}, and shown explicitly in Fig.~\ref{fig:nchiAllsmall}, for these couplings, the comoving number spectrum $q^2 f_{\chi}(q)$ is blue-tilted for $q<1$ and leads to an $m_{\chi}$-independent comoving number density for $m_{\chi}\ll m_{\phi}$. This simplifies the integration of the equations of motion, as one can simply neglect the bare mass of $\chi$ during reheating. Finally, as we demonstrate below, the relative absence of particle production during inflation suggests that the perturbative Boltzmann approximation provides an adequate approximation for both the PSD and the relic abundance of $\chi$.
Numerically, we find that the backreaction effects can be ignored for $\sigma/\lambda \lesssim 5\times 10^3$. At this limiting value, $\rho_{\chi}$ comprises at most 10\% of the total energy budget of the universe during reheating, and the PSD computed in the spectator approximation and the Hartree approximation (see the next subsection) are practically indistinguishable. 

\begin{figure*}[!t]
\centering
    \includegraphics[width=0.9\textwidth]{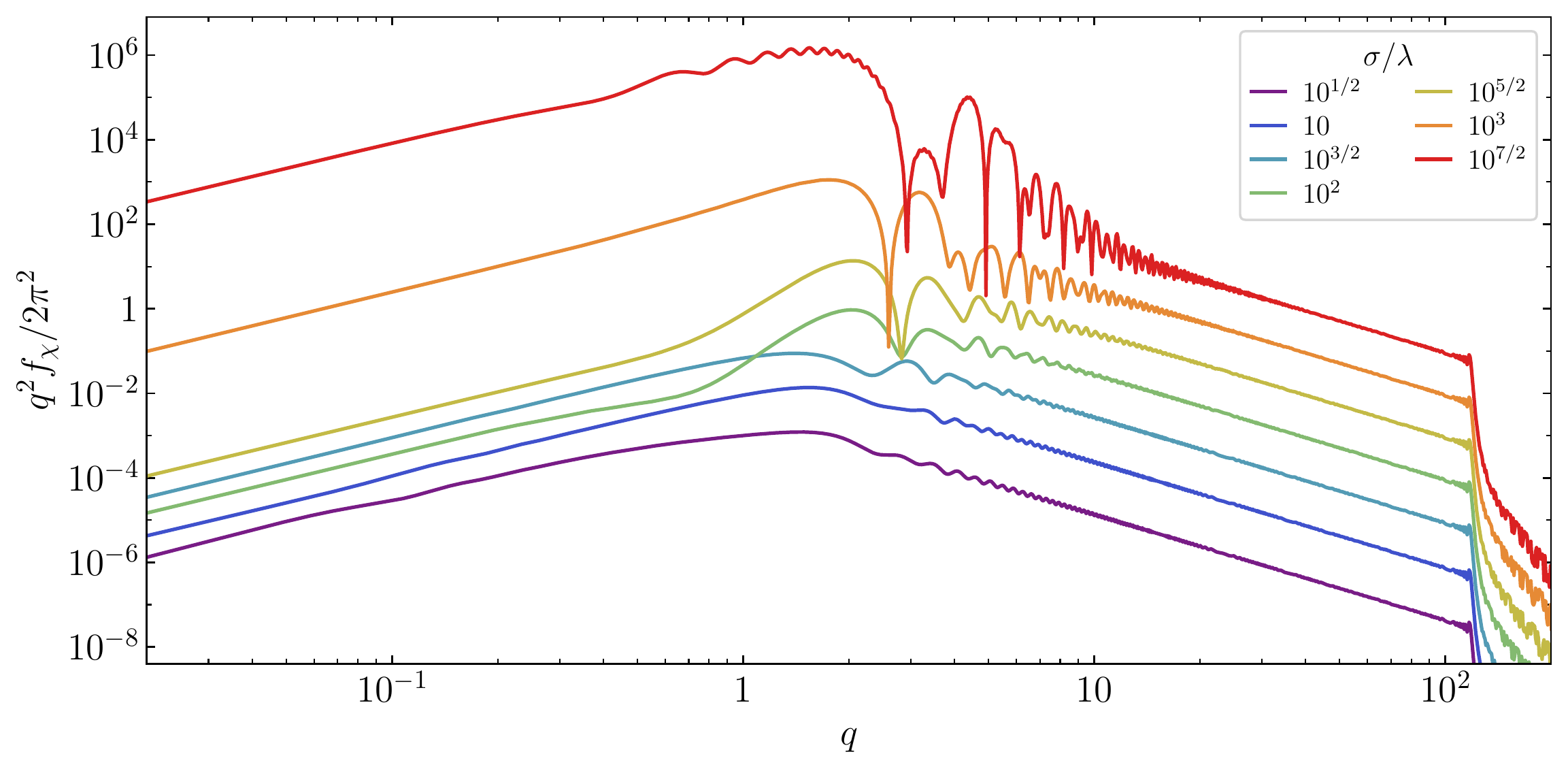}
    \caption{Phase space distribution of an excited scalar field evaluated at $a/a_{\rm end}=120$ for a selection of inflaton-DM couplings $\sigma/\lambda>1$ in the Hartree approximation. }
    \label{fig:PSDlargesigma}
\end{figure*}

Using the non-perturbative formalism of Section~\ref{sec:Hartree}, we show in Fig.~\ref{fig:PSDlargesigma} the result of the numerical integration of the phase space distribution $f_{\chi}(q,t)$ within the no backreaction regime for a selection of inflaton-DM couplings evaluated at $a/a_{\rm end}=120$. For all cases, the IR domain of the spectrum is blue-tilted, leading to a convergent value of $n_{\chi}$. In the UV, for $q\gg 1$, we immediately recognize the perturbative Boltzmann tail, with $f_{\chi}\sim q^{-9/2}$. We also note that the relative amplitudes of the tails follow the perturbative prediction, growing with the square of the coupling $\sigma$. Also, in all cases the predicted sharp drop at $q=a/a_{\rm end}$ can be explicitly observed. However, it must be emphasized that outside of the far UV regime, the amplitude of the PSD does not follow the trend predicted by Eq.~(\ref{eq:fchiappq}). For example, in the IR we  note that the amplitude of the spectrum does not grow monotonically with $\sigma$, and as the coupling increases, the PSD peak amplitude grows faster than the amplitude of the UV tail. These effects are a manifestation of the resonant growth of the $\chi$ fluctuations.

We now briefly review some important aspects of preheating. For $\sigma > \lambda$ we can neglect the gravitational dark matter production arising from the $R/6$ term and the bare mass term $m_{\chi}$ in the effective mass~(\ref{eq:effmass}). Therefore, we can apply the usual Floquet analysis used for the broad parametric resonance~\cite{Kofman:1994rk, Shtanov:1994ce, Kofman:1997yn}. We rewrite the equation of motion of the inflaton in Fourier space~(\ref{boson:eom1}) as
\begin{equation}
    \label{eq:eomcosmtime}
    \ddot{x}_p(t) + \bar{\omega}_p^2 x_p(t) \; = \; 0\, ,
\end{equation}
where here we use cosmic time with the field redefinition $x_p(t) = a^{3/2}(t) \chi_p(t)$, and the angular frequency is given by $\bar{\omega}_p^2 \simeq p^2/a^2 + \sigma \phi(t)^2$, not to be confused with $\omega_p$ defined in (\ref{eq:omS}). Since the early universe is dominated by coherent inflaton oscillations about a quadratic minimum, with $\phi(t) = \phi_0(t) \cdot \cos(m_{\phi} t)$, we can rewrite the mode equation~(\ref{eq:eomcosmtime}) as a well-known Mathieu equation
\begin{equation}
    \label{eq:mathieau}
    \frac{\diff^2 x_p}{\diff z^2} + (A_p - 2 \, \kappa \cos{2z}) x_p \; = \; 0\, ,
\end{equation}
where we introduced the variables
\begin{equation}
    z \; = \; m_{\phi}t + \frac{\pi}{2} \, , \quad A_p \; = \; \frac{p^2}{m_{\phi}^2 a^2} + 2 \kappa, \, \quad \kappa \; = \; \frac{\sigma \phi_0^2}{4 m_{\phi}^2} \, .
    \label{eq:Mathieu_parameters}
\end{equation}
Here the quantity $\kappa$ is known as the resonance parameter. The properties of the Mathieu equation are characterized by the Mathieu stability/instability chart, which can be found in~\cite{Garcia:2021iag}. When the resonance parameter is significantly large, $\kappa \gg 1$, the non-perturbative particle production is primarily characterized by the broad parametric resonance. In this case, $A_p > 2\kappa$, and the dominant particle production occurs when the mode solutions $x_p$ cross the unstable Mathieu bands and begin growing exponentially. In the exponentially unstable regions, the mode solutions grow as $x_p \propto \exp(\mu_p z)$~\cite{Kofman:1997yn}, and the characteristic exponent typically has the values $0 \lesssim \mu_p \lesssim 0.28$~\cite{Huang:2011gf}. The particle occupation number in the exponentially unstable regions grows as $n_p(t) \sim |x_p|^2 \sim \exp(2\mu_p z)$. Importantly, the total band number increases with the value of the resonance parameter $\kappa$, and for the fastest growing modes with zero momentum, the band number is approximately given by $n_{\rm{band}} \simeq \sqrt{A_p} \simeq \sqrt{2 \kappa}$. When the mode $x_p$ finally crosses the first instability band the parametric resonance and the explosive particle production stops. A more detailed discussion on the Floquet analysis is presented in Appendix A of~\cite{Garcia:2021iag}.

An increase in the effective coupling $\sigma/\lambda$ corresponds to an increase in the resonance parameter $\kappa$. In Fig.~\ref{fig:PSDlargesigma}, the main features of the parametric growth can be identified in the form of the PSDs shown there. The larger the coupling, the higher the maximum of the distribution, correlated with the mode function crossing over several, wide resonance bands. The modes that are most enhanced correspond to the IR modes, populated at the early stages of reheating. In contrast, the UV modes do not experience exponential growth, and for them, the distribution follows the perturbative result. Finally, at the evaluation time corresponding to $a/a_{\rm end}$, the resonance has already shut down, and the PSD is effectively frozen for $q<120$. Modes with a larger comoving momentum continue to be populated perturbatively until the end of reheating.

When the resonance parameter $\kappa$ is very large, the field quantum fluctuations  become significant, and the mode-mode couplings and non-linear effects can no longer be ignored. In such a case, the inflaton condensate gets fragmented into gradients. Therefore, we need to take into account the rescattering of the produced dark matter $\chi$ into $\phi$, which leads to the fragmented inflaton particle production and the scattering between the $\chi$ and $\phi$ particles. The backreaction regime with very large values of $\kappa$ can be treated in the simplest way using the Hartree approximation. However, such a simple picture is not always sufficient. From the Fourier analysis perspective, this could be understood in terms of mode-mode couplings between the inflaton fluctuations and scalar dark matter. To get around these issues, it is convenient to assume that these quantum fields behave as classical fields and solve a classical system of equations on a spatial lattice. As mentioned before, the backreaction effects become important when $\rho_{\chi}$ becomes larger than around 10\% of the total energy density of the universe during reheating~\cite{Enqvist:2015sua, Repond:2016sol}. Therefore, in this work, we use {\tt{CosmoLattice}}~\cite{Figueroa:2020rrl,Figueroa:2021yhd} to calculate the PSDs and number densities in the backreaction regime. We discuss it in detail in the next subsection.

\begin{figure*}[!t]
\centering
    \includegraphics[width=\textwidth]{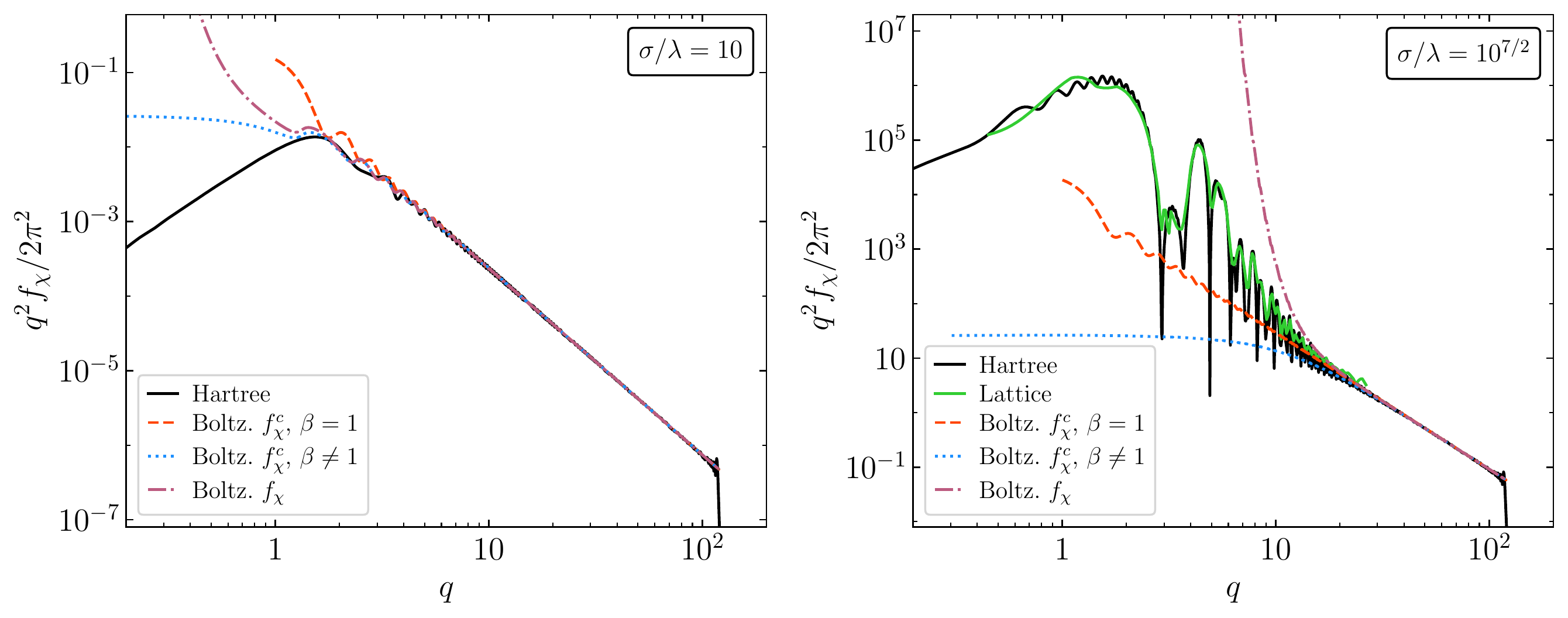}
    \caption{Comparison of the phase space distribution between the Hartree approximation (solid black) and the solution to the Boltzmann equation at different degrees of approximation, evaluated at $a/a_{\rm end}=120$. We show the numerical solution of (\ref{eq:boltzfc}) with $\beta=1$ (dashed red), $\beta\neq 1$ (dotted blue), and the full numerical solution of (\ref{eq:boltzalmostfull}) (dash-dotted purple). In the right panel, we include the PSD obtained from a lattice computation down to $f_{\chi}\simeq 0.1$ (green curve).}
    \label{fig:BoltzVSNonpert}
\end{figure*}

Fig.~\ref{fig:BoltzVSNonpert} shows a more detailed comparison between the perturbative and non-perturbative predictions for the phase space distribution of $\chi$. The left panel corresponds to the PSD generated during reheating with the coupling ratio $\sigma/\lambda=10$. The continuous curve corresponds to the non-perturbative calculation, labeled as ``Hartree", and shown also in Fig.~\ref{fig:PSDlargesigma}. The dashed red curve corresponds to the numerical solution of the Boltzmann equation in the absence of bosonic and kinematic effects, i.e., the solution to Eq.~(\ref{eq:boltzfc}) with $m_{\rm eff} = 0$. We note that this approximation is in good agreement with the ``exact" solution for $q\gtrsim 3$. However, for smaller values of $q$, this approximation overestimates the non-perturbative solution by up to an order of magnitude at $q=1$. This is partly due to the effect of the initial Hubble friction, which leads to suppression in the rate of non-perturbative $\chi$ production (c.f.~Eq.~(\ref{eq:omegaend})). The most important effect of this difference arises from the kinematic suppression during the first couple of oscillations. When $\sigma/\lambda=10$, $m_{\rm eff}^2/m_{\phi}^2\simeq 3.5$ at $a=a_{\rm end}$, decreasing below unity before the completion of the first $\phi$ oscillation. Therefore, the kinematic suppression factor $\beta$ is less than unity after the decay becomes kinematically allowed. We show in Fig.~\ref{fig:BoltzVSNonpert} the solution to the Boltzmann equation in the absence of the bosonic enhancement but including this kinematic suppression (dotted blue line). We note that this curve approximates the distribution better throughout the range $q\geq 1$. Also, this approximation extends into the IR regime due to the reduced value of the momentum for $\beta<1$ in (\ref{eq:boltzfc}). However, the blue tilt of the black curve is not matched, and instead, the distribution is flat.

The solution to the Boltzmann equation, which includes both the enhancement and kinematic effects, is shown in Fig.~\ref{fig:BoltzVSNonpert} as the dash-dotted purple curve. In the left panel, we note that this curve is also a good fit to the non-perturbative result for $q\geq 1$, albeit the peak of the PSD is slightly overestimated. However, we observe that for $q<1$ the distribution grows rapidly toward the IR. The mathematical reason for this is the interplay between the kinematic suppression and the bosonic growth: $f_{\chi}^c$ in Eq. (\ref{eq:boltzfc}) contains a factor of $\beta^{-2}$, which upon substitution into Eq. (\ref{eq:boltzfexp}) results in the unbounded enhancement of deep IR modes. To suppress this effect in the computation of number densities, in what follows we impose the condition $q\geq 1$ for the Boltzmann PSDs. 

The right panel of Fig.~\ref{fig:BoltzVSNonpert} corresponds to the PSD of $\chi$ with  a  coupling $\sigma/\lambda=10^{7/2}$. Such a coupling is close to the backreaction regime, but the Hartree approximation leads to the correct PSD and relic abundance. Similar to the left panel, the solution for $f_{\chi}^c$ without the kinematic suppression simply behaves as a power law with exponent $-9/2$, adequately matching the post-resonance tail, but in this case, underestimating the distribution for the resonantly-enhanced modes with $q\lesssim 30$. Including only the kinematic suppression leads to the dotted blue curve, which corresponds to a flat spectrum for $q\lesssim 10$, and it underestimates the distribution even more severely. On the other hand, the full Boltzmann solution with $f_{\chi}$, which contains the bosonic enhancement effect, is shown to vastly overestimate the distribution. Only for $q\gtrsim 20$ is this approximation close to behaving as the envelope of the non-perturbative solution. We conclude that as the coupling $\sigma$ increases, the perturbative Boltzmann solution in either of the successive degrees of approximation provides a poor approximation to the non-perturbative PSD of $\chi$. In the same panel, we also observe the PSD obtained from the classical lattice computation (green curve), described in the following subsection.

The failure of the Boltzmann approximation can also be observed in Fig.~\ref{fig:nchiAllsmall}. The range explored in this section corresponds roughly to the non-shaded area with $\sigma/\lambda>1$. We note that, compared to the black Hartree curve, the $f_{\chi}^c$ Boltzmann approximation gives the correct number density only for a discrete set of couplings. Note that in the plot this perturbative approximation does not match the non-perturbative trend followed by $n_{\chi}$ as a function of $\sigma/\lambda$. On the other hand, the full perturbative result, $f_{\chi}$, which contains the kinematic and bosonic effects, accurately predicts the value of the comoving number density in the range $10^{1/2}\lesssim \sigma/\lambda\lesssim 10^{3/2}$. Below this range, $n_{\chi}$ is underestimated, while above this range $n_{\chi}$ is severely overestimated. This is consistent with our previous discussion of Fig.~\ref{fig:BoltzVSNonpert}.

For the Hartree approximation, in Fig.~\ref{fig:nchiAllsmall} we observe a monotonically growing trend for couplings between $\sigma/\lambda =1$ and $\sigma/\lambda \lesssim 150$. For these inflaton-DM coupling strengths, the bulk of the DM relic abundance is produced after the first few oscillations of the inflaton field, with $\chi$ crossing a small number of resonance bands (typically one).
For larger $\sigma$, the global trend is still that of a growing $n_{\chi}(a/a_{\rm end})^3$, but the local trend is not monotonic. As discussed in~\cite{Garcia:2021iag} in the context of the energy density of $\chi$, we observe the appearance of localized peaks. These peaks correspond to comoving number densities that are saturated at later times, $a/a_{\rm end}\gtrsim 10$. The reason for it is that for large values of the coupling, the mode functions of $\chi$ can cross a larger number of (broad) instability bands as the resonance builds up. However, the parametric resonance is a quasi-stochastic process, which is mapped into a quasi-random steepness of the exponential growth of $n_{\chi}$ with the coupling $\sigma$. 

We next discuss briefly the effects of instantaneous preheating. During the first oscillation of the inflaton field $\phi$, the production of dark matter particles $\chi$ is strongly suppressed. However, when the coupling ratio is large, $\sigma/\lambda \gg 1$, the resonant dark matter production overcomes the Hubble damping, and one can readily estimate the occupation number for each mode $x_p$
\begin{equation}
    n_p \simeq \exp \left( - \frac{\pi(p^2 + m_\text{eff}^2)}{|\dot{m}_\text{eff}|}\right)\, ,
\end{equation}
after the first time when the field $\phi$ crosses the origin of the potential~\cite{Felder:1999pv, Chung:1999ve}. One can show that the produced energy density scales as $\rho_{\chi} \propto \sigma^{5/4}$, and the large values of the ratio $\sigma/\lambda$ lead to a strongly enhanced production of $\rho_{\chi}$ during the first few oscillations~\cite{Garcia:2021iag}. The maximum energy density in $\chi$ can in fact be reasonably well approximated by accounting only for the first zero-crossing of the inflaton up to $\sigma/\lambda\sim\mathcal{O}(10^2)$. On the other hand, the same cannot be said about the comoving number density. For all couplings, as we have mentioned above, saturation of $n_{\chi}a^3$ requires the accumulation of growth over several oscillations. The difference in behavior is owed to the fact that the energy density of light $\chi$ redshifts as $a^{-4}$, and thus any growth subsequent to the first instance of $\phi=0$ is very efficiently diluted by expansion. Only in the presence of strong, broad resonance, the instantaneous preheating approximation fails, and a fully numerical computation of the maximum of $\rho_{\chi}$ is necessary. The effect of the resonance becomes more important until the backreaction regime is reached, which occurs around $\sigma/\lambda \gtrsim 5\times 10^{3}$. We now proceed to explore DM production in this range of couplings.

\subsection{Backreaction regime   $(\sigma/\lambda>5\times 10^3)$}\label{sec:backreaction}

If the inflaton-DM coupling is above the $\sigma/\lambda \simeq 5\times 10^3$ threshold, backreaction effects need to be taken into account since the effect of particle production on the background dynamics cannot be ignored. At the lowest order, this can be accounted by including the contribution of $\rho_{\chi}$ to the expansion rate of the universe,
\beq\label{eq:friedmannfull}
\rho_{\phi} + \rho_{\chi} \;=\; 3H^2 M_P^2\,,
\eeq
and considering the effect of the growth of the fluctuations of $\chi$ in the homogeneous equation of motion of the inflaton,
\beq
\ddot{\phi} + 3H\dot{\phi} + V_{\phi} + \sigma \langle \chi^2\rangle \phi \;=\;0\,,
\eeq
where~\cite{Garcia:2021iag,Lebedev:2021tas}
\beq
\langle \chi^2\rangle  \;=\; \frac{1}{(2\pi)^3a^2}\int \diff^3\bp\, \left( |X_p|^2 - \frac{1}{2\omega_p}\right)\,.
\eeq
This approximation, which ignores the mode-mode coupling of the inflaton perturbations with the momentum modes of $\chi$, is known as the Hartree approximation~\cite{Kofman:1997yn}. The comoving spectra shown as the continuous curves in Figs.~\ref{fig:PSDlargesigma} and \ref{fig:BoltzVSNonpert} are computed using the Hartree formalism. For all couplings shown in the figures, the Hartree approximation accurately provides the late-time $\chi$ distribution. The inflaton field is mostly homogeneous, and hence perturbative particle production described by the decay of the $\phi$ condensate eventually populates the Boltzmann tail of the PSD.

Fig.~\ref{fig:nchiAllsmall} shows the comoving number density of DM using the Hartree approximation in the backreaction regime as the solid black curve. We observe that in this region, the effects of the broad resonance results in the presence of quasi-stochastic features, which depend on how many resonance bands were crossed during reheating. We also note that above $\sigma/\lambda =10^4$, the DM number density appears to be saturated, as already identified in Ref.~\cite{Lebedev:2021tas}. This maximum density may be estimated by noting that this quantity cannot exceed the inflaton comoving number density at the end of inflation, which is $n_\phi(a_\text{end})\simeq \rho_\text{end}/m_\phi \simeq 3.2 \times 10^{-6} M_P^3$. This number is in good agreement with the observed plateau in $n_{\chi}$.

Up to $\sigma/\lambda \lesssim 5\times10^3$, the Hartree approximation accurately describes the evolution of the inflaton-DM system throughout the duration of reheating. However, for larger values of $\sigma$, the resonance parameter $\kappa$ defined in Eq.~(\ref{eq:Mathieu_parameters}) can become very large and non-linear effects can no longer be ignored. The Hartree approximation only correctly accounts for the production of $\chi$ in the early stages of preheating. In the later stages of preheating, the inflaton condensate gets fragmented into gradients. Therefore, we need to take into account the rescattering of the produced dark matter $\chi$ into $\phi$, which leads to the fragmented inflaton particle production and the scattering between the $\chi$ and $\phi$ particles.

\begin{figure*}[t!]
    \centering
  \includegraphics[width=0.99\textwidth]{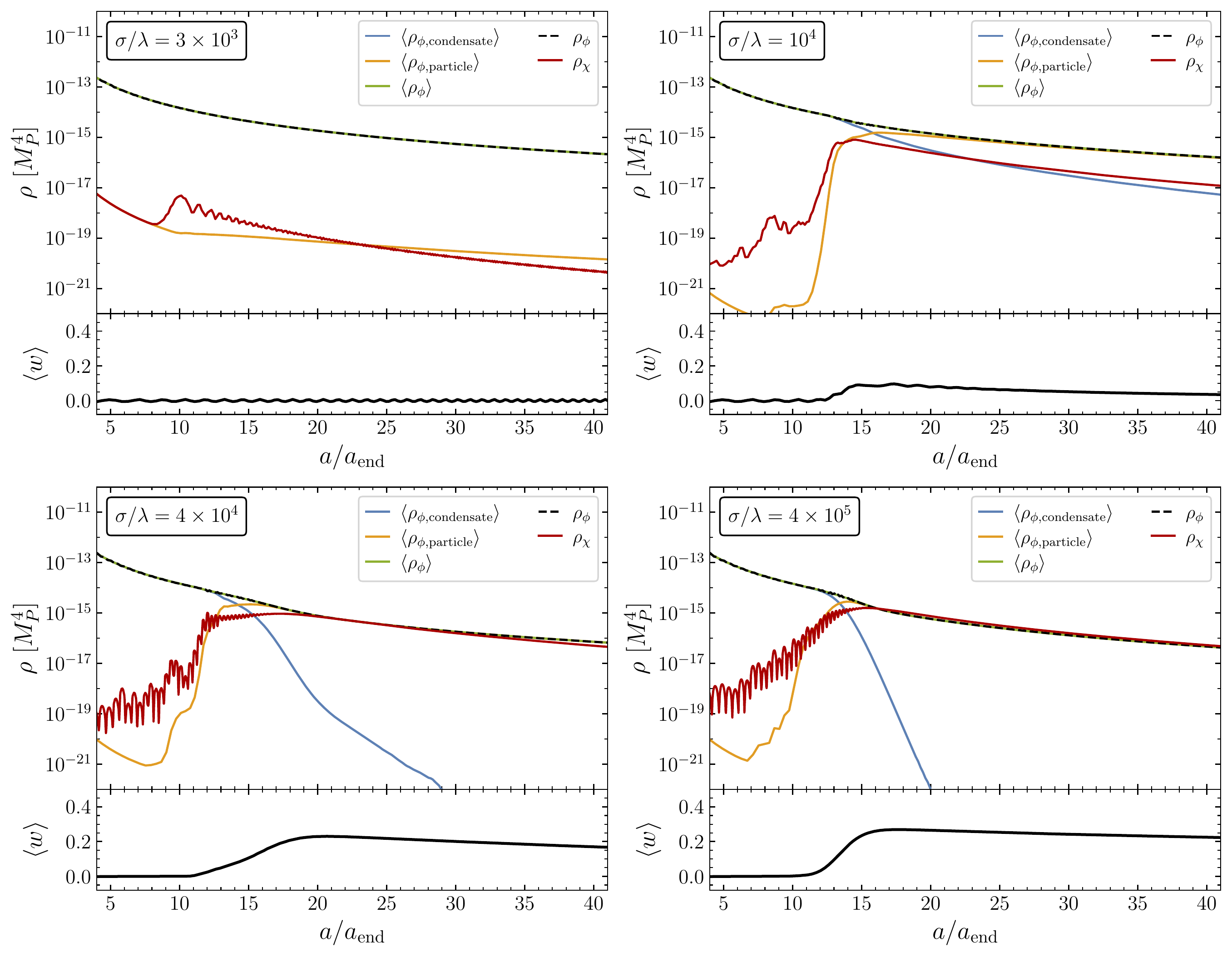}
    \caption{Energy density and oscillation-averaged equation of state parameter $\langle w \rangle$ as a function of the scale factor for a selection of couplings $3 \times 10^3 <\sigma/\lambda<4 \times 10^5$ obtained from {\tt CosmoLattice}. The quantities $\rho_\phi$  (dashed black) and $\rho_\chi$ (dark red) are the instantaneous inflaton and DM energy density, respectively. The quantities $\langle \rho_{\phi,\text{condensate}}\rangle$ (blue) and $\langle \rho_{\phi,\text{particle}}\rangle$ (orange) represent the condensate and particle contribution to the inflaton energy density, respectively, while $\langle \rho_{\phi}\rangle$ (green) corresponds to the sum of these two contributions. These quantities are defined in Sec.~\ref{sec:backreaction}. 
    }
    \label{fig:fragmentation}
\end{figure*}
In the Fourier formalism of Section~\ref{sec:Hartree}, accounting for these effects would result in mode-mode couplings that lead to a set of non-linear quantum-operator equations. Nevertheless, as it is clear from Fig.~\ref{fig:PSDlargesigma}, for sufficiently large couplings the occupation number of the momentum modes $n_p \gg 1$. If this is the case, the quantum fields can be treated as classical fields, and the inflaton-DM dynamics is amenable to be studied via the solution of the non-linear classical equations of motion
\begin{align}\label{eq:eomphifull}
\ddot{\phi} + 3H\dot{\phi} - \frac{\nabla^2\phi}{a^2} + V_{,\phi}(\phi,\chi) \;&=\;0\,,\\ \label{eq:eomchifull}
\ddot{\chi} + 3H\dot{\chi} - \frac{\nabla^2\chi}{a^2} + V_{,\chi}(\phi,\chi) \;&=\;0\,,
\end{align}
together with the Friedmann equation (\ref{eq:friedmannfull}). Non-spectral methods, which involve finite-difference techniques on a spatial lattice are particularly favored. To study scalar DM production in the strong backreaction regime, we use the  lattice numerical code {\tt CosmoLattice v1.0}~\cite{Figueroa:2020rrl,Figueroa:2021yhd}. We favor this code for its ease of use, in particular, to set up the system (\ref{eq:eomphifull})-(\ref{eq:eomchifull}) for arbitrary potentials and initial conditions, and the ease to extract the PSDs for both fields. However, it must be mentioned that this code is not complete due to the absence of metric fluctuations, which can be excited by the parametric resonance~\cite{Nambu:1996gf,Bassett:1998wg,Bassett:1999mt,Jedamzik:2010dq,Huang:2011gf,Giblin:2019nuv}, and the lack of dissipative effects~\cite{Repond:2016sol, Fan:2021otj}. In the following discussion, we present  our numerical results for the background energy density and for the DM phase space distribution.

\paragraph{Energy density.} As we have mentioned above, the inflaton field loses its coherence in the presence of rescattering. Following the treatment presented in Ref.~\cite{Garcia:2021iag}, one can estimate the energy density of $\phi$ in the form of a spatially homogeneous condensate $\rho_{\phi,\text{condensate}}$ by evaluating the quantity
\begin{equation}
    \rho_{\phi,\text{condensate}} \, \equiv \, \dfrac{1}{2} \bar{\dot{\phi}}^2+V(\bar \phi) \,,
\end{equation}
where $\bar \phi$ and $\bar{\dot{\phi}}$ are the spatial average of the field and its time derivative over the lattice volume, respectively. The energy density fraction of $\phi$ in the form of particles can be estimated as the remaining contribution to the $\phi$ energy density, given by $\rho_{\phi,\text{particle}}\equiv \rho_{\phi}-\rho_{\phi,\text{condensate}}$. We show in Fig.~\ref{fig:fragmentation} the corresponding oscillation-averaged quantities $\langle \rho_{\phi,\text{condensate}} \rangle$, $\langle \rho_{\phi,\text{particle}} \rangle$, and $\langle \rho_{\phi} \rangle$ determined from lattice simulations as well as the instantaneous $\rho_{\phi,\chi}$ energy densities in addition to the total oscillation-averaged equation of state (EOS) parameter\footnote{The equation of state parameter $w$ is defined as being the total background pressure divided by the total energy density.}
\begin{equation}
    \langle w(a) \rangle \, \equiv \, \dfrac{1}{\Delta a} \int_{a}^{a+\Delta a} w(\tilde a) \diff \tilde a \,,
\end{equation}
where $\Delta a$ is the time interval between the two successive zero-crossings of the field value. From Fig.~\ref{fig:fragmentation}, one can immediately see that for couplings as small as $\sigma/\lambda = 3 \times 10^3$, the produced dark matter density is insufficient to trigger a significant disruption of the condensate as the contribution $\langle \rho_{\phi,\text{particle}} \rangle$ to $\langle \rho_{\phi} \rangle$ is practically negligible. For such a coupling, the averaged EOS parameter remains unaffected and close to $\langle w \rangle \simeq 0$, as expected in the weak coupling regime. For a coupling $\sigma/\lambda= 10^{4}$, at around $a/a_\text{end}\simeq 10-15$, the produced dark matter density becomes sufficient to alter the dynamics of the inflaton and to disrupt partially the condensate as the contribution $\langle \rho_{\phi,\text{condensate}} \rangle$ becomes subdominant compared to $\langle \rho_{\phi,\text{particle}} \rangle$. For such couplings, the EOS parameter slightly differs from $\langle w \rangle \simeq 0$ as the contribution from the relativistic dark matter particles to the expansion rate of the universe becomes non-negligible, and the inflaton fragments. For large couplings such as $\sigma/\lambda=4 \times 10^{4}$, after $a/a_\text{end} \gtrsim 10-15$, a rapid dark matter production fragments the condensate efficiently, and the universe becomes dominated by an admixture of relativistic dark matter particles and a particle-like inflaton component, which causes the EOS parameter to reach the values of $\langle w \rangle \simeq 1/3$ before relaxing to a pressureless value as the expansion proceeds. As the coupling is pushed towards larger values $\sigma/\lambda=4 \times 10^{5}$, a state of quasi-equilibrium is reached and the total produced dark matter density is saturated.

\paragraph{Phase space distribution.}

Below $\sigma/\lambda \simeq 5\times10^3$, the lattice code reproduces the PSD for the dark matter particle for sufficiently large occupation numbers. As an example, the right panel of Fig.~\ref{fig:BoltzVSNonpert} compares the Hartree distribution for $\sigma/\lambda=10^{7/2}$ (black) with the lattice PSD (green). For $q\lesssim 20$, both curves are very similar to each other, including the peaked features due to the resonant jumps in the production of particles for $q\gtrsim 2$. One visible difference is a smoother distribution maximum in the lattice case. However, the most important difference is the absence of the Boltzmann tail for the lattice distribution. It can be seen from the figure how all five estimates to $f_{\chi}$ approximately converge at $q\simeq 20$. For larger values of $q$, the kinematic blocking and bosonic enhancement effects can be disregarded. Since in this case the PSD is $f_{\chi}\ll 1$, the lattice approximation cannot be expected to be reliable. What we obtain as an output from the lattice code is an unphysical, growing UV distribution, which we have therefore cut out from the figure. As the $\phi$ field can be verified to maintain its homogeneity, for $q\gtrsim 20$ the Boltzmann result is expected to accurately describe the population of the dark matter UV modes.

\begin{figure*}[!t]
\centering
    \includegraphics[width=1\textwidth]{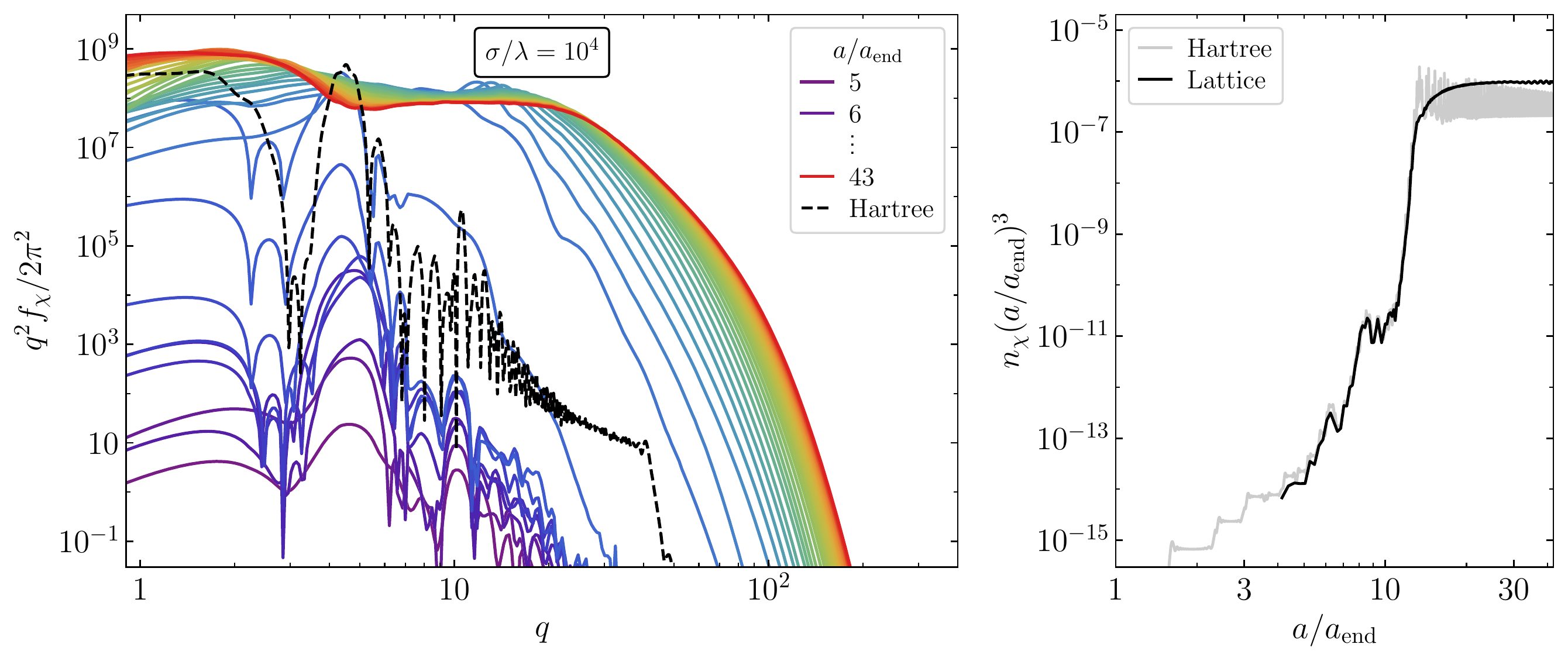}
    \includegraphics[width=1\textwidth]{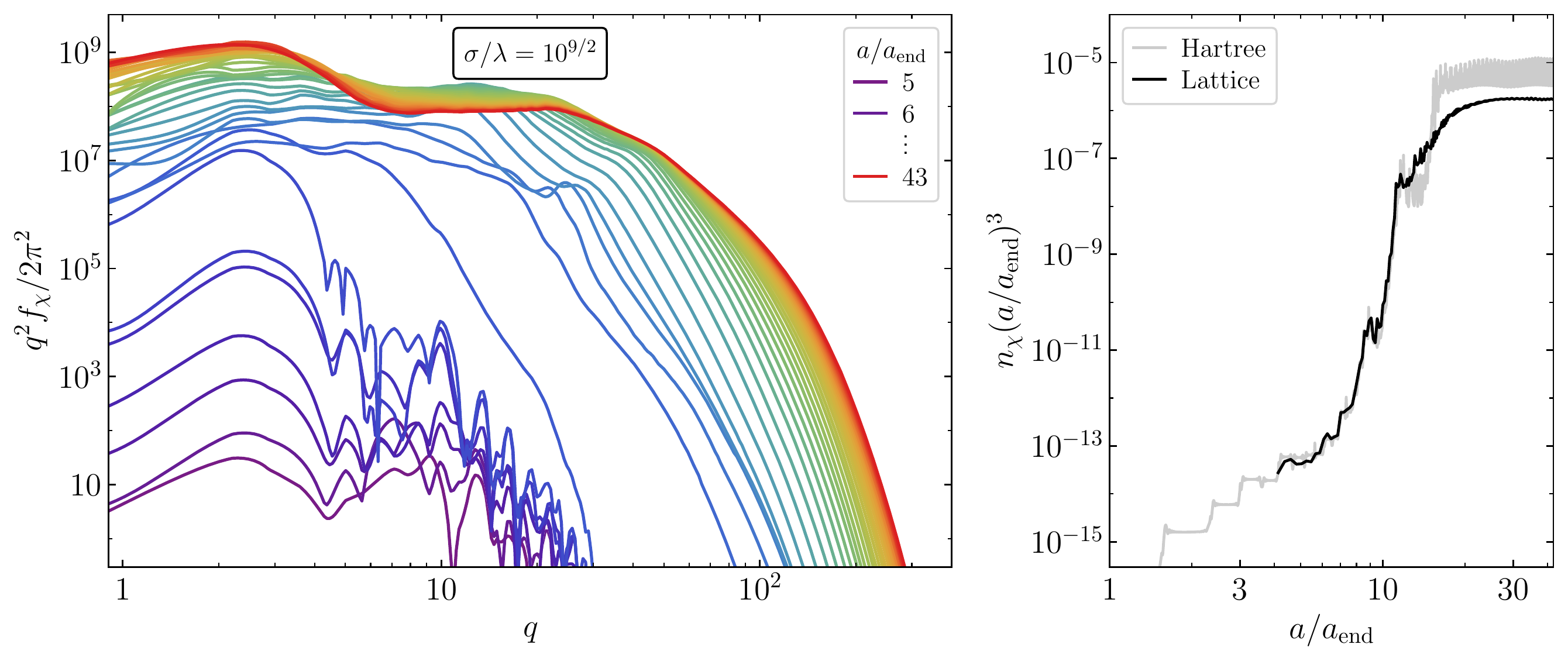}
    \includegraphics[width=1\textwidth]{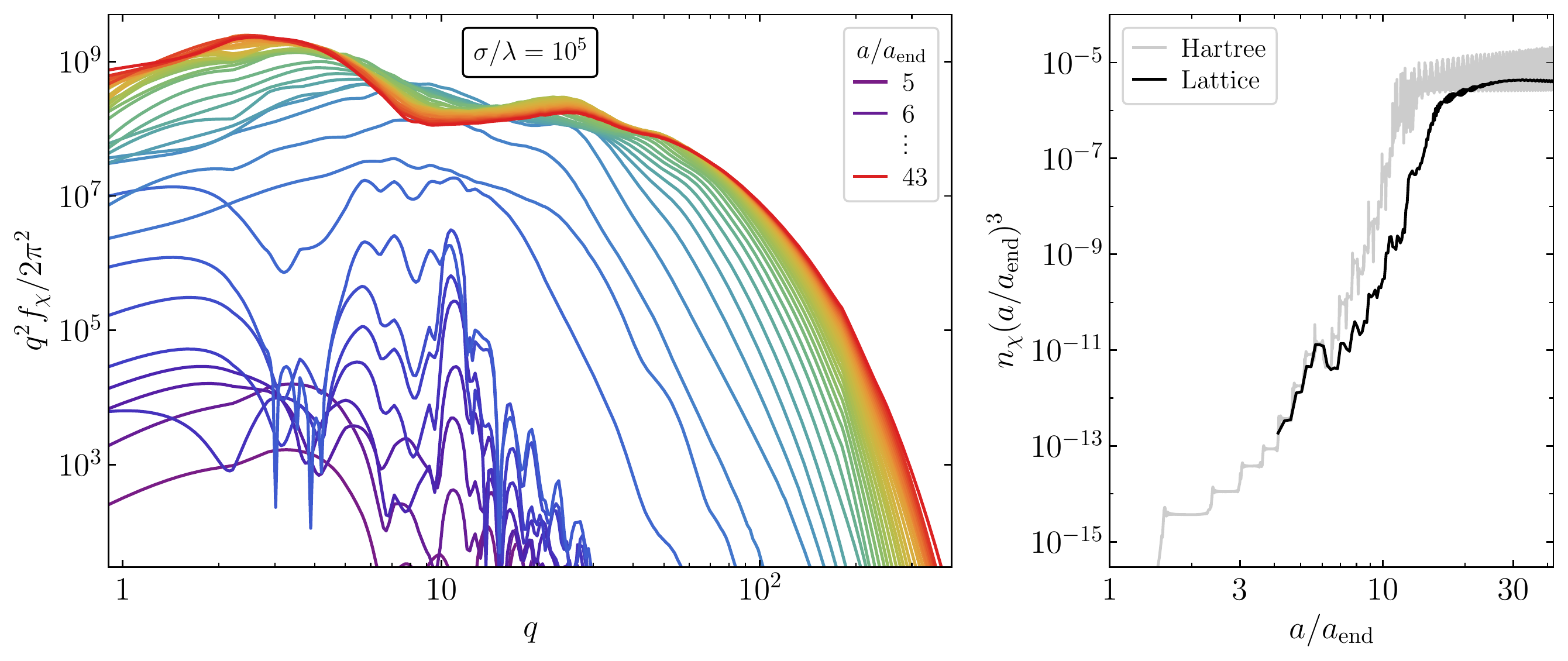}
    \caption{The phase space distribution as a function of the comoving momentum $q$ evaluated at different values of $a/a_{\rm{end}}$ (left), and the comoving number density as a function of the scale factor $a/a_{\rm{end}}$ (right), for selected values of the inflaton-DM coupling $\sigma/\lambda$ in the backreaction regime. Obtained from {\tt CosmoLattice} with $N=512^3$ lattice points and $k_{\rm{IR}} = 1$.}\label{eq:latticePSDS}
\end{figure*}

On the other hand, for $\sigma/\lambda \gtrsim 5\times10^3$, the inflaton loses its spatial homogeneity, and its PSD widens away from its initial Dirac delta form~(\ref{eq:psdinflaton}). To better visualize the effect of the backreaction on the evolution of the PSD of $\chi$, we show a series of snapshots of $f_{\chi}$ for a sequence of scale factor values $a/a_{\rm{end}}$ in the left panels of Fig.~\ref{eq:latticePSDS}. The right panels show the growth of the comoving number density as a function of $a/a_{\rm end}$ in the Hartree (solid black) and the lattice (solid gray) approximations. The first row corresponds to $\sigma/\lambda=10^4$. For this relatively low value of the coupling, as it is near the backreaction limit in Fig.~\ref{fig:nchiAllsmall}, the main effects of the rescattering can already be observed. At early times, $a/a_{\rm end}\lesssim 13$, the growth of the distribution of $n_{\chi}$ agrees with the Hartree result. The number density goes through several stages of exponential growth, and the last exponential growth occurs around $a/a_{\rm end}\simeq 10$ and boosts the number density by about five orders of magnitude. 

After this sudden growth, the energy density peaks at $\rho_{\chi}/\rho_{\phi} \simeq 0.4$, which affects the expansion rate $H$. Most importantly, it also affects the dynamics of the inflaton through the backward process $\chi\chi\rightarrow \phi\phi$. This depletes the condensate in favor of free $\chi$ and $\phi$ particles, as shown in Fig.~\ref{fig:fragmentation}. The energy exchange erases the resonance peaks of the distribution, and a rapid transfer of energy toward the UV modes is observed. For $q\lesssim 30$, the spectrum evolves toward having only two plateau-like regions, with $f_{\chi}\sim q^{-3/2}$, separated by a step-like feature, indicative of Kolmogorov-like turbulence~\cite{Micha:2002ey,Micha:2004bv}. This turbulent plateau propagates toward the UV, and beyond it, the PSD appears exponentially suppressed with a quasi-thermal tail $f_{\chi}\sim e^{-\alpha(t) q}$ (and similarly for $\phi$). We note that rescattering populates relativistic modes more efficiently than the perturbative decay of the $\phi$ condensate. This can be verified by comparing the lattice results with the Hartree PSD, shown in the left panel of Fig.~\ref{eq:latticePSDS} as the dashed black curve, evaluated at $a/a_{\rm end}=43$. 

Due to the efficient UV cascade during backreaction, the energy density of $\chi$ is dominated by relativistic modes, and the energy density redshifts as radiation, $\rho_{\chi}\propto a^{-4}$. This rapid dilution eventually terminates the strong coupling between the $\phi$ and $\chi$ sectors, and backreaction ends. After this point, the shape of $f_{\chi}$ at low momentum is frozen. The UV cascade can now only proceed through the perturbative scattering of free inflatons. The quasi-thermal tail of $f_{\phi}$ is mapped into the growth in the UV of the exponential tail of $\chi$ via the Boltzmann equation~\cite{Ballesteros:2020adh}. The result is a frozen comoving number density, a frozen PSD for $f_{\chi}\gtrsim 1$, and a frozen RMS value for $q$, which we discuss in the next section. Particle production eventually stops when the inflaton is fully depleted into visible matter and radiation. 

The middle and the bottom rows of Fig.~\ref{eq:latticePSDS} show the scale factor dependence of the PSD and the comoving number density of $\chi$ for two additional couplings in the backreaction regime, $\sigma/\lambda=10^{9/2}$ and $10^5$. Regarding the dynamics of the system, for these two couplings, our discussion is similar to the previous discussion for the $\sigma/\lambda=10^4$ case. The key differences are the following: As the coupling strengths vary, so does the pattern of resonant growth of the field. For $10^{9/2}$, the Hartree and lattice evolution decouple around $a/a_{\rm end}\simeq 11$, while for $10^5$ it occurs earlier, at $a/a_{\rm end}\simeq 7$. We also note that the final number density can be lower or higher than the corresponding Hartree result. Most importantly, the UV cascade gets stronger as the coupling increases. 

The comoving number density for a wide range of couplings $\sigma/\lambda \gg 1$, evaluated using lattice methods, is shown as the dashed blue line in Fig.~\ref{fig:nchiAllsmall}. We note that below the backreaction regime, the lattice results align with the Hartree results. On the other hand, in the presence of backreaction, the Hartree approximation does not agree with the lattice results, and for our analysis we use the more accurate lattice calculations. We have abstained from exploring arbitrarily large couplings, as for $\sigma/\lambda \gtrsim 10^6$ reaching the end of backreaction requires significant time and computational resources. Moreover, due to the lack of control over the formation of configuration space structures, such as oscillons, Q-balls, or black holes~\cite{Copeland:1995fq,Salmi:2012ta,Amin:2010dc,Amin:2011hj,Coleman:1985ki,Lee:1991ax,Kusenko:2008zm,Jedamzik:2010dq,Torres-Lomas:2014bua,Martin:2019nuw}, dissipative effects, and the growth of metric fluctuations, we do not extrapolate our results beyond this range.

\subsection{Dark matter abundance}

\begin{figure*}[!t]
\centering
    \includegraphics[width=0.9\textwidth]{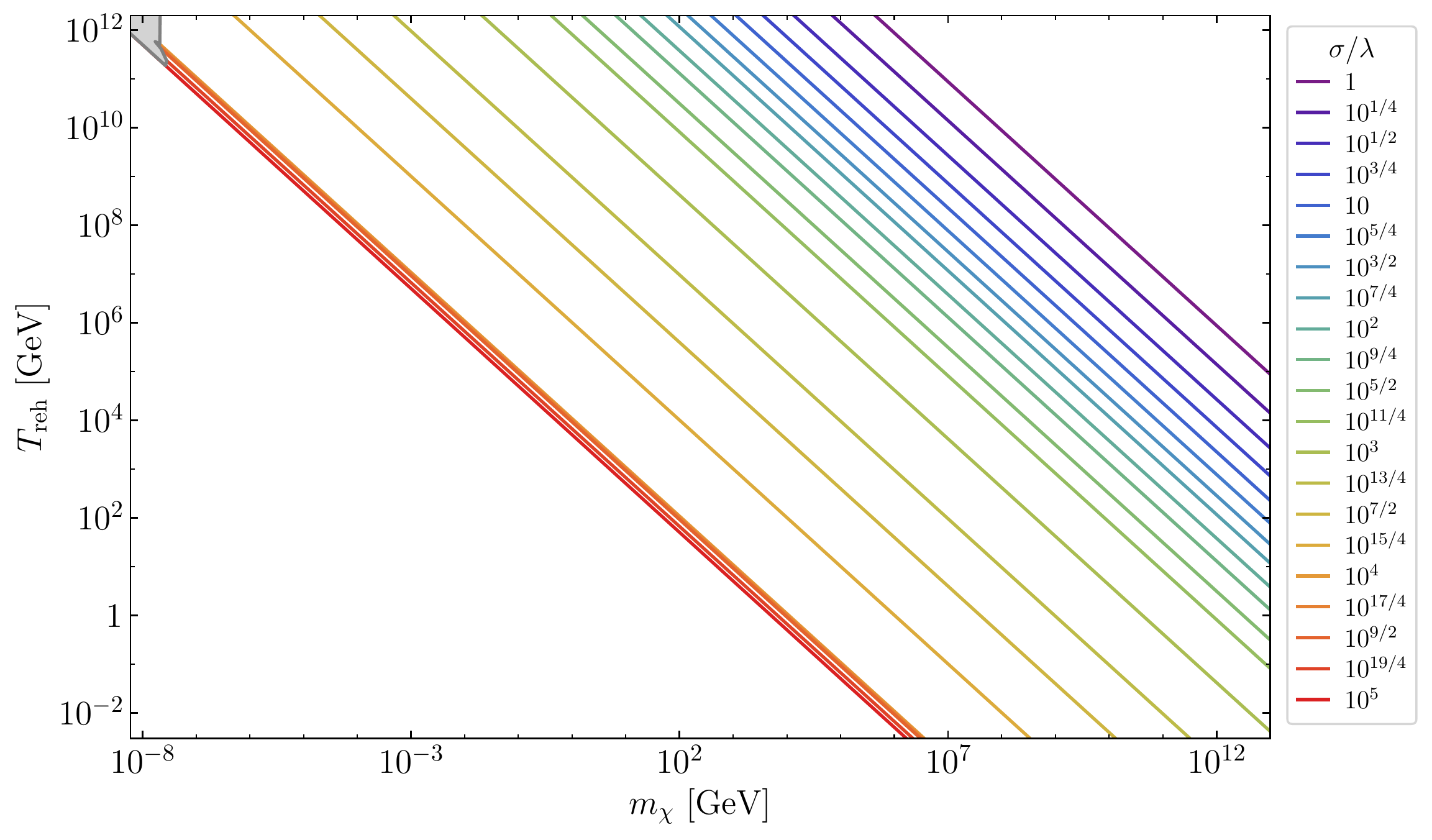}
    \caption{Parameter space for light scalar dark matter ($m_{\chi}<m_{\phi}$) with a strong coupling to the inflaton, $\sigma/\lambda\geq 1$. Along each line we use the observed dark matter relic abundance $\Omega_{\chi} h^2 \;=\; 0.12$. For $\sigma/\lambda>10^{7/2}$, we use the lattice results. Shaded in gray (top left corner) is the region excluded by the Lyman-$\alpha$ measurement of the matter power spectrum.}
    \label{fig:abundance_large}
\end{figure*}

In the absence of mass and IR-cutoff dependences of the comoving number density of $\chi$ for $\sigma>\lambda$, the DM relic abundance (\ref{eq:omegagrav}) can be rewritten as
\begin{align} \notag
    \Omega_\chi h^2 \; &\simeq \; 0.12  \left( \dfrac{2.05 \times 10^{-11}}{\lambda} \right)\left( \dfrac{n_\chi (a/a_\text{end})^3}{1.8 \times 10^{-12}\,M_P^3} \right) \\
    &\qquad \times \left( \dfrac{m_\chi}{1~\text{GeV}} \right)  \left( \dfrac{T_\text{reh}}{10^{10}~\text{GeV}} \right)  \,.
\end{align}
The reheating temperature necessary to saturate the observed abundance scales inversely with the DM mass and the comoving number density. This relation is shown explicitly in Fig.~\ref{fig:abundance_large} for couplings in the intermediate regime and beyond. We have included the line for $\sigma=\lambda$, which is also shown in Fig.~\ref{fig:abundance_small}. The parameter space above this line is excluded, as this coupling yields the smallest possible dark matter number density due to the gravitational interference. As the coupling $\sigma$ increases, the $\Omega_{\chi}h^2=0.12$ contours are displaced to the left (or downwards) in a manner determined by the comoving number densities, shown in Fig.~\ref{fig:nchiAllsmall}. We note that for $\sigma/\lambda \gtrsim 10^{7/4}$, it is possible to saturate the abundance at reheating temperatures lower than that for the case of gravitational-strength coupling.

\section{Lyman-$\alpha$ constraints}
\label{sec:lymalpha}

Contrary to the standard WIMP scenario, dark matter candidates produced in an out-of-equilibrium state with a generic momentum distribution could possess a non-negligible pressure component and depart substantially from the cold dark matter paradigm. This non-negligible pressure component, related to the energy density $\rho_\chi$ via the equation of state $P_\chi=w_\chi \rho_\chi$, would result in a suppression of clustered structure overdensities on the galactic distances. The suppression corresponds to a cutoff in the present matter power spectrum $\mathcal{P}(k)$ in Fourier space for modes with the wavenumber $k$ larger than the free-streaming horizon wavenumber $k_{\rm H}(a)$. This horizon scale can be expressed as~\cite{Heimersheim:2020aoc,Ballesteros:2020adh}
\begin{equation}
  k_\text{H}(a)\, \simeq \, \left[ \int_0^a \dfrac{\sqrt{10 w_\chi(\tilde a)}}{3\mathcal{H}(\tilde{a})}  \dfrac{\diff \tilde{a}}{\tilde{a}} \right]^{-1} \, ,
  \label{eq:FSH}
\end{equation}
where $\mathcal{H}\equiv aH$ is the conformal Hubble rate. The non-observation of this cutoff in the matter power spectrum allows to set constraints on non-thermal dark matter scenarios. In particular, the Lyman-$\alpha$ forest measurements constrain the cutoff scale to be $k_\text{H}(a=1)\simeq 15 \,h\,\text{Mpc}^{-1}$. This constraint is typically presented as a lower bound on the mass of warm dark matter (WDM) candidates $m_\text{WDM}>m_\text{WDM}^{\text{Ly}\mbox{-}\alpha}$ produced via thermal freeze-out~\cite{Narayanan:2000tp,Viel:2005qj,Viel:2013fqw,Baur:2015jsy,Irsic:2017ixq,Palanque-Delabrouille:2019iyz,Garzilli:2019qki},
\beq \label{boundWDM}
m_\text{WDM}^{\text{Ly}\mbox{-}\alpha}\;\simeq\; (1.9-5.3)~\text{keV at 95\% C.L.} 
\eeq
As can be seen from Eq.~(\ref{eq:FSH}), the suppression of the power spectrum at large $k$ is controlled by the parameter $w_\chi$ in the limit $w_\chi\ll 1$.\footnote{For a detailed discussion, see~\cite{Ballesteros:2020adh}} In this case, the equation of state parameter can be expressed as
\begin{equation}
 w_\chi \,  \simeq   \, \dfrac{T_{\star}^2}{3m_{\chi}^2}\frac{\langle q^2 \rangle}{a^2}\, ,
    \label{eq:eos_ncdm}
\end{equation}
where $T_{\star}$ (denoted also by $T_{\rm NCDM}$ in the literature) is a characteristic energy scale used to define the dimensionless comoving momentum $q$. In this work, $T_{\star}$ has been defined via Eqs.~(\ref{eq:qdef}) and~(\ref{eq:tstar}). The normalized second moment of the distribution function $ \langle q^2\rangle $ is computed for a generic distribution $f(q)$ as
\begin{equation}
    \langle q^2\rangle \, \equiv \, \dfrac{\int \diff q\, q^4 f(q) }{\int \diff q\, q^2 f(q) } \,.
    \label{eq:secondmoment}
\end{equation} 
One can estimate the Lyman-$\alpha$ lower bound on the dark matter mass $m_\chi \, >m_{\chi}^{\text{Ly}\mbox{-}\alpha}$ by matching the EOS parameter of Eq.~(\ref{eq:eos_ncdm}) to the EOS parameter of a WDM candidate whose mass corresponds to the constraint of Eq.~(\ref{boundWDM}). This condition yields~\cite{Ballesteros:2020adh}
\beq\label{eq:lyalphaconst}
 \,m_{\chi}^{\text{Ly}\mbox{-}\alpha} \;=\; m_{\rm WDM}^{\text{Ly}\mbox{-}\alpha} \left(\frac{T_{\star}}{T_{\rm WDM,0}}\right)\sqrt{\frac{\langle q^2\rangle}{\langle q^2\rangle_{\rm WDM}}}\,.
\eeq
Therefore, the constraint of Eq.~(\ref{eq:lyalphaconst}) only depends on the first and second moments of the corresponding PSD. Having the phase space distributions for the complete range of couplings, we now proceed to discuss the effect of the power spectrum constraint on the parameter space of the scalar dark matter $\chi$. We emphasize that one fundamental assumption about the procedure used here is an initial (early times of radiation domination era) adiabatic DM overdensity~\cite{Ballesteros:2020adh}. A discussion regarding the nature of perturbations is postponed to Sec.~\ref{sec:discussion}. In the remaining part of this section, we assume adiabatic initial conditions for the dark matter perturbations.

\subsection{The perturbative DM mass bound}

We begin by deriving the Lyman-$\alpha$ bound on the production of $\chi$ quanta in perturbation theory. From our computation of $n_{\chi}$, we expect the validity of this result to hold for a very limited range of the inflaton-DM couplings. Nevertheless, it is instructive to compute this bound as it can be determined analytically. Moreover, we also find that for almost all values of the effective couplings $\sigma/\lambda \lesssim 5\times10^3$, for sufficiently large $q$, the UV tail of the distribution is populated perturbatively. As we show below, this makes the Boltzmann result an indispensable benchmark to extract the power spectrum constraint from the numerically computed PSD of $\chi$. 

We have found that perturbatively, $f_{\chi}\sim f_{\chi}^c\sim q^{-9/2}$ for $q\gg 1$,
and this result does not depend on the kinematic suppression or bosonic enhancement effects (see, e.g., Fig.~\ref{fig:BoltzVSNonpert}). The comoving number density of dark matter, which is proportional to the denominator of Eq.~(\ref{eq:secondmoment}), is practically insensitive to the end of reheating, as it rapidly converges to an asymptotic value for $a_{\rm reh}\gg a_{\rm end}$. On the other hand, the numerator of Eq.~(\ref{eq:secondmoment}) grows without a bound if the cascade to higher $q$ with $ f_{\chi}\sim q^{-9/2}$ proceeds until arbitrarily late times. However, this does not occur as the production of $\chi$ is heavily suppressed after reheating. The Gaussian tail in Eq.~(\ref{eq:fchiappreh}) then plays the role of the regulator required to compute the RMS momentum.

In order to obtain a closed form expression for the structure formation constraint on $m_{\chi}$, we substitute the approximation for the PSD of dark matter at present times (\ref{eq:fchiappreh}) into the definition of $\langle q^2\rangle$, given by Eq. (\ref{eq:secondmoment}). In this limit, that does not include the kinematic and statistical effects, we find 
\begin{align} \notag
    \langle q^2\rangle \;&\simeq\; 0.641 \left( \frac{a_{\rm reh}}{a_{\rm end}} \right)^2\; \dfrac{\Gamma\left(1/4,1.56 (a_{\rm end}/a_{\rm reh})^2 \right)}{\Gamma\left(-3/4,1.56 (a_{\rm end}/a_{\rm reh})^2 \right)}\\ \label{eq:q2boltz}
&\simeq\;2.433\sqrt{\frac{a_{\rm reh}}{a_{\rm end}}}\,,
\end{align}
where in the last step we used $a_{\rm{reh}} \gg a_{\rm{end}}$.
Combining the expression for $T_{\star}$, defined in (\ref{eq:tstar}), and using the ratio $a_{\rm end}/a_0$, given by (\ref{eq:fulla0}), with
\beq
\frac{a_{\rm reh}}{a_{\rm end}} \;\simeq\; \left(\frac{\rho_{\rm end}}{\rho_{\rm reh}}\right)^{1/4}\left(\frac{H_{\rm end}}{\Gamma_{\phi}}\right)^{1/6}\,,
\eeq
the substitution into Eq.~(\ref{eq:lyalphaconst}) yields
\begin{align}
m_{\chi} &> 15.78\,{\rm keV}\,\left(\frac{m_{\rm WDM}^{\text{Ly}\mbox{-}\alpha}}{3\,{\rm keV}}\right)^{4/3} \hspace{-10pt} m_{\phi}\, \rho_{\rm end}^{-1/4}g_{\rm reh}^{-1/12}\\ \notag
& \simeq\; 32.4\,{\rm eV} \left(\frac{m_{\rm WDM}^{\text{Ly}\mbox{-}\alpha}}{3\,{\rm keV}}\right)^{4/3}\left(\frac{\lambda}{2.05\times 10^{-11}}\right)^{1/4}\\ \label{eq:mdmbound}
&\qquad \times \left(\frac{427/4}{g_{\rm reh}}\right)^{1/12} \, .
\end{align}
Note that all the dependence on the duration of reheating and on the inflaton-DM coupling $\sigma$ cancels out. The very early decoupling of the scalar field $\chi$ results in a dark matter candidate that can mimic a cold dark matter relic even for masses as low as a few eV.%
\begin{figure}[!t]
\centering
    \includegraphics[width=\columnwidth]{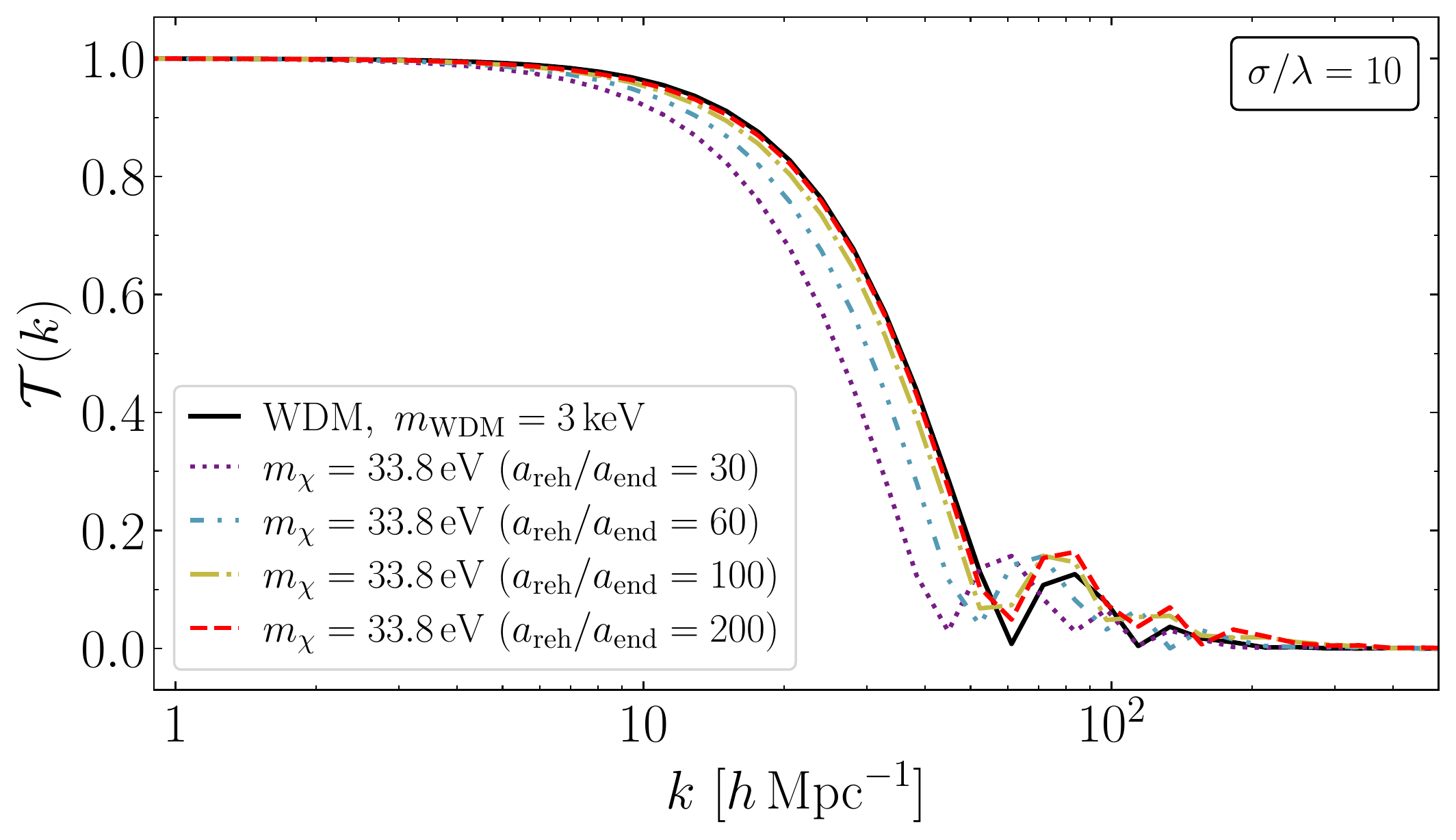}
    \caption{Linear transfer function, defined in Eq.~(\ref{eq:definetransferfunction}), for scalar DM production with $\sigma/\lambda=10$. The result was computed with {\tt CLASS} and based on the numerical distribution shown in Fig.~\ref{fig:PSDlargesigma}, completed with the perturbative Boltzmann tail~(\ref{eq:fchiappreh}) near and beyond the end of reheating. Different curves correspond to different duration of the reheating epoch, demonstrating the convergence for $a_{\rm reh}\gg a_{\rm end}$. The transfer function for the WDM case is shown for comparison with a black line. }
    \label{fig:1e1Tk}
\end{figure}
We show in Fig.~\ref{fig:1e1Tk} the form of the transfer function
\beq
\mathcal{T}(k) \;\equiv \; \left(\frac{\mathcal{P}(k)}{\mathcal{P}_{\Lambda \rm CDM}(k)}\right)^{1/2}\,
\label{eq:definetransferfunction}
\eeq
where $\mathcal{P}(k)$ is the linear matter power spectrum evaluated for $\sigma/\lambda=10$ and computed with {\tt CLASS}~\cite{Blas:2011rf,Lesgourgues:2011rh}, and $\mathcal{P}_{\Lambda \rm CDM}(k)$ is the linear power spectrum expected in the $\Lambda $CDM model. As we discussed in Section~\ref{sec:largecoupling}, for this value of the effective coupling, the perturbative approximation $f_{\chi}^c$ is a good fit for the exact PSD calculated using the Hartree formalism. In order to properly mimic the end of reheating, the distribution shown in Fig.~\ref{fig:BoltzVSNonpert} has been completed with the corresponding Gaussian tail, as given by Eq.~(\ref{eq:fchiappreh}).

We show five lines in Fig.~\ref{fig:1e1Tk}. The solid black line corresponds to the WDM transfer function for $m_{\rm WDM}=3\,{\rm keV}$. The remaining color-coded lines correspond to the numerical evaluation of $\mathcal{T}(k)$ for different duration of reheating, with mass set to the computed lower bound, $m_{\chi}=33.8\,{\rm eV}$. This mass is obtained via a second re-scaling of Eq.~(\ref{eq:mdmbound}), as described below (see Eq.~(\ref{eq:mdmrescale})). We note that the curve corresponding to $a_{\rm reh}=30\,a_{\rm end}$ has a different cutoff scale than the WDM case. Nevertheless, as reheating is delayed, the WDM and $\chi$ bounds are almost identical, coinciding for $a_{\rm reh}/a_{\rm end}>100$. This asymptotic behavior is consistent with the corresponding saturation of the comoving number density at $a(t)\gg  a_{\rm end}$. The coincidence of the transfer functions for our light scalar DM and WDM validates the use of Eq. (\ref{eq:lyalphaconst}) to derive Eq. (\ref{eq:mdmbound}).

In the following (sub)sections we derive the form of the Lyman-$\alpha$ bound in the weak and strong coupling regimes based on the matching formula (\ref{eq:lyalphaconst}) and using the numerically computed phase space distributions.

\subsection{Bounds on DM production at weak coupling}

We first derive the bounds that structure formation imposes on the mass of the DM scalar $\chi$ using the numerically computed phase space distributions in the Hartree approximation. In this section, we consider small couplings $\sigma/\lambda\leq 1$. For all couplings in this range, we expect a weaker bound than the perturbative bound (\ref{eq:mdmbound}). For $\sigma/\lambda<1$ this occurs due to the dominance of the cold IR modes in the PSD. For $\sigma=\lambda$, this happens because of the suppression of the UV modes due to the interference between gravitational and decay production of $\chi$.

For $\sigma/\lambda\ll 1$, the PSD form is shown in Figs.~\ref{fig:PSDgrav_masses} and \ref{fig:PSDsmallsigma}. In this regime, the integral of $q^2f_{\chi}$, which determines the comoving number density, is sensitive to the horizon scale cutoff $q_0$. However, the integral of $q^4f_{\chi}$ is not IR sensitive, and the dominant contribution comes from the perturbative UV tail. Therefore, to calculate the RMS comoving momentum, we consider the following steps: (1) We extend the PSD into the IR, as described in Section~\ref{sec:three}, given that $q_0\ll 1$ for the reheating temperatures that saturate $\Omega_{\rm DM}$. This IR extension is mass-dependent, and we construct it for a wide range of masses $m_{\chi}\ll m_{\phi}$. (2) We also extend the distribution into the UV, including the perturbative Gaussian tail in (\ref{eq:fchiappreh}) that appears for particles produced beyond the end of reheating. This UV extension is mass-independent, as shown in Fig.~\ref{fig:PSDgrav_masses}. (3) We use this doubly-extended distribution to numerically evaluate $\langle q^2\rangle$ from Eq. (\ref{eq:secondmoment}). (4) For all cases, we find that $\langle q^2\rangle \propto m_{\chi}^{\alpha}  \sqrt{a_{\rm reh}/a_{\rm end}}$ for $a_{\rm reh}\gg a_{\rm end}$, with $\alpha<1/2$. The mass dependence arises from the contribution of the IR modes to the normalization of the RMS integral. (5) This mass-dependent prefactor for $\langle q^2\rangle$ allows us to re-scale the perturbative $m_{\chi}$ bound (\ref{eq:mdmbound}) to the corresponding non-perturbative value by solving the following equation
\beq\label{eq:mdmrescale}
\left( m_{\chi}^{\text{Ly}\mbox{-}\alpha} \right)_{\rm non\mbox{-}pert} \;=\; \left( m_{\chi }^{\text{Ly}\mbox{-}\alpha}  \right)_{\rm pert} \sqrt{\frac{\langle q^2\rangle_{\rm non\mbox{-}pert}}{\langle q^2\rangle_{\rm pert}}}\,.
\eeq

\begin{figure*}[!t]
\centering
    \includegraphics[width=0.9\textwidth]{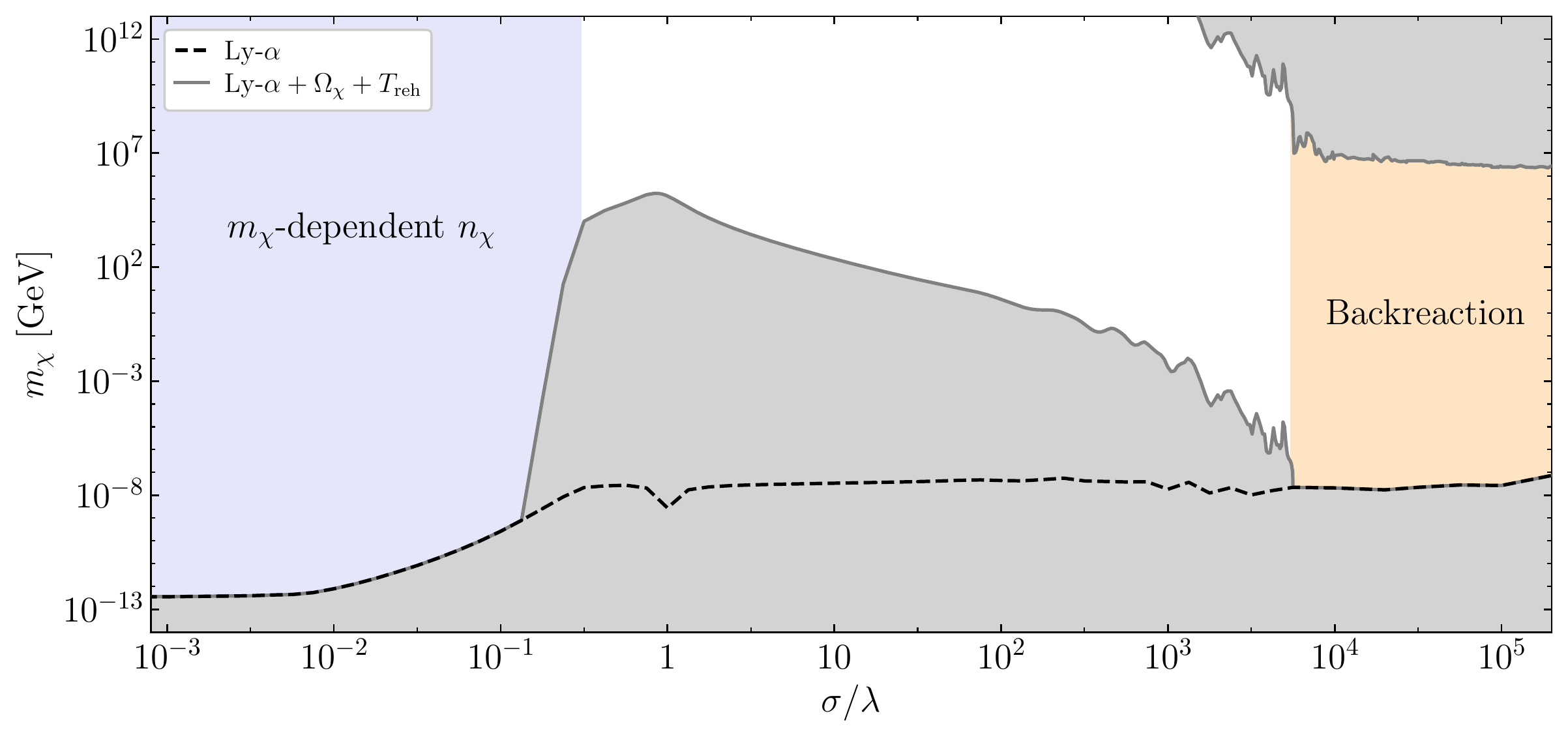}
    \includegraphics[width=0.9\textwidth]{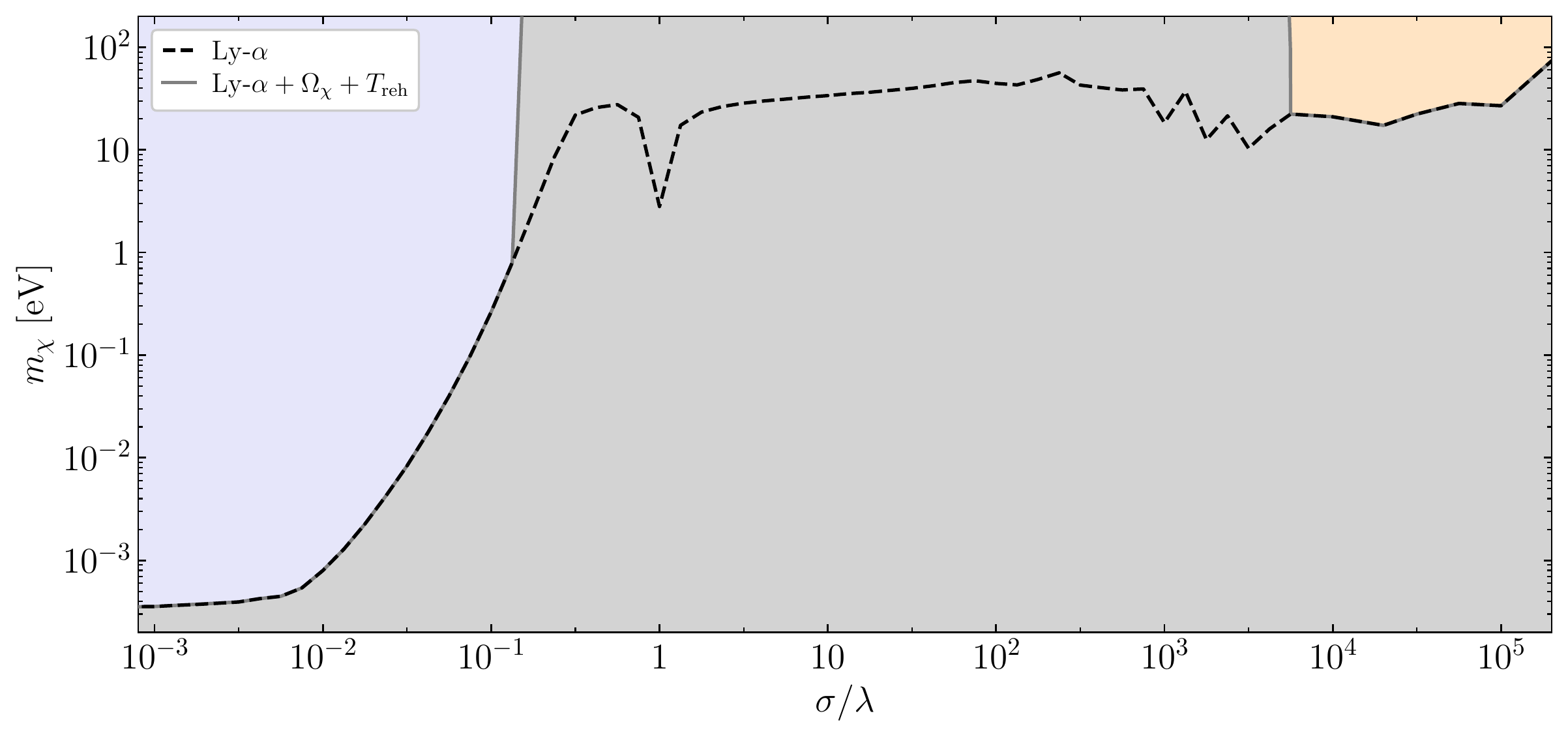}
    \caption{Combined relic abundance and structure formation constraints on the mass of a scalar DM candidate with action (\ref{eq:actionchi}) for the complete range of couplings considered in this work (in gray). The black dashed line corresponds exclusively to the matter power spectrum lower bound on $m_{\chi}$. In those regions constrained more strongly by $\Omega_{\chi}h^2=0.12$, the bound comes from the perturbative reheating assumption, $T_{\rm reh} < m_{\phi}$ (lower bound on $m_{\chi}$) or $T_{\rm reh}>T_{\rm BBN}$ (upper bound on $m_{\chi}$). The top panel shows the full range of masses up to $m_{\chi}<m_{\phi}$. The bottom panel shows the detail of the matter power spectrum constraint.}
    \label{fig:massbounds}
\end{figure*}

As expected, the general result is the weakening of the Lyman-$\alpha$ constraint at small couplings due to the dominance of the cold IR modes in the PSD. The saturation of the relic abundance and the structure formation bounds are shown simultaneously in Fig.~\ref{fig:massbounds} and in Fig.~\ref{fig:abundance_small}. Notably, the lower bound on the mass for the gravitationally-produced scalar dark matter corresponds to $m_{\chi} > 3.4\times 10^{-4}\,{\rm eV}$. Therefore, super-light\footnote{Not to confuse it with coherent {\em ultra-light} scalar DM, with $m_{\chi}\gtrsim 10^{-22}\,{\rm eV}$~\cite{Marsh:2015xka,Hui:2016ltb}.} scalar dark matter is allowed in the absence of a direct coupling to $\phi$, although requiring a reheating temperature below the electroweak scale. The Ly-$\alpha$ constraint strengthens as the coupling $\sigma$ increases. Here one must recall that for $\sigma/\lambda \lesssim 10^{-1/2}$, the PSD and the comoving number density depend on the DM mass. Hence, the power spectrum bound is saturated for masses many orders of magnitude beyond our numerical capabilities, and only in this range of couplings we are not able to cross-check our claimed value for $m_{\chi}^{\text{Ly}\mbox{-}\alpha}$ with {\tt CLASS}.

For larger couplings, $10^{-1/2}\lesssim \sigma/\lambda <1$, the comoving number density is independent of $m_{\chi}$ and the previously discussed re-scaling scheme for the Lyman-$\alpha$ bound is simplified, as $\langle q^2\rangle_{\rm non\mbox{-}pert}$ in (\ref{eq:mdmrescale}) is $m_{\chi}$-independent. Moreover, the consistency check with {\tt CLASS} is feasible, and in all cases we find a good agreement, up to a few percent in the worst cases. The corresponding Lyman-$\alpha$ bounds are also shown in Fig.~\ref{fig:massbounds} as the black dashed line. In this range of couplings, we note that the strongest constraint on $m_{\chi}$ comes not from $\mathcal{T}(k)$, but from the requirement that $\Omega_{\chi}h^2=0.12$. More specifically, the simultaneous saturation of the Ly-$\alpha$ and the relic abundance constraints would require reheating temperatures of magnitude far beyond the bound $T_{\rm reh}<m_{\phi}$, as can be inferred from Fig.~\ref{fig:abundance_small}. This constraint is artificial, imposed by the assumption of a perturbative decay of $\phi$ into the visible sector, as described in Section~\ref{sec:model}. The end result is lower bounds as large as $m_{\chi}\gtrsim 1.7\times 10^{5}\,{\rm GeV}$ for $\sigma/\lambda \lesssim 1$.

The sole weak coupling for which the re-scaling program based on Eq.~(\ref{eq:mdmrescale}) is not applicable, is $\sigma=\lambda$. In this case, as it is discussed in Section~\ref{sec:weakcoupling}, the PSD decreases at a faster rate than the generic Boltzmann distribution, with $f_{\chi}\sim q^{-15/2}$. The steeper decrease in the UV translates in a RMS comoving momentum that becomes independent of the duration of reheating for $a_{\rm reh}\gg a_{\rm end}$. Therefore, in this case, the Ly-$\alpha$ bound by itself carries a dependence on the reheating temperature from $T_{\star}$,
\begin{align}\label{eq:mdmsigma1}
m_{\chi} \;&>\; 9.58\,{\rm keV} \,\left(\frac{m_{\rm WDM}^{\text{Ly}\mbox{-}\alpha}}{3\,{\rm keV}}\right)^{4/3} \sqrt{\langle q^2\rangle} \frac{m_{\phi} T_{\rm reh}^{1/3}}{\rho_{\rm end}^{1/3}}\,.
\end{align}
In principle, the dependence on $T_{\rm reh}$ can be eliminated assuming the DM abundance is saturated. If this were the case, we could make use of the relation (\ref{eq:phenofit}) between $T_{\rm reh}$ and $m_{\chi}$, with $\gamma=-1$ (see Fig.~\ref{fig:fit_small}), and substitute in the previous equation. However, the solution for $m_{\chi}$ that saturates both bounds would require $T_{\rm reh}\gg M_P$. If we instead substitute $T_{\rm reh}\simeq m_{\phi}$ in (\ref{eq:mdmsigma1}), we obtain $m_{\chi}\gtrsim 2.8\,{\rm eV}$, which is the value shown in the dashed line in Fig.~\ref{fig:massbounds}. We note, nevertheless, that this mass is far from saturating the relic abundance with $T_{\rm reh}=m_{\phi}$, and instead, the smallest DM mass consistent with the mechanism studied in this paper corresponds to $m_{\chi}\simeq 1.2\times 10^5\,{\rm GeV}$.

\subsection{Bounds on DM production at large coupling}

For $\sigma/\lambda>1$, the Hartree approximation correctly accounts for the production of scalar DM during reheating in the absence of strong backreaction. In this Hartree range, the re-scaling of the perturbative bound (\ref{eq:mdmbound}) via (\ref{eq:mdmrescale}) is feasible, and straightforward, since the second moment of the PSD is mass-independent, with a perturbative dependence on the duration of reheating. That is, $\langle q^2\rangle_{\rm non\mbox{-}pert}\propto \sqrt{a_{\rm reh}/a_{\rm end}}$, and the constant of proportionality alone determines the re-scaling of the Lyman-$\alpha$ bound. A graphical demonstration of the robustness of this program is shown in Fig.~\ref{fig:1e1Tk}, discussed in detail above.

The resulting values for $m_{\chi}$ can be seen in Fig.~\ref{fig:massbounds}, in the range corresponding to $1<\sigma/\lambda \lesssim 5\times10^3$. We note that, unsurprisingly, the mass constraint does not deviate significantly from the Boltzmann value (\ref{eq:mdmbound}). However, throughout this range, the saturation of the DM density parameter $\Omega_{\chi}$ with $T_{\rm reh}<m_{\phi}$ imposes a stricter bound by several orders of magnitude. In the top panel of the figure, we also observe that for $\sigma/\lambda \gtrsim 1.5\times 10^3$, an upper bound on $m_{\chi}$ appears. Particle production at strong couplings enters the broad resonance range and super-heavy DM could easily overclose the universe. The increase in $\rho_{\chi}$ would need to be compensated with a reheating temperature lower than that necessary for successful Big Bang Nucleosynthesis (BBN), $T_{\rm BBN}\simeq 2\,{\rm MeV}$.

Veering into the backreaction regime, for couplings $\sigma/\lambda \gtrsim 5\times10^3$, the effect of the broad parametric resonance is a copious production of DM, and the fragmentation of the inflaton condensate. As we discussed in Section~\ref{sec:backreaction}, the lack of coherence of the inflaton results in the loss of the Boltzmann tail in the UV, and the UV cascade proceeds instead via the population of an exponential, quasi-thermal tail. When backreaction ends and the pure redshift regime begins, the shape of this tail freezes and the UV cascade can proceed perturbatively via the scattering of inflatons, $\phi\phi \rightarrow \chi\chi$, completing the exponential tail of the distribution in the UV. The absence of a power-law tail implies then that $\langle q^2\rangle \rightarrow {\rm const.}$ at the end of backreaction. Therefore, similarly to the $\sigma=\lambda$ case, the RMS value of the rescaled comoving momentum converges to an $a_{\rm reh}$-independent constant for $a_{\rm reh}/a_{\rm end}\gg 1$, and the Lyman-$\alpha$ bound is dependent on the reheating temperature, as shown in Eq.~(\ref{eq:mdmsigma1}). 

\begin{figure}[!t]
\centering
    \includegraphics[width=\columnwidth]{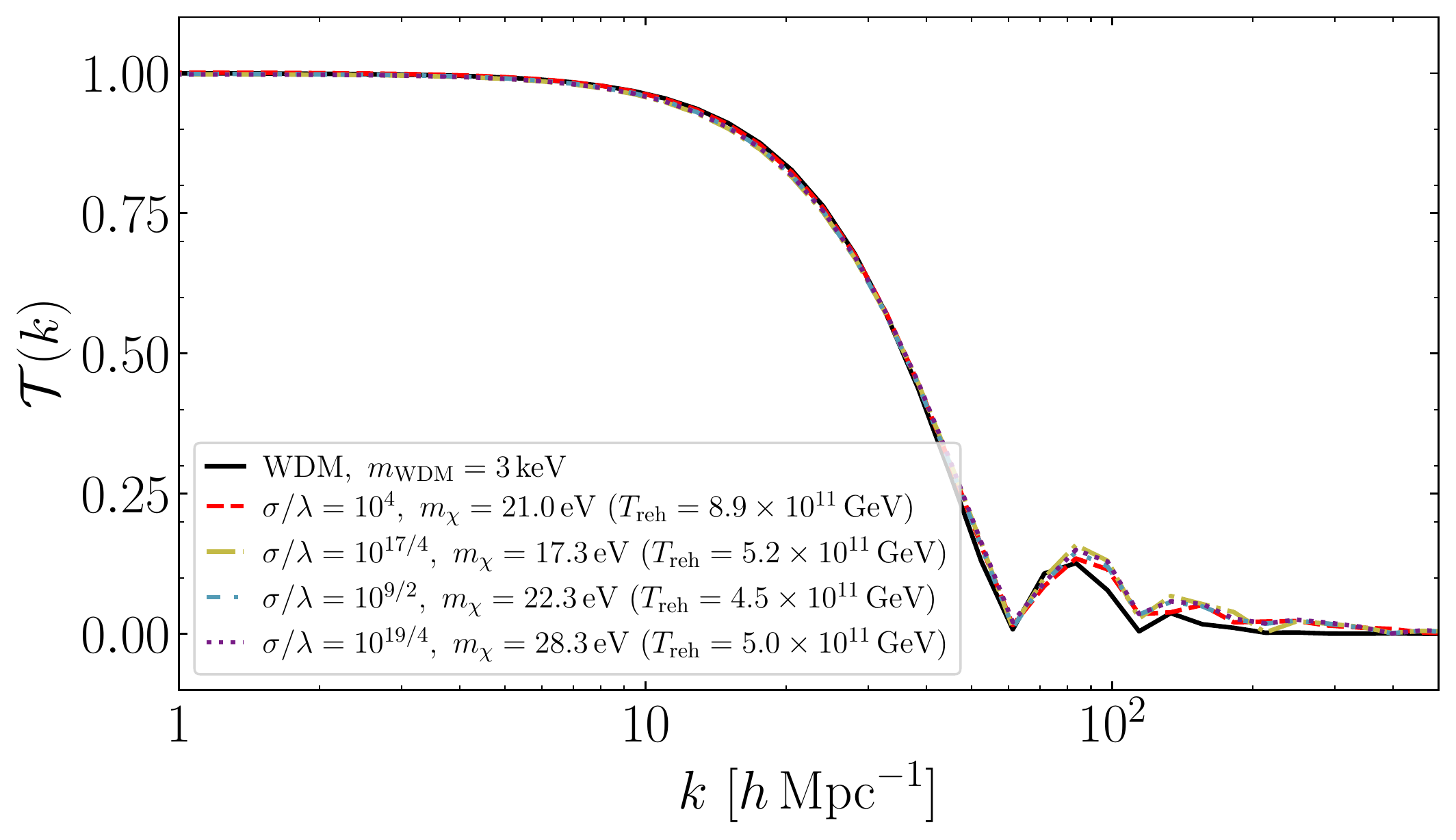}
    \caption{Linear transfer function, defined in Eq.~(\ref{eq:definetransferfunction}), computed with {\tt CLASS}, for scalar DM production in the backreaction regime. For each coupling, the combined $\Omega_{\rm DM}$ and Lyman-$\alpha$ constraints are saturated for the smallest possible scalar mass (maximum reheating temperature), as given by Eq.~(\ref{eq:mdmbackreaction}). The transfer function for the WDM case is shown for comparison with a black line.}
    \label{fig:latticeTk}
\end{figure}

For large couplings, we can then substitute into (\ref{eq:mdmsigma1}) the relation (\ref{eq:phenofit}) between $T_{\rm reh}$ and $m_{\chi}$ (with $\gamma=-1$) without violating the assumption that $T_{\rm reh}<m_{\phi}$. Algebraic manipulation leads to the following expression for the power spectrum constraint,
\begin{align} \notag
m_{\chi} \;&>\; (0.98\,{\rm MeV})^{5/4} \left(\frac{m_{\rm WDM}^{\text{Ly}\mbox{-}\alpha}}{3\,{\rm keV}}\right) \\
&\qquad \times\langle q^2\rangle^{3/8} \tau^{1/4} m_{\phi}^{3/4} \rho_{\rm end}^{-1/4}\,.
\end{align}
Specializing to T-attractor inflation, we can equivalently write
\begin{align} \notag
m_{\chi}\;&>\; 0.26\,{\rm eV}\,\left(\frac{m_{\rm WDM}^{\text{Ly}\mbox{-}\alpha}}{3\,{\rm keV}}\right) \\ \label{eq:mdmbackreaction}
&\qquad \times \langle q^2\rangle^{3/8} \tau^{1/4} \left(\frac{\lambda}{2.05\times 10^{-11}}\right)^{1/8}\,.
\end{align}
The RMS comoving momentum and $\tau$ can be directly calculated from the distributions obtained from the lattice code. Upon inputting the corresponding DM mass bound and the PSD for a set of four couplings in the backreaction regime into {\tt CLASS} one obtains Fig.~\ref{fig:latticeTk}. In all cases, the free-streaming scale is a good match to the WDM one at $m_{\rm WDM}=3\,{\rm keV}$. We note that the growth of the bound with the coupling is not monotonous, due to the quasi-stochastic nature of the parametric resonance during reheating. We also note that $\mathcal{T}(k)$ are evaluated at the reheating temperatures for which the Lyman-$\alpha$ and the $\Omega_{\chi}$ bounds are simultaneously saturated. A smaller $T_{\rm reh}$ requires a larger DM mass to saturate the density parameter constraint. 

The dashed black line in Fig.~\ref{fig:massbounds} shows the Lyman-$\alpha$ bound as a function of the inflaton-DM coupling strength. We observe that, for $\sigma/\lambda \gtrsim 5\times10^3$, the effect of fragmentation and the corresponding cascade populates UV modes in an efficient manner. Therefore, the lower bound on $m_{\chi}$ is determined solely by the absence of suppression of structure below $k\simeq 15 \,h\,\text{Mpc}^{-1}$ scales. Interestingly though, we note that the numerical value of the minimum DM mass is not far from that predicted by the Boltzmann approximation, being $m_{\chi}\gtrsim 27\,{\rm eV}$ for $\sigma/\lambda = 10^5$.

\section{Effective number of relativistic species}
\label{sec:Neff}

As stated in the Introduction, in this work we explore the production of scalar dark matter in the purely gravitational, weak direct coupling (perturbative), and strong direct coupling (non-perturbative) regimes. We find a weakening of the Lyman-$\alpha$ constraint on the DM mass with respect to WDM, 
which can be explained by the very early production of DM during reheating (or during inflation). However, such a light dark particle could contribute to the effective number of degrees of relativistic species, $N_{\rm eff}$, at early times, during BBN, or recombination. After the end of reheating, the total energy density of the universe can be written as
\beq
\rho \;=\; \left[1 + \frac{7}{8}\left(\frac{T_{\nu}}{T}\right)^4N_{\rm eff} \right]\rho_{\gamma} + \rho_{\chi} + \cdots\,,
\eeq
where $\rho_{\gamma}$ corresponds to the energy density in photons, $T_{\nu}$ is the effective neutrino temperature, satisfying the constraint $T_{\nu}/T=(4/11)^{1/3}$ after the electron-positron annihilation, and the dots correspond to the contribution of all other contents of the universe. For couplings $\sigma/\lambda \gtrsim 10^{-1/2}$, that we discussed in Section~\ref{sec:weakcoupling}, the form of the PSD of the dark species $\chi$ is independent of its mass. Therefore, we can maximize the contribution to $N_{\rm eff}$ by identifying $\chi$ not with a dark matter component, but with a {\em dark  radiation} one, with $m_{\chi}\rightarrow 0$. In this case, the correction to the Standard Model value, $N_{\rm eff}=3.046$, is given by
\beq\label{eq:deltaNeff}
\Delta N_{\rm eff} \;=\; \frac{8}{7}\left(\frac{T}{T_{\nu}}\right)^4 \frac{\rho_{\chi}}{\rho_{\gamma}}\,.
\eeq
The energy density of $\chi$ can be evaluated at late times by finding its maximum value during preheating before $\chi$ decouples and begins redshifting like radiation. This maximum energy density, denoted by $\rho_{\chi,{\rm max}}$, can be evaluated in the Hartree approximation, or with the use of lattice codes~\cite{Garcia:2021iag}. Similarly, we denote the scale factor at which this maximum is reached by $a_{\rm max}$ . Evaluating the photon energy density in terms of the temperature at sufficiently late times, with the ratio $T/T_{\nu}$ fixed to its present value, we can then rewrite Eq. (\ref{eq:deltaNeff}) as 
\begin{align}\notag
\Delta N_{\rm eff} \;&=\; \frac{120}{7\pi^2} \left(\frac{11}{4}\right)^{4/3} \frac{\rho_{\chi,{\rm max}}}{T^4}\left(\frac{a_{\rm max}}{a}\right)^4\\ \notag
&\simeq\; \frac{4}{7} \left[ g(T)^4\left(\frac{11}{4}\right)^4 \frac{\pi^2}{30}\left(\frac{m_{\phi}^4}{\rho_{\rm end}}\right)\left(\frac{T_{\rm reh}}{m_{\phi}}\right) ^{4}\right]^{1/3}\\ 
& \qquad \times \left(\frac{\rho_{\chi,{\rm max}}}{\rho_{\rm end}}\right)\left(\frac{a_{\rm max}}{a_{\rm end}}\right)^4\,.
\end{align}
To derive the second line, we have assumed adiabatic expansion after the end of reheating. Using the T-attractor model of inflation values, with $m_{\phi}$ given by Eq.~(\ref{eq:infmass}) and $\rho_{\rm{end}}$ given by Eq.~(\ref{eq:rhoend}), we find
\begin{align} \notag
\Delta N_{\rm eff} \;&\simeq\; 6.7\times 10^{-4} \left( g(T) \frac{T_{\rm reh}}{m_{\phi}}\right)^{4/3}\\ \notag
&\qquad\qquad\qquad \times \left(\frac{\rho_{\chi,{\rm max}}}{\rho_{\rm end}}\right)\left(\frac{a_{\rm max}}{a_{\rm end}}\right)^4 \\
&\lesssim\; 2.4\times 10^{-3} \left( g(T) \frac{T_{\rm reh}}{m_{\phi}}\right)^{4/3}\, ,
\end{align}
where to calculate the second inequality we used the numerical values from~\cite{Garcia:2021iag}.\footnote{We used $\rho_{\chi,\rm{max}} \simeq 7.1 \times 10^{-16} \, M_P$  and $a_{\rm{max}}/a_{\rm{end}} \simeq 17.8$.}
In the range of couplings shown in Fig.~\ref{fig:nchiAllsmall}, the maximum value for $\Delta N_{\rm eff}$ corresponds to $\sigma/\lambda\simeq 10^{9/2}$, as computed with {\tt CosmoLattice}. In the Hartree approximation, the limit is stronger, $\Delta N_{\rm eff} \;\lesssim\; 7\times 10^{-3} \left( g(T)  T_{\rm reh}/m_{\phi}\right)^{4/3}$. Using the former, we note that the constraint $T_{\rm reh}/m_{\phi}$, which is necessary for the validity of our results, implies that BBN does not restrict the production of dark $\chi$ radiation since at this scale $2.3< N_{\rm eff}< 3.4$~\cite{Lisi:1999ng,Cyburt:2015mya, Fields:2019pfx}. At the CMB temperature, $\Delta N_{\rm eff}\lesssim 1.5\times 10^{-2} (T_{\rm reh}/m_{\phi})^{4/3}$, and the required sensitivity for $T_{\rm reh}\sim m_{\phi}$ is below the current detectability threshold, and slightly smaller than the projected sensitivity~\cite{Planck:2018vyg,DiValentino:2019dzu,Abazajian:2019eic}. Moreover, the lower reheating temperatures or weaker $\phi$-$\chi$ couplings would render this dark component undetectable in the CMB. For example, when $\sigma/\lambda=1$, the contribution to $\Delta N_{\rm eff}$ is weakened by $\sim 15$ orders of magnitude. We conclude that the strong scalar preheating, followed by the perturbative decay of the inflaton, does not lead to a measurable contribution to the amount of non-photonic relativistic species.

In the weak coupling regime, with $\sigma/\lambda\lesssim 10^{-1/2}$, the PSD of $\chi$ depends on the scalar mass. We compute the contribution to $\Delta N_{\rm eff}$ for pure gravitational production, from the smallest mass compatible with the Ly-$\alpha$ constraint, $m_{\chi}\simeq 3.4\times 10^{-4}\,{\rm eV}$. As we discussed in Section~\ref{sec:puregravity}, saturation of the DM relic abundance fixes $T_{\rm reh}\simeq 34\,{\rm GeV}$. Given that the equation of state in this case is not that of pure radiation, we follow the discussion presented in~\cite{Baldes:2020nuv,Merle:2015oja,Baumholzer:2019twf,Ballesteros:2020adh} and include only the contribution to $N_{\rm eff}$ from the relativistic modes,
\beq\label{eq:deltaneffnonrel}
\Delta N_{\rm eff} \;\simeq\; \frac{8}{7}\left(\frac{T}{T_{\nu}}\right)^4\frac{\rho_{\chi}-m_{\chi}n_{\chi}}{\rho_{\gamma}}\,,
\eeq
where we separated the DM energy density into relativistic and non-relativistic parts. After some algebraic manipulation, it can be shown that in terms of the rescaled comoving momentum $q$, the energy density of relativistic modes can be computed as
\begin{align}\notag
\rho_{\chi}-m_{\chi}n_{\chi} \;=\; &\frac{m_{\phi}^4}{2\pi^2}\left(\frac{a_{\rm end}}{a}\right)^4\\
&\times  \int_0^{\infty}\diff q\,q^2\left[ \sqrt{q^2 + q_R^2}- q_R \right] f_{\chi}(q)\,.
\label{eq:relativisticchi}
\end{align}
Here the relativistic scale $q_R$, mirroring the derivation of Eq.~(\ref{eq:fulla0}), can be written as
\begin{align} \notag
q_R \;&\equiv\; \frac{m_{\chi}a}{m_{\phi}a_{\rm end}}\\
&\simeq\; \left(\frac{392040 g_0^2}{1849\pi^4 g(T)^2}\right)^{1/6}\left( \frac{H_{\rm end}^2 M_P^2}{T_{\rm reh} m_{\phi}^3}\right)^{1/3} \left(\frac{m_{\chi}}{T}\right)\,.
\end{align}
At $T_{\rm CMB}\simeq 0.26\,{\rm eV}$, we find that $q_R\simeq 2\times 10^4$ for $\sigma=0$. Therefore, this large scale has the effect of suppressing the contribution of all the non-relativistic, sub-horizon modes of the phase space distribution. The computation can then be completed using the Boltzmann form of the PSD (\ref{eq:fchiappreh}). Expanding (\ref{eq:deltaneffnonrel}), we obtain
\begin{align} \notag
\Delta N_{\rm eff} \;&\simeq\; \frac{43}{84} \left(\frac{215}{36\pi^4}\right)^{1/3}\left(\frac{g(T)}{g_0} \frac{m_{\phi}^3T_{\rm reh}}{H_{\rm end}^2 M_P^2}\right)^{4/3}\\
&\quad\times \int_0^{\infty}\diff q\,q^2\left[ \sqrt{q^2 + q_R^2}- q_R \right] f_{\chi}(q) \nonumber\\
&\simeq\; 5\times 10^{-33}\, .
\end{align}
Hence, a light DM scalar produced gravitationally during inflation and reheating has no measurable effect on the effective number of degrees of freedom. 

\section{Discussion}
\label{sec:discussion}

\subsection{Comparison with previous work}

The authors of Ref.~\cite{Herring:2019hbe} analyzed the non-adiabatic excitation of a light spectator scalar field during inflation, minimally coupled to gravity and not coupled to $\phi$. To analytically track its evolution, the authors approximated the inflationary phase by using an exact de Sitter era followed by a period of radiation domination after instantaneous reheating. They found a distribution function $f(q)\sim q^{-3}$ at low momentum for a light scalar field, $m_{\chi}\ll m_{\phi}$, which grows due to the tachyonic instability during inflation. Our numerical analysis shows that for all light masses, $m_{\chi}\ll H_{\rm end}\simeq 0.4 m_{\phi}$, down to $m_\chi \sim 10^{-3} H_\text{end}$,\footnote{Results for smaller masses demand more computational power and precision, as the mode functions must be tracked down deep inside the horizon in the quasi-de Sitter era, and far into the reheating epoch due to the oscillatory relaxation shown in Fig.~\ref{fig:PSDgrav_a}.} the phase space distribution at low momentum behaves as $f(q)\sim q^{-3}$ for modes that cross the horizon deep inside the inflation era. Therefore, we find that our numerical results are consistent with the analytical estimate of~\cite{Herring:2019hbe}. The authors of Ref.~\cite{Ling:2021zlj} found a similar scaling for the dark matter number density in the regime of DM masses much lighter than the inflaton mass. 

A key difference between our treatment and Ref.~\cite{Herring:2019hbe} is the reheating epoch. Ref.~\cite{Herring:2019hbe} assumes an instantaneous transition between the quasi-de Sitter era and radiation domination, whereas we account for the delay in the decay of the inflaton. A simplified treatment of instantaneous reheating leads to the need to introduce a UV cutoff in the DM phase space distribution. 
Since this cutoff results in a loss of relativistic modes, it was concluded in Ref.~\cite{Herring:2019hbe} that DM mass as low as $m_{\chi} \sim 10^{-5}\,{\rm eV}$ can saturate the DM relic abundance, while being compatible with the structure formation constraints. On the other hand, our careful computation and analysis of the form and amplitude of the PSD during reheating allows us to determine both constraints without the UV cutoff. We find that the dominant contribution to the dark matter number density comes from the long physical wavelengths with $q<1$, which strongly redshift during preheating resulting in a population of cold DM particles. However, the UV tail of the distribution is affected by the perturbative gravitational production during preheating. This UV tail controls the equation of state parameter, and by extension determines the Lyman-$\alpha$ constraint. For this reason, we find that only $m_{\chi} \gtrsim 0.3\,{\rm meV}$ leads to a sufficiently cold dark relic.

Additionally, in this work we have quantitatively explored the degree at which the excitation of super-horizon modes during inflation is suppressed as the effective coupling $\sigma/\lambda$ is dialed toward unity and beyond. However, we did not discuss the models with a non-minimal coupling to gravity, $\mathcal{L}\supset - (1/2) \xi \chi R^2$, that were discussed in~\cite{Herring:2019hbe}. The presence of this coupling could suppress the non-adiabatic growth of long wavelength modes during inflation. In the conformal limit, with $\xi=1/6$, the adiabaticity of super-horizon modes during inflation is not violated, resulting in a suppressed DM number density production.

On the perturbative side, gravitational production of dark matter with a graviton mediator has been studied in~\cite{Mambrini:2021zpp, Haque:2021mab, Clery:2021bwz, Clery:2022wib, Barman:2021ugy, Barman:2022tzk, Haque:2022kez}, where it was shown that such processes play a dominant role during reheating. Similar processes involving the inflaton condensate scattering to thermal bath particles were studied in~\cite{Clery:2021bwz, Haque:2021mab}. However, we argue here that such a simple perturbative picture of gravitational production of dark matter is insufficient since it does not account for the tachyonic growth of the superhorizon modes. Additionally, a simple perturbative calculation ignores potential isocurvature constraints (see the discussion below). Therefore, to calculate the full dark matter relic abundance arising from the gravitational production one needs to take to adopt the full non-perturbative approach that we considered in Section~\ref{sec:puregravity}. We leave a complete analysis of the gravitational particle production and the superhorizon mode contribution for future work.

\subsection{Isocurvature perturbations}

\paragraph{During inflation.}
As it is commonly encountered in the literature, one typically finds that the presence of a light scalar state during inflation might induce large isocurvature perturbations. Dark matter isocurvature perturbations relative to photon are constrained by \textit{Planck} measurements, given by $ \beta_{\rm{iso}} \; \equiv \; P_S(k_*)/(P_{\zeta}(k_*) + P_S(k_*)) < 0.038$ at 95\% C.L. with the pivot scale $k_*=0.05~\text{Mpc}^{-1}$~\cite{Planck:2018jri}, where $P_{\zeta}(k)$ and $P_{S}(k)$ is the curvature and isocurvature perturbation power spectrum, respectively. This bound constrains the dark matter perturbations to be almost completely adiabatic at $k_*$.

In this work, we assumed that the light scalar field $\chi$ does not obtain a vacuum expectation value during inflation, and it is characterized by the Bunch-Davies initial state which corresponds to vanishing occupation numbers. As extensively argued in~\cite{Herring:2019hbe,Gordon:2000hv}, under the lack of a DM misalignment, the geodesic nature of the inflationary trajectory in the scalar field manifold precludes the growth of isocurvature modes at linear order.\footnote{See~\cite{Bartolo:2001rt} for a related discussion arising from scalar fields during inflation.} Moreover, for large couplings, $\sigma\gg 1$, the growth of DM modes during inflation is negligible due to the large positive effective mass acquired by $\chi$ due to its coupling to the inflaton. As a consequence, under our working assumptions, we do not expect a significant linear isocurvature perturbation generation during the inflationary phase for the entire spectrum of couplings $\sigma / \lambda$ considered in this work.

For small couplings, $\sigma\ll \lambda$, isocurvature could in principle still be induced at second order due to the tachyonic instability of DM modes~\cite{Chung:2004nh}. A straightforward computation of this contribution seemingly leads to an overproduction of isocurvature in the pure gravitational limit if $m_{\chi}\ll H_{\rm end}$~\cite{Ling:2021zlj}. Nevertheless, it has been argued that the consistent regularization of the DM energy density during inflation, which requires the unequivocal identification of the zero-point energy, necessarily requires that the non-linear isocurvature vanishes~\cite{Herring:2019hbe}. In the light of this lack of consensus we postpone the detailed exploration of isocurvature constraints beyond the linear order for future work.

\paragraph{During preheating.} During reheating, the pivot scale $k_\star$ remains superhorizon, reentering the horizon only during the radiation-domination era. This implies that at this scale, the power spectrum fluctuations are only affected by the non-adiabatic pressure component~\cite{Kodama:1984ziu}.

For small couplings $\sigma/\lambda < 1$, dark matter is primarily produced before the end of inflation, and therefore, one should not expect to generate the isocurvature perturbations during reheating~\cite{Gordon:2000hv}. For couplings $1<\sigma/\lambda < 10^3$, DM is produced during the early stage of reheating, but its energy density is always negligible compared to that of the inflaton. Therefore the same conclusion that the isocurvature contribution is negligible applies. For large couplings $\sigma/\lambda > 10^3$, the backreaction on the inflaton condensate can no longer be ignored and the dark matter energy density is comparable to that of the inflaton, as illustrated in Fig.~\ref{fig:fragmentation}. To determine the isocurvature contribution in this regime one would need to perform a detailed numerical analysis, including metric perturbations and configuration-space clumping, which lies beyond the scope of this work.

\subsection{Generality and extent of our results}

 \noindent\textbf{Beyond dark matter.} The results presented in this work show that the production of scalar particles during preheating is generic, and go beyond the dark matter framework. In principle, they can be applied to any light ($m_{\chi}\ll H_{\rm end}$), uncoupled massive spectator scalar for models of inflation with a quadratic minimum.

\noindent
\textbf{Modification of the inflaton potential.}
For any effective coupling strength explored in this paper, we expect the results to be robust under the modification of the inflaton potential away from the minimum, assuming plateau-like inflation. For T-model inflation , $\lambda\propto H_{\rm inf}^2\approx H_{\rm end}^2\approx m_{\phi}^2$, and $\sigma/\lambda \simeq 0.22 \sigma(M_P/H_{\rm end})^2 $. Thus, our findings can be mapped to other plateau-like models (e.g.~Starobinsky-like inflation) by determination of the Hubble scale at the end of inflation. Results in the small coupling regime $\sigma/\lambda <1$ would be qualitatively unchanged, provided that the inflaton sector reproduces the effects of typical slow-roll inflation with a single inflaton field, i.e.,~a quasi-de Sitter phase of constant expansion acceleration.
The non-overclosure constraint $T_{\rm reh}\lesssim 34\,{\rm GeV}$ is expected to remain of the same order. For scenarios in which $H$ changes in value significantly during slow-roll, the efficiency of the tachyonic growth during inflation will be modified, leading to potentially significant deviations from our analysis in the small coupling regime. For $\sigma/\lambda \gtrsim 10^{-1/2}$, DM is produced at the end of inflation or during reheating, and our results would also be robust under the modification of the inflationary sector, as long as the potential possesses a quadratic minimum. A modification of this assumption would alter the expansion history during reheating and affect the production rate~\cite{Garcia:2020wiy}. Moreover, the inflaton is generically more prone to fragmentation for minima flatter than $V\propto \phi^2$~\cite{Lozanov:2016hid,Lozanov:2017hjm,Antusch:2020iyq,Antusch:2021aiw}. 

\noindent
\textbf{Interactions with the visible sector.}
The presence of a direct dark-visible sector coupling, or preheating in the visible sector, could also affect our general results.  In particular, here we do not consider the processes and effects of inflaton-mediated dark-visible sector interactions, see e.g.~\cite{Ghosh:2022hen}.

{\noindent\color{black}{\textbf{Minimum dark matter production.}}} It is worth emphasizing that, under the assumption of a minimal coupling to gravity, there is a minimum amount of scalar particles that are produced before the radiation domination era. The suppression is determined by the size of the interference between the gravitational and direct inflaton-scalar coupling, which is maximal for $\sigma\simeq \lambda$. This minimal abundance could have consequences for other particle production scenarios which assume a vanishing DM population at the beginning of the radiation domination era, such as freeze-in scenarios. 

{\noindent\color{black}{\textbf{Lyman-$\alpha$ and matching procedure.}}} Another important result of this work concerns the analytical matching procedure used to estimate the Lyman-$\alpha$ constraints, adapted from~\cite{Ballesteros:2020adh}. This matching procedure relies on the assumption that higher multipoles $\ell$ in the Legendre expansion of the perturbation of the dark matter PSD can be neglected. This results in an equation of state parameter $w_\chi$ that can be expressed in terms of the second moment of the distribution function. In Ref.~\cite{Ballesteros:2020adh}, this procedure was tested assuming ``smooth'' DM distributions of the form $f_\chi(q)\sim q^\alpha e^{-\beta q^\gamma}$. In this work we have shown that the matching procedure is validated even for non-smooth distributions, i.e.,~distributions featuring peaks over several orders of magnitude, as is the case of PSDs computed in the parametric resonance regime below the backreaction threshold.

{\noindent\color{black}{\textbf{Pivot scale.}}}
We must emphasize the fact that our results are given in terms of the ratio $\sigma / \lambda$. We have specialized for definiteness to a coupling $\lambda \sim \mathcal{O}(10^{-11})$ as given in Eq.~(\ref{eq:lambda}) in order to reproduce Planck measurements for T-models of inflation with $N_*=55$. However, we expect our conclusions to be robust upon the modification of the pivot scale normalization of $\lambda$. 

{\noindent\color{black}{\textbf{Higher dimensional operators.}}}
As our analysis is performed in the range of couplings $\sigma/\lambda \in [0,10^{-5}]$, for the largest coupling considered in this work $\sigma^2 \sim \lambda$ which implies that for larger couplings, higher orders in perturbation theory would become relevant and would affect the inflaton potential.  The role of higher dimensional operators has recently been considered in Ref.~\cite{Lebedev:2022ljz}.

{\noindent\color{black}{\textbf{Dark matter self-interactions.}}}
It is important to note that when deriving our results, we neglected a possible dark matter self-interaction coupling $\mathcal{L}\supset \lambda_\chi \chi^4$. The presence of a sizable coupling $\lambda_\chi $ could drive dark matter to self-thermalize, in which case the phase space distribution would take the form of a Bose-Einstein distribution. If such quartic coupling is larger than $\lambda_\chi \gtrsim\mathcal{O}(10^{-8})$, thermalization could be achieved at late times, when dark matter becomes non-relativistic~\cite{Egana-Ugrinovic:2021gnu}. The corresponding change in the structure formation bounds would require a dedicated treatment, which goes beyond the goals of this work. The requirement for this coupling to be negligible at the tree-level is sufficient for our analysis to remain valid, since the quantum corrections typically would generate smaller values, which can be readily seen from the arguments discussed in the previous paragraph.

\section{Conclusions}
\label{sec:conclusions}

The inflationary paradigm, outside of Higgs-inflation models, relies on the addition of at least one extra scalar degree of freedom to the Standard Model that is needed to correctly set the initial conditions for the hot, thermal universe. Therefore, it is natural to investigate the consequences of the presence of further additional spin-0 fields in the early stages of the universe. In some scenarios, their presence can modify the predicted spectrum of scalar fluctuations via multifield effects~\cite{Gordon:2000hv,Byrnes:2006fr,Easther:2013rva,Starobinsky:2001xq,DiMarco:2002eb,Choi:2007su,Mukhanov:1997fw,Langlois:2008mn,Peterson:2010np,Ellis:2014opa}, allowing a mechanism for the creation of the matter-antimatter asymmetry~\cite{Affleck:1984fy,Dine:1995kz,Kasuya:2006wf,Dutta:2010sg,Marsh:2011ud,Higaki:2012ba,Enqvist:2003gh,Dine:2003ax,Allahverdi:2012ju,Gaillard:1995az,Campbell:1998yi,Gherghetta:1995dv,Garcia:2013bha}, or provide a compelling mechanism for the origin of dark matter through the coherent excitation of such scalars from the quantum fluctuations that are present during inflation~\cite{Marsh:2015xka}. In this work, we have studied the presence of an additional scalar degree of freedom, denoted by $\chi$, which couples minimally to gravity, and only directly to the inflaton field via the four-legged vertex $\propto \phi^2\chi^2$. The absence of other couplings makes $\chi$ stable and therefore a natural dark matter candidate. Due to the absence of a dark matter coupling with the visible sector, constraints and signatures in the late universe mainly come from gravitational interactions and as a result, mainly come from structure formation.
.

The accelerated expansion of the universe during reheating is a ``hard reset'' on the matter and radiation energy densities compared to the Big Bang. The universe is driven into a cold, empty state, and it is only repopulated with Standard Model particles during the coherent oscillation of the inflaton and its simultaneous decay. A natural expectation is to have a dark sector that can couple only to the inflaton that is also populated during reheating. Nevertheless, for scalar dark matter, this need not be the case, and the bulk of the relic abundance can be populated earlier, during the inflationary epoch.

\begin{figure*}[!t]
\centering
    \includegraphics[width=0.95\textwidth]{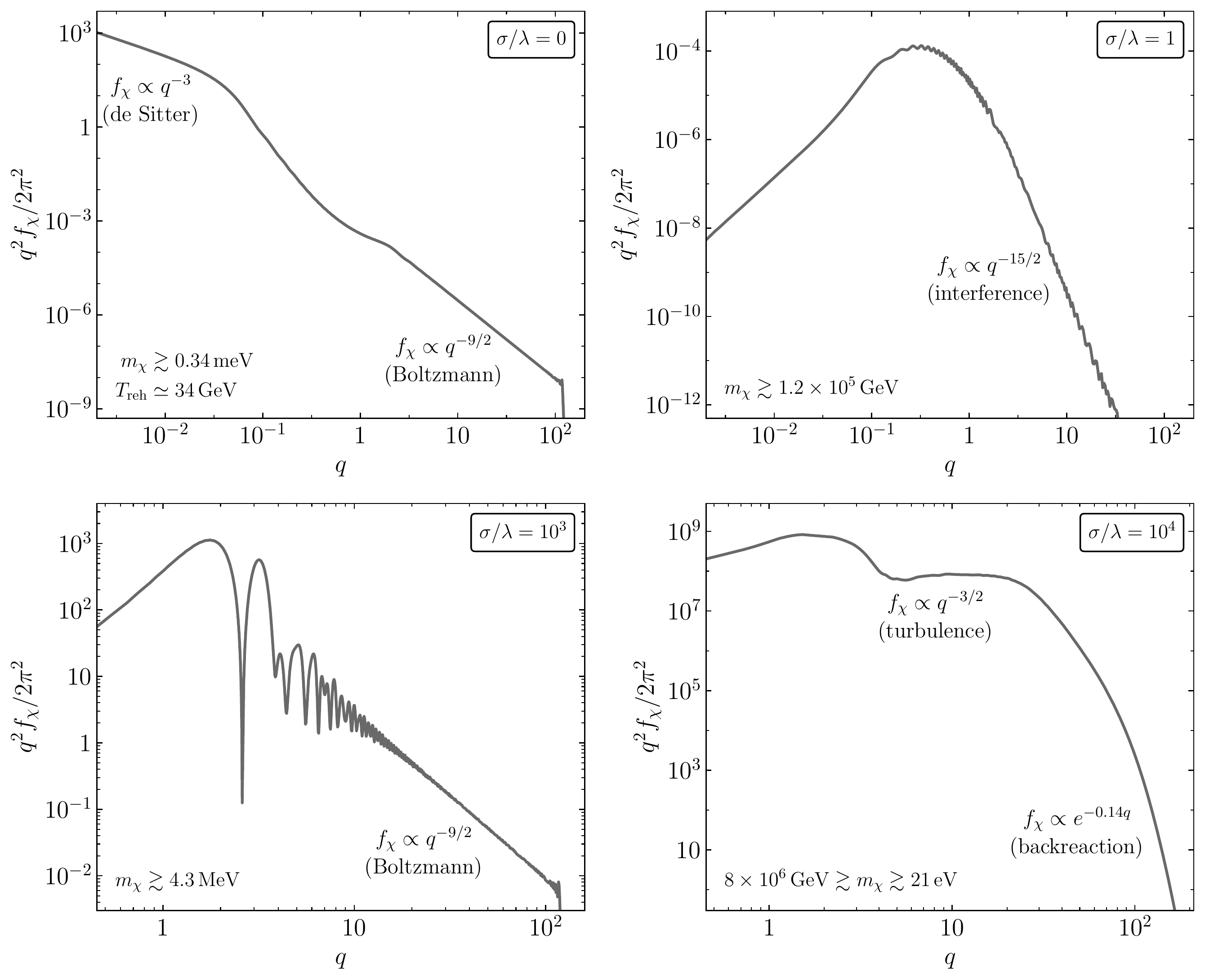}
    \caption{Dark matter phase space distribution for a selection of couplings covering the landscape of regimes considered in this work: gravitational production $\sigma/\lambda=0$ (Sec.~\ref{sec:puregravity}), strong interference  $\sigma/\lambda=1$ (Sec.~\ref{sec:weakcoupling}), broad resonance  $\sigma/\lambda=10^3$ (Sec.~\ref{sec:nobackreaction}), and backreaction  $\sigma/\lambda=10^4$ (Sec.~\ref{sec:backreaction}) from top-left to bottom-right panels. Each panel shows the characteristic scaling of the distribution function at particular $q$ regimes. In addition, for each panel, the Lyman-$\alpha$ constraint on the DM mass is shown. For $\sigma=0$, we also show the only value of the reheating temperature that saturates $\Omega_{\rm DM}$. For $\sigma/\lambda=10^4$, an upper bound on $m_{\chi}$ is also presented, arising from the requirement that $T_{\rm reh}>T_{\rm BBN}$.}
    \label{fig:summary}
\end{figure*}

Depending on the value of the direct coupling, dark matter production is driven by different processes. For each production regime discussed in this work, the main results are summarized below and the corresponding dark matter distribution is illustrated in Fig.~\ref{fig:summary}.

\paragraph{Pure gravitational production.}
In the case of a vanishing inflaton-dark matter coupling, the phase space distribution is primarily populated during inflation. The momentum modes of a scalar with a mass below the Hubble scale experience a tachyonic growth when they leave the horizon. While outside the horizon, these long-wavelength modes cannot be identified as particles, but upon their re-entry during reheating, or the matter/radiation-dominated eras, they will contribute to the non-vanishing momentum population of dark particles. We find that the distribution for modes produced during inflation is red-tilted, and therefore corresponds to the dominant contribution to the dark matter density contrast. Particle production continues during reheating, populating a tail of relativistic particles that are well described by the solution to the perturbative Boltzmann equation. This high-momentum population would provide a sufficiently averaged free-streaming velocity and lead to the suppression of structure formation at small scales. Applying the matching formalism of~\cite{Ballesteros:2020adh}, we find that masses larger than 
\begin{equation}
    m_\chi \, \gtrsim 0.34~\text{meV} \,,~~~ [\sigma/\lambda \ll 1]
\end{equation}
are consistent with the Lyman-$\alpha$ constraint on the amplitude of the matter power spectrum and that the observed relic abundance is saturated only if the reheating temperature is $\sim 34\,{\rm GeV}$.

\paragraph{Weak coupling.}
Increasing the inflaton-dark matter coupling, we find that for $\sigma/\lambda<1$, the production of dark relics is dominated by the gravitational effects and suppressed by the direct coupling due to an interference effect in the amplitudes. Up to $\sigma/\lambda\lesssim 10^{-1/2}$, the superhorizon long-wavelength modes dominate the momentum distribution. Therefore, small dark matter masses between 
\begin{equation}
 0.3\,{\rm meV} \lesssim  \,  m_\chi \, \lesssim 30\,{\rm eV} \,,~~~~ [\sigma/\lambda\lesssim 10^{-1/2}]
\end{equation}
are allowed by the structure formation constraints. Saturation of $\Omega_{\rm DM}$ occurs only for relatively small values of $T_{\rm reh}$ below the electroweak scale for couplings smaller than $10^{-2}$. For larger couplings, the abundance may be saturated for temperatures between the electroweak scale and the inflationary scale. 

\paragraph{Strong interference.}
A minimum in the dark matter production rate occurs for $\sigma/\lambda\simeq 1$. For this value, the interference between gravity and the four-legged coupling is maximal. The Boltzmann equation is not capable of adequately approximating the phase space distribution, and non-perturbative methods are necessary. In this case, most dark matter particles are created at the end of inflation and the smallest reheating temperature that can lead to the measured density parameter is given by $T_{\rm reh}\simeq 10^5\,{\rm GeV}$. The smallest mass for which our assumptions about the mechanism of reheating are valid, and which saturates $\Omega_{\rm DM}$, coincidentally corresponds to $m_{\chi}\simeq 10^5\,{\rm GeV}$.

\paragraph{Strong coupling, no backreaction.}
For couplings $1\lesssim \sigma/\lambda\lesssim 5\times 10^3$, we find that most of the dark matter is produced during the first oscillations of the inflaton during reheating. Parametric resonance becomes the dominant particle production process at this stage. This resonant growth eventually stops after a few inflaton oscillations, and particle production continues through the perturbative decay of the inflaton. In this regime, we find that the saturation of the relic abundance also guarantees a cold-like matter power spectrum. A wide range of masses
\begin{equation}
    m_\chi \, \gtrsim ~10^6 \,\left( \dfrac{\sigma}{\lambda} \right)^{-7/3} {\rm GeV}\,,~~~ [1\lesssim \sigma/\lambda\lesssim 5\times 10^3]
\end{equation}
is allowed, dependent on the strength of the coupling. 

\paragraph{Strong coupling, backreaction.}
Finally, in the range of strong couplings, $\sigma/\lambda\gtrsim 5\times 10^{3}$, the picture of inflaton decaying into dark matter particles needs to be modified. The interaction leads to rescattering effects, which deplete the homogeneous inflaton condensate in favor of free inflaton particles. The result is a distribution that is populated first through the parametric resonance, and later through the non-linear rescattering, when the energy densities of the inflaton and dark sectors become comparable. The final outcome is a distribution that lacks the perturbative Boltzmann UV tail, and instead exhibits a quasi-thermal tail. The saturation of the dark matter relic abundance is possible with temperatures as high as the inflaton mass scale, and as low as the lower bound imposed by primordial nucleosynthesis. On the other hand, the Lyman-$\alpha$ constraint requires
masses 
\begin{equation}
 21~\text{eV} \lesssim  \,  m_\chi \, \lesssim ~8 \times 10^5 \, {\rm GeV}\,,~~~ [\sigma/\lambda > 5\times 10^3]
\end{equation}
to avoid the overclosure of the universe.\par\medskip

One of our main conclusions is the following: except for a narrow domain in the range of the effective coupling, centered at $\sigma/\lambda\simeq 10$, the Boltzmann approximation always fails at predicting the correct relic abundance of scalar dark matter coupled only to the inflaton. Implementing the Bose-enhancement effects in the perturbative Boltzmann equation results in a relic abundance far exceeding the value predicted by the non-perturbative analysis. In the pure gravitational regime, the perturbative analysis cannot account for the excitation of the superhorizon modes during inflation and therefore underestimates by at least several orders of magnitude the total amount of dark matter particles that are produced.

Given that our estimate of scalar production is not limited to the non-relativistic dark particle case, we have also explored the possibility of identifying $\chi$ with a dark radiation component. We find that, unless the reheating temperature is at the upper perturbativity bound, and the scalar is produced in the presence of strong backreaction, the impact on the effective number of non-photonic relativistic degrees of freedom is beyond our current experimental sensitivity.

Shortly after the completion of this work, another paper conducting a similar analysis appeared, that partially overlaps with our analysis of gravitational
particle production~\cite{Kaneta:2022gug}. The authors found similar conclusions and results using the Bogoliubov approach, and they found analytically that for the pure gravitationaly production, the PSD at low momentum behaves as~$f(q)\sim q^{-3}$, whereas at high momentum it scales as $f(q)\sim q^{-9/2}$, which agrees with our numerical and analytical analysis.

\begin{acknowledgments}

We would like to thank Mustafa Amin, Yohei Ema, Daniel G. Figueroa, and Adrien Florio for helpful discussions. MP acknowledges support by the Deutsche Forschungsgemeinschaft (DFG, German Research Foundation) under Germany's Excellence Strategy – EXC 2121 “Quantum Universe” – 390833306. MG acknowledges the support from the Instituto de F\'isica, UNAM in procuring computational resources. SV is supported by the University of Minnesota Doctoral Dissertation Fellowship. This work was made possible by with the support of the Institut Pascal at Université Paris-Saclay during the Paris-Saclay Astroparticle Symposium 2021, with the support of the P2IO Laboratory of Excellence (program “Investissements d’avenir” ANR-11-IDEX-0003-01 Paris-Saclay and ANR-10-LABX-0038), the P2I axis of the Graduate School Physics of Université Paris-Saclay, as well as IJCLab, CEA, IPhT, APPEC, the IN2P3 master projet UCMN and EuCAPT ANR-11-IDEX-0003-01 Paris-Saclay and ANR-10-LABX-0038). Non-perturbative numerical results in the Hartree approximation were obtained from a custom Fortran code utilizing the thread-safe arbitrary precision package MPFUN-For~\cite{mpfun}.
\end{acknowledgments}

\appendix

\section{Perturbative dark matter production}
\label{app:A}
In this Appendix, we discuss the perturbative production of scalar dark matter. We first perform the calculation by treating the inflaton as a decaying condensate~\cite{Nurmi:2015ema, Garcia:2020wiy} and considering dark matter production through inflaton scattering induced by gravity-mediated $s$-channel diagrams and via direct quartic coupling~\cite{Mambrini:2021zpp, Clery:2021bwz}. In a second approach, we compute the resulting dark matter production by treating the inflaton as a collection of vanishing-momentum quanta. We show that both approaches lead to identical results. The total action can be expressed as
\begin{equation}
    \mathcal{S} \; = \; \mathcal{S}_\text{EH} + \mathcal{S}_\phi + \mathcal{S}_\chi \,,
    \label{eq:actiontot}
\end{equation}
where the Einstein-Hilbert action is given by
\begin{equation}
    \mathcal{S}_\text{EH} \; = \; - \frac{1}{2}\int \diff^4 x \sqrt{-g} M_P^2 R \, ,
    \label{eq:actioneh}
\end{equation}
and $\mathcal{S}_\phi $ and $\mathcal{S}_\chi$ are given by Eqs.~(\ref{eq:actionphi}) and~(\ref{eq:actionchi}), respectively. One can expand the space-time metric by considering a small deviation with respect to the Minkowski metric $g_{\mu \nu} \simeq \eta_{\mu \nu}+2 h_{\mu \nu}/M_P$, where $h_{\mu \nu}$ is the canonically-normalized perturbation, and from Eq.~(\ref{eq:actiontot}) we find the following interaction terms 
\begin{equation}
    \label{eq:intLag}
    \mathcal{L}_I\,=\,-\dfrac{1}{M_P} h_{\mu \nu}\Big( T^{\mu \nu}_{\phi} +T^{\mu \nu}_{\chi}   \Big) - \dfrac{\sigma }{2} \phi^2 \chi^2\, .
\end{equation}
$T^{\mu \nu}_{\phi}$ is the energy-momentum tensor of the inflaton, given by Eqs.~(\ref{app:infcond1}) and (\ref{app:infpart1}) when the inflaton is treated as a condensate and a collection of particles, respectively, and $T^{\mu \nu}_{\chi}$ is
\begin{equation}
    \label{app:enmomchi1}
    T^{\mu \nu}_{\chi} \; = \; \partial^{\mu} \chi \partial^{\nu} \chi - g^{\mu \nu} \left[\frac{1}{2} \partial^{\alpha}\chi \partial_{\alpha} \chi - \frac{1}{2} m_{\rm{eff}}^2 \chi^2 \right] \, ,
\end{equation}
where $m_{\rm{eff}}$ is the effective dark matter mass given by Eq.~(\ref{eq:effectiveDMmass})
that accounts for mass contribution induced by the oscillating inflaton. 
The field $h_{\mu \nu}$ describes a massless spin-2 graviton whose propagator carrying momentum $k$ in de Donder gauge is~\cite{Giudice:1998ck}
\begin{equation}
    \label{app:gravprop}
    \Pi^{\mu \nu \rho \sigma }(k)\,=\, \dfrac{\eta^{\rho \nu} \eta^{\sigma \mu } + \eta^{\rho \mu} \eta^{\sigma \nu } - \eta^{\rho \sigma} \eta^{\mu \nu }   }{2k^2}\,.
\end{equation}
To calculate the matrix element corresponding to the pair production of scalar dark matter quanta, described by the final state $| f\rangle= |\chi \chi \rangle$, we can treat the inflaton either as an oscillating condensate that decays to dark matter or as a collection of particles within a condensate that lead to dark matter production through an $s$-wave scattering process. We discuss two different approaches in the following subsections.

\subsection{Inflaton treated as a condensate}
At the beginning of reheating, the inflaton oscillates about a quadratic minimum and behaves like matter.\footnote{Here we discuss only the quadratic case, and more general models with quartic or higher minima are discussed in~\cite{Garcia:2020wiy}.} The energy density of the inflaton is governed by the Friedmann-Boltzmann equation~(\ref{eq:phidecay1}), and the right-hand side of this equation characterizes the energy transfer per space-time volume ($\rm{Vol_4}$),
\begin{equation}
    \label{app:gamma1}
    \Gamma_{\phi}\rho_{\phi} \; \equiv \; \frac{\Delta E}{\rm{Vol_4}} \, ,
\end{equation}
with
\begin{align} \notag
    \Delta E \; \equiv \; \int &\frac{\diff^3{\bf{p}}_A}{(2\pi)^3 2p_A^0} \frac{\diff^3{\bf{p}}_B}{(2\pi)^3 2p_B^0} \left(p_A^0 + p_B^0 \right)\\   \label{app:deltaE}
&\times  |\langle  f|\exp \left( i \int \diff^4 x \, \mathcal{L}_I \right)  |0 \rangle|^2 \, ,
\end{align}
where $\langle f | = \langle A, B|$ is the final two-particle state, $|i\rangle=|0\rangle$ is the initial state assumed to be a vaccuum state, and $\mathcal{L}_I$ is the interaction Lagrangian. One can express the transition amplitude squared from the inflaton condensate to the final state $|A, B \rangle$ by summing over each inflaton oscillation mode,
\begin{align} \notag
    &|\langle f |\exp \left( i \int \diff^4 x \, \mathcal{L}_I \right)  |0 \rangle|^2 \\
&\qquad =\, {\rm{Vol}_4} \sum_{n=-\infty}^\infty |\mathcal{M}_n|^2 (2 \pi)^4 \delta^{(4)}(p_n -p_A-p_B)\,.
\end{align}
Here $|\mathcal{M}_n|^2$ is the transition amplitude squared of the $n$-th oscillation mode and $p_n = \left(E_n, \bf{0} \right)$ is the inflaton four-momentum, where $E_n = n \, m_{\phi}$ denotes the energy of the oscillation mode. If we combine this expression with Eqs.~(\ref{app:gamma1}) and (\ref{app:deltaE}), we obtain~\cite{Kainulainen:2016vzv, Ichikawa:2008ne}
\begin{equation}
    \label{app:gammaphi1}
    \Gamma_{\phi} \; = \; \frac{1}{8\pi \rho_{\phi}} \sum_{n=1}^{\infty} |\mathcal{M}_n|^2 E_n \beta_n \left(m_A, m_B \right) \, ,
\end{equation}
with the kinematic factor
\begin{align} \notag
    &\beta_n \left(m_A, m_B \right) \\
&\qquad \equiv \; \sqrt{\left(1 - \frac{(m_A + m_B)^2}{E_n^2} \right)\left(1 - \frac{(m_A - m_B)^2}{E_n^2} \right)} \, .
\end{align}
The creation of scalar dark matter particles $\chi$ from the coherent inflaton oscillation modes can be derived by treating the inflaton as a classical field (perfect fluid). In this case, the inflaton energy-momentum tensor is given by
\begin{align}
    \label{app:infcond1}
    T^{\mu \nu}_{\phi} \, &= \,  (\rho_\phi+P_\phi)u^\mu u^\nu - P_\phi g^{\mu \nu} \nonumber \\
    &= \,  \left[2 \frac{(1+w_{\phi})}{(1-w_{\phi})} u^{\mu} u^{\nu} - \frac{2w_{\phi}}{1 - w_{\phi}} g^{\mu \nu}\right]V(\phi)\,,
\end{align}
with $u^{\mu} u_{\mu} \; = \; -1$, where $u_\mu = (1,\, 0,\, 0,\, 0)$ is the inflaton four-velocity, and in the second line we expressed the energy-momentum tensor as a function of the scalar potential $V(\phi)$. 
 The transition amplitude is calculated using the interaction Lagrangian~(\ref{eq:intLag}) together with Eqs.~(\ref{app:enmomchi1}) and~(\ref{app:infcond1}):
 \begin{align} \notag
     \mathcal{M} \; &= \; \sigma \phi^2 - \frac{1}{M_P^2} \left[1 + \frac{2m_{\rm{eff}}^2}{s} \right] V(\phi)\\
&=\; \phi^2 \left(\sigma - \lambda \left[1 + \frac{2m_{\rm{eff}}^2}{s} \right] \right) \, .
 \end{align}
Note that this expression does not depend on the equation of state parameter, $w_{\phi}$. With $\mathcal{M} \propto \phi^2(t) = \phi^2_0(t) \cdot \mathcal{P}^2(t)$, where we used Eq.~(\ref{eq:phioft}), the periodic function $\mathcal{P}(t)$ that characterizes rapid inflaton oscillations can be expanded in terms of the Fourier series as
\beq
\mathcal{P}(t) \;=\; \sum_{n=-\infty}^{\infty} \mathcal{P}_n e^{-in\omega_{\phi} t}
\eeq
and
\beq
\mathcal{P}^2(t) \;=\; \sum_{n=-\infty}^{\infty} \hat{\mathcal{P}}_n e^{-in\omega_{\phi} t} \, ,
\eeq
where for a quadratic minimum, $\mathcal{P}(t) = \cos (m_{\phi} t)$ and $\mathcal{P}^2(t) = \cos^2(m_{\phi} t)$, with the angular frequency of oscillation given by $\omega_{\phi} = m_{\phi}$. One can readily show that only the second mode $n = 2$ contributes to the Fourier series of $\mathcal{P}^2(t)$, with $\hat{\mathcal{P}}_2=1/4$, and the corresponding matrix element is given by
\begin{equation}
   \mathcal{M}_2\,=\, \phi_0^2 \hat{\mathcal{P}}_2\left[ \sigma - \lambda \left(1+ \dfrac{m_{\rm{eff}}^2}{2 m_\phi^2} \right)  \right]  \,,
   \label{eq:ampsq_condensate}
\end{equation}
where we used $\sqrt{s} = E_n = n \, m_{\phi}$. If we take into account the symmetry factor associated with two identical final states,\footnote{We clarify that there is no symmetry factor associated with the initial vacuum state.} the transition amplitude squared can be expressed as
\begin{align} \notag
    |\overline{\mathcal{M}_2}|^2 \,&=\, \frac{1}{2} \times \dfrac{\phi_0^4}{16} \left[ \sigma - \lambda \left(1+ \dfrac{m_{\rm{eff}}^2}{2 m_\phi^2} \right)  \right]^2  \\
    &\equiv \dfrac{\phi_0^4}{32} \hat \sigma^2 \,= \, \dfrac{1}{8} \dfrac{\rho_\phi^2}{m_\phi^4}  \hat \sigma^2  \,,
    \label{eq:M2squarred}
\end{align}
where we used the mean inflaton energy density (\ref{eq:aveosc}) and introduced the effective coupling
\begin{equation}
    \label{eq:sigmahat}
    \hat \sigma \, \equiv \,   \sigma - \lambda\left(1+ \dfrac{m_{\rm{eff}}^2}{2 m_\phi^2} \right) \, .
\end{equation}
Notice that the relative minus sign leads to destructive interference at tree level.\footnote{Similar models with destructive interference at tree level were recently discussed in the context of gravitational particle production with non-minimal couplings to gravity~\cite{Clery:2022wib}.} 

One can combine the transition amplitude squared~(\ref{eq:M2squarred}) with Eq.~(\ref{app:gammaphi1}), and find
 \begin{equation}   
     \label{app:gammaphi2}
     \Gamma_{\phi \rightarrow \chi \chi} \; = \; \frac{\rho_{\phi}}{m_{\phi}} \frac{\hat{\sigma}^2}{32\pi m_{\phi}^2} \sqrt{1 - \frac{m_{\rm{eff}}^2}{m_{\phi}^2}}\, .
 \end{equation}

\subsubsection{Production rate}
We first calculate the phase space distribution of the produced dark matter scalar field, $f_{\chi}$. Disregarding the Bose enhancement effects for the dark matter products $\chi$ and assuming that there are no backreaction effects that would produce inflatons in the condensate, the Boltzmann transport equation~(\ref{eq:boltzfull}) becomes:
\begin{widetext}
\begin{align}
    \frac{\partial f_{\chi}}{\partial t} - H|\bP|\frac{\partial f_{\chi}}{\partial |\bP|}  = \; &\frac{1}{P^0} \sum_{n=1}^{\infty} \int \frac{\diff ^3 \bK}{(2\pi)^3 n_{\phi}} \frac{\diff^3 \bP'}{(2\pi)^32P^{\prime 0}} (2\pi)^4 \delta^{(4)}(K_n-P-P')|\overline{\mathcal{M}_n}|^2 f_{\phi}(K) \, ,
\end{align}
\end{widetext}
where $K_n = \left(n \, m_{\phi}, \bf{0} \right)$. If we use the inflaton phase space distribution~(\ref{eq:psdinflaton}) and integrate over $\diff^3 \bP/(2 \pi)^3 $, we find the Boltzmann equation for the number density:
\begin{widetext}
\begin{equation}
    \dfrac{\diff n_\chi}{\diff t}+3 H n_\chi \, = \, 2 \sum_{n=1}^\infty \int  \dfrac{\diff^3 \bP }{(2 \pi)^3 2 P^0 } \dfrac{\diff^3 \bP^\prime }{(2 \pi)^3 2 P^{ \prime 0} }(2\pi)^4 \delta^{(4)}(K_n-P-P^\prime){|\overline{\mathcal{M}_n}|^2} \, \equiv R(t) \,,
\end{equation}
\end{widetext}
where $R(t)$ is the time-dependent production rate (per unit volume per unit time). Integration over the final states leads to the following rate:
\begin{align}   \notag
R(t) \,&=\, \dfrac{1}{4 \pi} \sum_{n=1}^\infty |\overline{\mathcal{M}_n}|^2 \beta_n \\ \notag &= \dfrac{|\overline{\mathcal{M}_2}|^2}{4 \pi}   \sqrt{1 - \frac{m_{\chi}^2}{m_{\phi}^2}} \\ \label{app:rate1}
& = \;  \frac{\rho_{\phi}^2}{m_{\phi}^4} \frac{\hat{\sigma}^2}{32\pi} \sqrt{1 - \frac{m_{\rm{eff}}^2}{m_{\phi}^2}}  
\,,
\end{align}
where we used Eq.~(\ref{eq:M2squarred}). Since the inflaton energy density can be expressed as $\rho_{\phi} \simeq m_{\phi} n_{\phi}$, which implies that $R(t) = \frac{\rho_{\phi}}{m_{\phi}} \Gamma_{\phi \rightarrow \chi \chi}$, using Eqs.~(\ref{app:gammaphi2}) and~(\ref{app:rate1}) we see that this relationship is satisfied and the results are consistent.

\subsection{Inflaton treated as a collection of particles}
We repeat the same exercise and calculate the dark matter production rate assuming that dark matter is produced via an $s$-channel graviton exchange. If we treat the inflaton as a collection of particles, its stress-energy tensor is given by
\begin{equation}
    \label{app:infpart1}
    T^{\mu \nu}_{\phi} \; = \; \partial^{\mu} \phi \partial^{\nu} \phi - g^{\mu \nu} \left[\frac{1}{2} \partial^{\alpha}\phi \partial_{\alpha} \phi - V(\phi) \right] \, .
\end{equation}
The scattering process $\phi(p_1) + \phi(p_2) \rightarrow \chi(p_3) + \chi(p_4)$ can be parametrized by\footnote{Similar $s$-scattering processes with a graviton mediator involving production of scalar dark matter were considered in~\cite{Bernal:2018qlk}.}
\begin{equation}
    \mathcal{M}_{\phi \phi \rightarrow \chi \chi} \propto M_{\mu \nu}^{\chi} \Pi^{\mu \nu \rho \sigma} M_{\rho \sigma}^{\phi} \, ,
\end{equation}
where the graviton propagator is given by~(\ref{app:gravprop}), and the partial amplitudes arising from the energy-momentum tensors~(\ref{app:enmomchi1}) and (\ref{app:infpart1}) can be expressed as
\begin{align} \notag
    &M^{\phi, \, \chi }_{\mu \, \nu} \\ 
    &\quad = \; \frac{1}{2} \left[p_{1 \mu} p_{2 \nu} + p_{1 \nu} p_{2 \mu} - \eta_{\mu \nu} p_1 \cdot p_2 - \eta_{\mu \nu} m_{\phi, \, {\rm{eff}}}^2 \right] \, .
\end{align}
The transition amplitude squared for the process $\phi \phi \rightarrow \chi \chi$, with the center-of-mass energy $s=4m_\phi^2$,\footnote{Since in the scattering rate we average over non-thermal inflaton phase space distribution~(\ref{eq:psdinflaton}), the average momentum of the collection of particles within a condensate is zero.} is given by
\begin{equation}
   |\overline{\mathcal{M}}|_{\phi \phi \rightarrow \chi \chi}^2\, = \,  \frac{1}{2} \hat{\sigma}^2 \,,
\end{equation}
where we included the symmetry factors associated with identical final states, and the effective coupling $\hat{\sigma}$ is given by Eq.~(\ref{eq:sigmahat}). The produced dark matter number density can be calculated from the following equation:
\begin{widetext}
\begin{equation}
    \dfrac{\diff n_\chi}{\diff t}+3 H n_\chi \, = \, 2 \int  \diff \Pi_1 \diff \Pi_2 \diff \Pi_3 \diff \Pi_4 (2\pi)^4 \delta^{(4)}(p_1 + p_2 - p_3 - p_4) |\overline{\mathcal{M}}|_{\phi \phi \rightarrow \chi \chi}^2 f_\phi(p_1,t) f_\phi(p_2, t) \, \equiv R(t) \,,
\end{equation}
\end{widetext}
where $\diff\Pi_i \equiv \frac{\diff^3 \bf{p_i}}{(2\pi)^3 2E_i}$, and if we use the inflaton phase space distribution~(\ref{eq:psdinflaton}), we recover the production rate~(\ref{app:rate1}).

\bibliography{references}

\end{document}